\newif\ifhighlightedits
\highlighteditstrue

\documentclass[aip,cha,
reprint,
secnumarabic,
nofootinbib, tightenlines,
nobibnotes, showkeys, showpacs,
superscriptaddress,
]{revtex4-1}
\usepackage{float}
\usepackage[utf8]{inputenc}
\usepackage{color}
\usepackage{amsmath}
\usepackage{amsfonts}
\usepackage{amssymb}
\usepackage{graphicx}
\usepackage{algpseudocode}
\usepackage{amsbsy}
\usepackage{hyperref}
\usepackage{listings} 
\usepackage[percent]{overpic}
\usepackage{array}

\newcommand{\rpos}{\ensuremath{r}}
\newcommand{\pos}{\mathbf{\rpos}}
\newcommand{\ssp}{\ensuremath{a}}
\newcommand{\sspRed}{\ensuremath{\hat{\ssp}}}
\newcommand{\slicep}{\ensuremath{\hat{\ssp}'}}
\newcommand{\velRed}{\ensuremath{\hat{\vel}}}    
\newcommand{\StabMatRed}{\ensuremath{\hat{\StabMat}}}
\newcommand{\LieEl}{\ensuremath{g}}  
\newcommand{\Lg}{\ensuremath{\mathcal{T}}}  
\newcommand{\groupTan}{\ensuremath{t}}    
\newcommand{\sliceTan}{\ensuremath{t'}}    
\newcommand{\zeit}{\ensuremath{\tau}}
\newcommand{\slicePhase}{\ensuremath{\hat{\phi}}}
\newcommand{\inprod}[2]{\left\langle #1 ,\, #2 \right\rangle}

\newcommand\flowRed[2]{{\hat{f}^{#1}(#2)}}

\newcommand{\vel}{\ensuremath{v}}
\newcommand{\StabMat}{\ensuremath{A}}
\newcommand{\beq}{\begin{equation}}
\newcommand{\continue}{\nonumber \\ }

\newcommand{\eeq}{\end{equation}}
\newcommand{\ee}[1] {\label{#1} \end{equation}}

\newcommand{\bea}{\begin{eqnarray}}
\newcommand{\eea}{\end{eqnarray}}
\newcommand{\barr}{\begin{array}}
\newcommand{\earr}{\end{array}}

\renewcommand\Im{\ensuremath{{\rm Im}\,}}
\renewcommand\Re{\ensuremath{{\rm Re}\,}}
\newcommand{\memory}{\textit{Me}}

\newcommand\REQ[1]{\ensuremath{{\rm REQ{#1}}}}
\newcommand\RPO[1]{\ensuremath{{\rm RPO{#1}}}}
\newcommand\PO[1]{\ensuremath{{\rm PO{#1}}}}
\newcommand{\CHAOS}[1]{\ensuremath{{\rm C{#1}}}}

\newcommand{\etal}{{\em et al.}}    
\newcommand{\ie}{{i.e.}}        

\newcommand{\rf}     [1] {~\cite{#1}}

\newcommand{\refref} [1] {ref.~\cite{#1}}

\newcommand{\refeq}  [1] {(\ref{#1})}
\newcommand{\refeqs} [2]{(\ref{#1}--\ref{#2})}

\newcommand{\reffig} [1] {figure~\ref{#1}}

\newcommand{\refFig} [1] {Figure~\ref{#1}}

\newcommand{\reftab} [1] {table~\ref{#1}}

\newcommand{\refsect}[1] {section~\ref{#1}}

\newcommand{\refappe}[1] {appendix~\ref{#1}}

\newcommand{\refAppe}[1] {Appendix~\ref{#1}}

\newcommand\flow[2]{{f^{#1}(#2)}}

\newcommand{\PoincS}{{\cal P}}     

\begin{document}

\title[]{State space geometry of the chaotic pilot-wave hydrodynamics}

\author{Nazmi Burak Budanur}
\affiliation{Nonlinear Dynamics and Turbulence Group,
             IST Austria,
             3400 Klosterneuburg, Austria}
\email{burak.budanur@ist.ac.at}                
\author{Marc Fleury}%
\affiliation{
	3344 Peachtree Rd,
	Atlanta, GA 30326
}

\date{\today}

\begin{abstract}

We consider the motion of a droplet bouncing on a vibrating bath of the 
same fluid in the presence of a central potential. We formulate a 
rotation symmetry-reduced description of this system, which allows for 
the straightforward application of dynamical systems theory tools. 
As an illustration of the utility of the symmetry reduction, we apply it to a model of 
the pilot-wave system with a central harmonic force. We begin our analysis 
by identifying local bifurcations and the onset of chaos. 
We then describe the emergence of chaotic regions and their merging 
bifurcations, which lead to the formation of a global attractor. In this final 
regime, the droplet's angular momentum spontaneously changes its sign as observed in
the experiments of Perrard \etal\ 
(\textit{Phys. Rev. Lett.}, 113(10):104101, 2014).

\end{abstract}

\maketitle

\begin{quotation}
	During the quantum physics' infancy, Louis de Broglie\rf{deBroglie1}  
	imagined it as the 
	outcome of the dynamics of a point-like particle that is
	interacting with a continuous background 
	field. This viewpoint was to a large extent forgotten 
	during the second half of the twentieth century due to the great success of
	the Copenhagen interpretation. In recent years, a resurgent 
	interest in de Broigle's ``wave-particle duality'' has developed as a result 
	of its discovery in a completely different field: fluid mechanics.
	In a series of experiments pioneered by Couder \etal \rf{CFGB2005,CPFB2005},
	several phenomena that were once thought to be exclusive to quantum  
	physics were demonstrated in the macroscopic setting of a droplet of 
	silicon oil bouncing on vertically vibrating bath of the same fluid. 
	We examine a class of these systems with rotational symmetry and formulate
	a novel method for their analysis. We demonstrate the utility of our approach 
	in a numerical study where we describe the chaotic dynamics of a hydrodynamic
	pilot-wave model through a geometrical approach, in which we identify 
	solutions with qualitative differences and intermittent transitions in 
	between. 
\end{quotation}

\section{Introduction}

In 2005,
Couder \etal \rf{CFGB2005} showed that a droplet of silicon oil can 
indefinitely bounce on the bath of the same fluid when the bath is 
vibrating vertically with an acceleration close to but below the 
Faraday instability point, at which surface waves appear spontaneously.
In the same year, Couder \etal \rf{CPFB2005} 
also reported that a transition from ``bouncing'' to 
``walking'' takes place when the bath's oscillation amplitude is
large enough.
In 
this latter state, the droplet becomes a ``walker'', and 
the system exhibits a macroscopic ``wave-particle duality'':
The force that the droplet experiences at each bounce is determined 
by the shape of the bath's surface, which itself is determined by the waves 
generated at the droplet's previous bounces. A year later, 
Couder and Fort reported\rf{CouderFort2006} 
that when these walkers were realized in single- or double-slit 
geometries, their final positions obeyed a 
statistical distribution akin to the single-particle diffraction 
experiments of quantum particles\footnote{
      The single- and double-slit experiments with walkers
	  were revisited by Andersen \etal \rf{AMRRLT2015} and Pucci \etal
	  \rf{PHFB2018} and the outcomes of these experiments appear to be 
  	  very sensitive to the experimental conditions.
}. 

Early experiments with bouncing droplets were followed by 
others that demonstrated various analogies to the
quantum mechanics\rf{Bush2015}. 
Fort \etal \rf{FEBMJC2010} and Harris and Bush\rf{HarrisBush2014} 
studied
walkers in rotating frames and
reported quantization 
of droplet trajectories. Similarly quantized orbits were also 
found\rf{PLMFC2014} when the two-dimensional pilot-wave system was 
realized with a central harmonic potential, which was generated 
by applying magnetic field on a droplet filled with a ferromagnetic 
fluid. Furthermore, the dynamics of bouncing droplets under 
						  central harmonic force
was shown to become chaotic\rf{PLFC2014}. 
These systems are of particular interest to us
due to their rotation symmetry:
Assuming the bath's 
walls are of circular 
shape (or far away), 
governing laws of motion preserve their shape under the rotation of 
coordinates. 
As we shall illustrate in this paper, the rotation symmetry 
translates to a redundant degree of freedom in the system's state 
space.

Experimental developments in bouncing droplets were 
accompanied by modeling efforts. Couder \etal \rf{CPFB2005} already predicted
the transition from bouncing to walking as a pitchfork bifurcation by
postulating a sinusoidal force to be exerted on the droplet by the 
bath during the bounces. Through a detailed analysis, 
Oza \etal\ \rf{ORB2013} reached an integro-differential equation of 
motion for the bouncing droplets and coined the term 
``pilot-wave hydrodynamics''. Numerical investigations of this 
model with confining central-potential terms 
(Coriolis, harmonic, or Coulomb type) demonstrated different 
routes to chaos\rf{THORB2016} and rich subsequent dynamics. Similar 
to their experimental counterparts, these systems also exhibit 
rotation symmetry and are at the focus of the current paper. 

    The references that we cited so far are the key developments 
	in the pilot-wave
hydrodynamics literature which are relevant to our work. 
For a thorough 
review, we refer the reader to the recent article by 
Bush\rf{Bush2015}. We are now going to turn our attention to a 
different branch of literature, where chaotic dynamics take place
in the presence of continuous symmetries.

In recent years, 
numerical studies of nonlinear partial differential equations 
such as 
the complex Ginzburg--Landau equation\rf{Luce95}, 
the Kuramoto--Sivashinsky equation\rf{SCD07,BudCvi15},
and the Navier--Stokes equations\rf{ACHKW11,WFSBC15},
demonstrated that degeneracies due to the continuous symmetries 
in these systems tend to obscure their dynamics. Take, for example, 
the fluid flow through a cylindrical pipe. This system is symmetric 
under axial translations and azimuthal rotations. 
Each generic (non-symmetric) 
solution of pipe flow has 
infinitely many copies that can be obtained by continuous symmetry 
transformations, all of which correspond to the same 
physical solution. Budanur \etal\rf{BudCvi14} showed that a simple 
polar-coordinate transformation in the Fourier space representation
of these systems can be used to eliminate these redundant degrees 
of freedom and such a transformation can be interpreted as a 
``slice'', that is a codimension-1 hyperplane in the state 
space of the system. This 
reformulation allowed for the straightforward use of the established 
geometrical methods of the slicing literature, which were mostly 
developed within the equivariant bifurcation theory\rf{ChossLaut00}. 
Budanur \etal\rf{BudCvi14} 
named this method ``first Fourier mode slice'' and applied it
to the Kuramoto--Sivashinsky equation. Later on, this method was 
successfully adapted for the
simulations of two-dimensional 
Kolmogorov flow\rf{Faraz15} and three-dimensional 
pipe flow\rf{BudHof17,BudHof18}. In this paper, we  
formulate this method for pilot-wave systems and demonstrate its 
utility in a numerical study. 

The rest of this paper is organized as follows. 
In the next section, we formulate a symmetry-reduced representation 
of the pilot-wave hydrodynamics with rotation symmetry.
At this stage, our 
treatment will be general and applicable to both 
experimental and numerical data. In \refsect{s-harmonic}, 
we describe the trajectory equation of Oza \etal \rf{ORB2013} 
with an additional central harmonic potential and apply our 
symmetry-reduction scheme to this system. We analyze the 
system's local bifurcations, the transition to chaos via period doubling 
cascades, and global bifurcations of the system's chaotic sets. 
We discuss our results in 
\refsect{s-conclusion}.

\section{Pilot-wave hydrodynamics as an infinite-dimensional dynamical system}
\label{s-DynSys}

\begin{figure}
	\centering
	\begin{overpic}[height=0.21\textwidth]{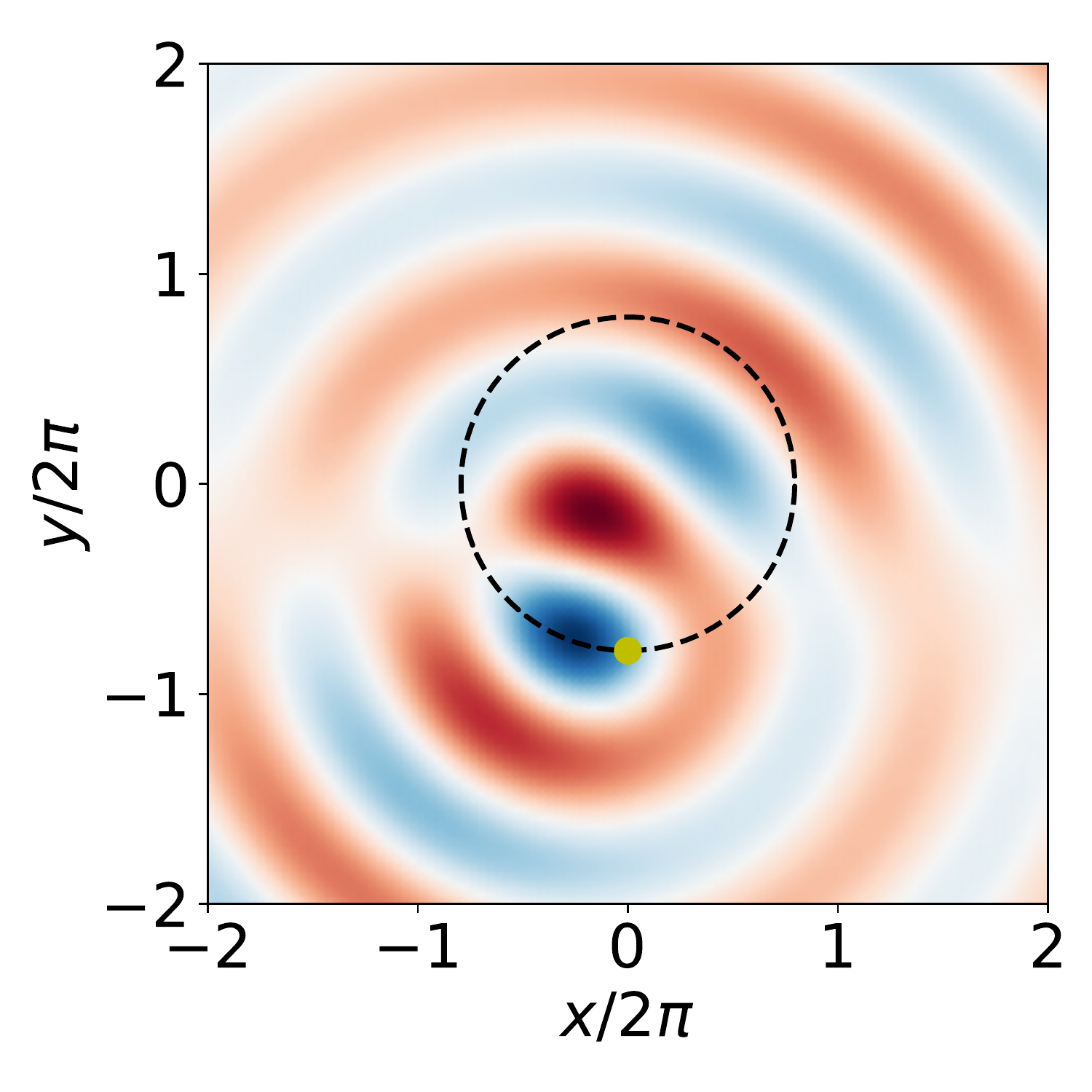}
		\put (0,0.0) {(a)}
	\end{overpic}                             
	\begin{overpic}[height=0.21\textwidth]{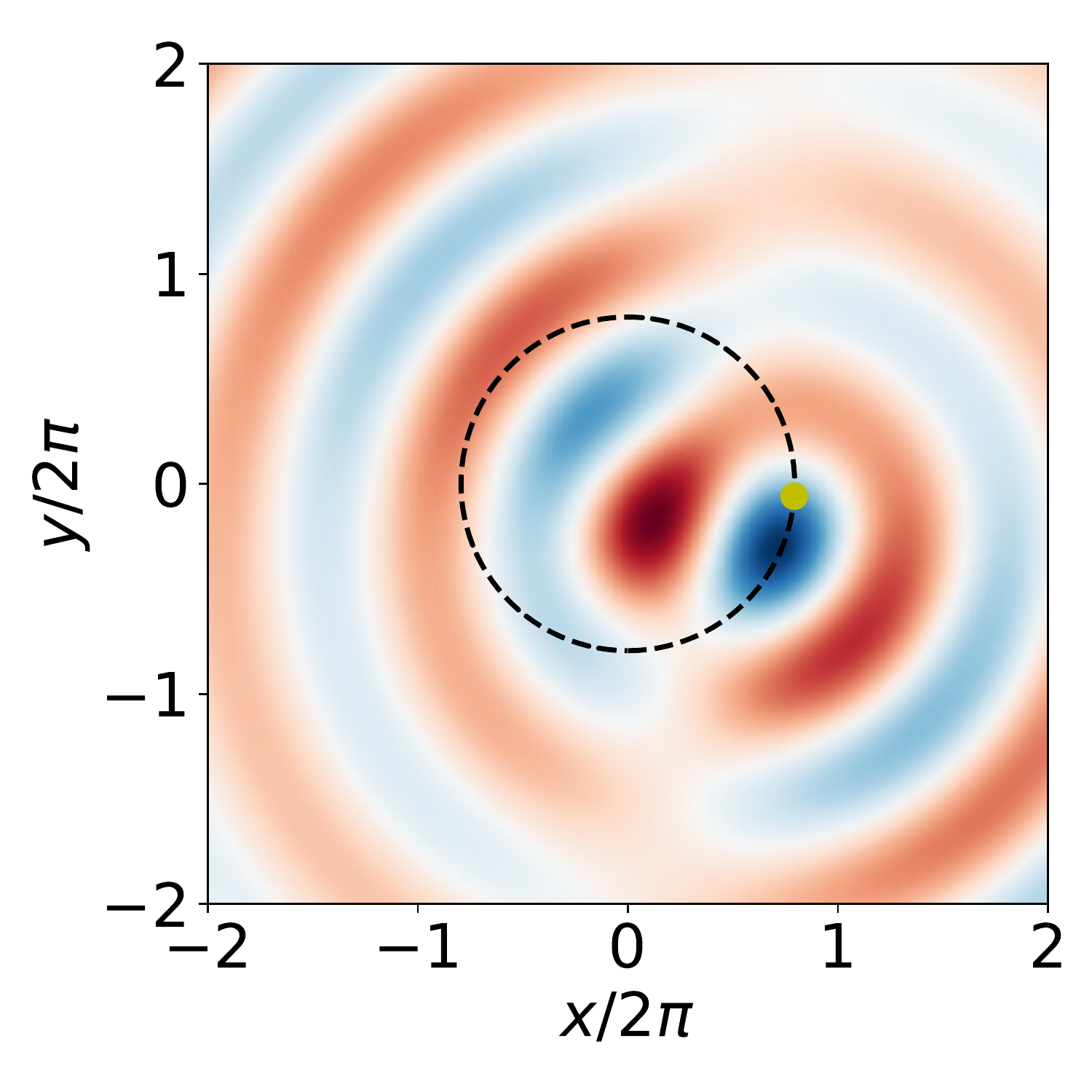}
		\put (0,0.0) {(b)}
	\end{overpic}
	\includegraphics[height=0.2\textwidth]{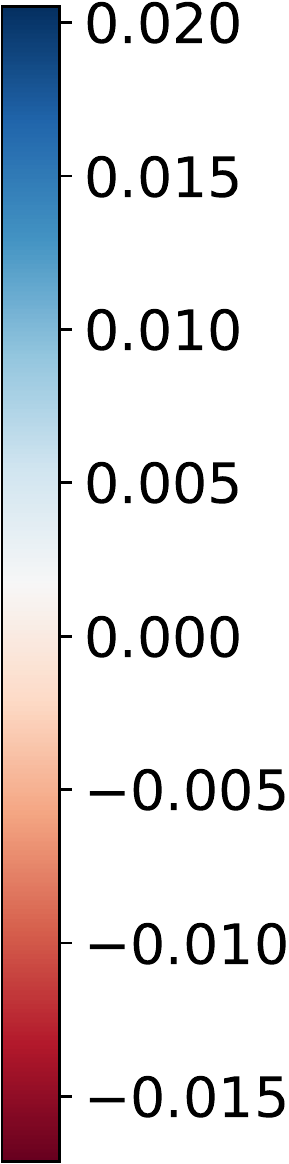}
	\caption{
		Two snapshots of a simulated circular trajectory where the 
		trace of the droplet is drawn as a dashed circle
		and its instantaneous position at the respective snapshot is 
		indicated with a yellow dot in each figure. 
		The magnitude of the wave field color coded in each snapshot.
		(a) $\zeit = 0$, (b) $\zeit = 20$ 
		in nondimensional units as explained in 
		\refsect{s-harmonic}. \label{f-circle}
	}
\end{figure}

We are going to investigate the dynamics of a point-like droplet 
as it bounces on the vertically vibrating bath of the 
same fluid in the presence of a 
central force. Let us denote the two-dimensional position and 
velocity of the droplet at time $\zeit$ with 
${\bf r (\zeit)} = (x (\zeit), y (\zeit))$ and 
${\bf v (\zeit)} = (v_x (\zeit), v_y (\zeit))$ respectively, and the fluid
height on the bath surface
as a function of polar coordinates $r$ and $\theta$ at time $\zeit$ 
with $h(r, \theta; \zeit)$. We assume that the future time-evolution of 
the system is completely determined by the current position and velocity 
of the droplet and the shape of the bath's surface. Under this assumption, 
we can define a state space vector
\beq
	\ssp(\zeit) = \left({\bf r}(\zeit), {\bf v}(\zeit), h(r, \theta; \zeit) \right) 
	\label{e-statevec}
\eeq
and a finite-time flow mapping $\flow{\zeit}{\ssp}$, which maps an initial
state $\ssp(0)$ to a final state $\ssp(\zeit)$ as
\beq
	\ssp(\zeit) = \flow{\zeit}{\ssp(0)} \, .
	\label{e-flow}
\eeq
Definitions 
\refeq{e-statevec} and \refeq{e-flow} make up an infinite-dimensional
dynamical system since $h(r, \theta; \zeit)$ takes values in a function 
space. In numerical applications, $h(r, \theta; \zeit)$ would be 
expressed as a finite-dimensional approximation such as 
a discretization on a grid, or a spectral expansion. 
In order to illustrate what we mean by these definitions, we show 
two snapshots of a simulated 
(details will be presented in \refsect{s-harmonic}) circular trajectory
of the pilot-wave system with harmonic potential in \reffig{f-circle}. 
Snapshots in this figure are separated by $20$
nondimensional units of time; and each panel contains the 
instantaneous information necessary to determine 
the future evolution of the system. 
According to our definitions, panel (a) ($\ssp(0)$) is mapped to 
panel 
(b) ($\ssp(20)$)  by the flow 
as $\ssp(20) = \flow{20}{\ssp(0)}$. 

If we further assume 
that the flow \refeq{e-flow} is smooth, then we can also define the 
state space velocity
\beq
	\dot{\ssp} = \vel (\ssp) 
	= \lim_{h\rightarrow0} \left(\flow{h}{\ssp} - \ssp \right) / h 
	\label{e-ODEform}
\eeq
and represent the system as an infinite set of ordinary differential 
(ODE) equations. In general, especially for experimental data, 
it may not be possible to obtain an explicit ODE form 
\refeq{e-ODEform}. Nevertheless, 
      if the dynamics is sufficiently smooth, we can assume that 
	  the state space velocity 
	  $\vel(\ssp)$ \refeq{e-ODEform} exists
     and we are going to use its 
definition \refeq{e-ODEform} in what follows.

\subsection{The rotation symmetry and the relative invariant solutions}
\label{s-symmetry}

While we have not specified the exact physics that governs the 
motion of a bouncing droplet yet, the conditions that we described
in the previous section along with the
additional assumption that the boundaries are far away (or circular) 
imply rotation symmetry. Since the shape of the bath's surface is 
determined by the past trajectory and the forcing is only radially 
dependent; the physics of the problem stays the same under the rotation
of coordinates. This, however, does not imply that the 
solutions would be symmetric under rotations. In fact, it is already visible
on the example of \reffig{f-circle} that the individual snapshots of 
the circular trajectory do not have rotation symmetry. However, for
each generic solution of the system, there exist infinitely many 
physically equivalent copies that can be obtained by continuous 
rotations. 
In the dynamical systems theory, this is called ``equivariance'', which we will now
formalize for the pilot-wave system. 

Let $\LieEl(\phi)$ be the rotation operator which acts on a state
vector \refeq{e-statevec} as
\bea
	\ssp' &=& \LieEl(\phi) \ssp \,, \nonumber \\
		  &=& \left(R(\phi) {\bf r}(t), 
			R(\phi) {\bf v}(t), 
			h(r, \theta - \phi; t) \right) \,,
	\label{e-Rotation}
\eea
where $R(\phi)$ are the standard $2\times2$ rotation matrices
\beq
R(\phi) = 
\begin{pmatrix}
	\cos \phi & - \sin \phi \\
	\sin \phi &   \cos \phi
\end{pmatrix} \,. \label{e-RotMat}
\eeq
If for every trajectory $\ssp (t)$ with $t \in [0, t_f]$ there exists 
another one $\LieEl(\phi) \ssp(t)$ obtained by the symmetry action, 
then the finite-time flow (or equivalently the state space 
velocity, when it exists) commutes with the symmetry action, \ie 
\beq
	\flow{\zeit}{\LieEl(\phi) \ssp} = \LieEl(\phi) \flow{\zeit}{\ssp} \,. 
	\label{e-EqvarCond}
\eeq
We are going to refer to \refeq{e-EqvarCond} as the 
``equivariance condition''. Note that the equivariance condition
does not imply that the flow is ``invariant'' under the action of the
symmetry, but rather, it transforms as does the state space 
coordinates themselves.
In our treatment, we are going to assume that the pilot-wave systems 
that we consider satisfy the equivariance condition 
\refeq{e-EqvarCond} with the rotation operator defined as in 
\refeq{e-Rotation}.

One way of elucidating the state space geometry of chaotic systems
is identifying the time-invariant solutions (equilibria, periodic 
orbits, invariant tori \ldots) that are embedded in the system's 
chaotic set\rf{DasBuch}. In systems with continuous symmetry, roles
of these solutions are taken over by their higher-dimensional 
equivalents, which we will refer to as ``relative invariant 
solutions''. Specifically, a ``relative equilibrium'' 
is a solution that evolves only in the symmetry direction at a
constant phase speed $c$. Let $\ssp_{\REQ{}}$ be a state space point
on a relative equilibrium, then its time-evolution satisfies
\beq
	\ssp_{\REQ{}} = \LieEl (-c \zeit) \flow{\zeit}{\ssp_{\REQ{}}} \label{e-reqv} \, .
\eeq
Note that a time-invariant set is formed by all solutions on the 
relative equilibrium, \ie\ the one-dimensional curve 
$\{g(c\zeit) \ssp_{\REQ{}} : \zeit \in [0, 2 \pi / c)\}$. The second type of 
relative invariant solutions we are going to consider here are 
``relative periodic orbits'', which have trajectories that 
recur at a symmetry-transformed location after a finite period $T_{\RPO{}}$, 
\ie\
\beq
	\ssp_{\RPO{}} = \LieEl(-\phi_{\RPO{}}) 
					\flow{T_{\RPO{}}}{\ssp_{\RPO{}}}\,, 
	\label{e-rpo}
\eeq
where $\ssp_{\RPO{}}$ is a point on a relative periodic orbit, 
$T_{\RPO{}}$ is its period, and $\phi_{\RPO{}}$ is its phase-shift.
Similar to the relative equilibrium, all points on a relative 
periodic orbits, \ie\ the 2-torus
$\{g(\phi) \flow{\zeit}{\ssp_{\RPO{}}} : 
 \phi \in [0, 2 \pi ) \, , \zeit \in [0, T_{\RPO{}}) \}$, form a 
time-invariant set. In other words, the relative invariant solutions
that appear as a consequence of a single-parameter continuous 
symmetry correspond to manifolds with dimensions 1 higher than
those of their standard counterparts. 
Notice that the special case $c=0$ in \refeq{e-reqv} corresponds 
to an equilibrium solution, and similarly $\phi_{\RPO{}} = 0$, in
\refeq{e-rpo} defines a periodic orbit. 

It is possible to define the linear stability of relative invariant
solutions by defining Jacobians of \refeq{e-reqv} and 
\refeq{e-rpo}\rf{ChossLaut00,DasBuch}. However, we are going to skip
these definitions here as they will be redundant once we introduce
the symmetry reduction in the next section.

\subsection{Continuous symmetry reduction}
\label{s-slices}

\begin{figure}
	\centering
	\setlength{\unitlength}{0.23\textwidth}
	\begin{picture}(1,0.70685575)%
	\put(0,0){\includegraphics[width=\unitlength,page=1]{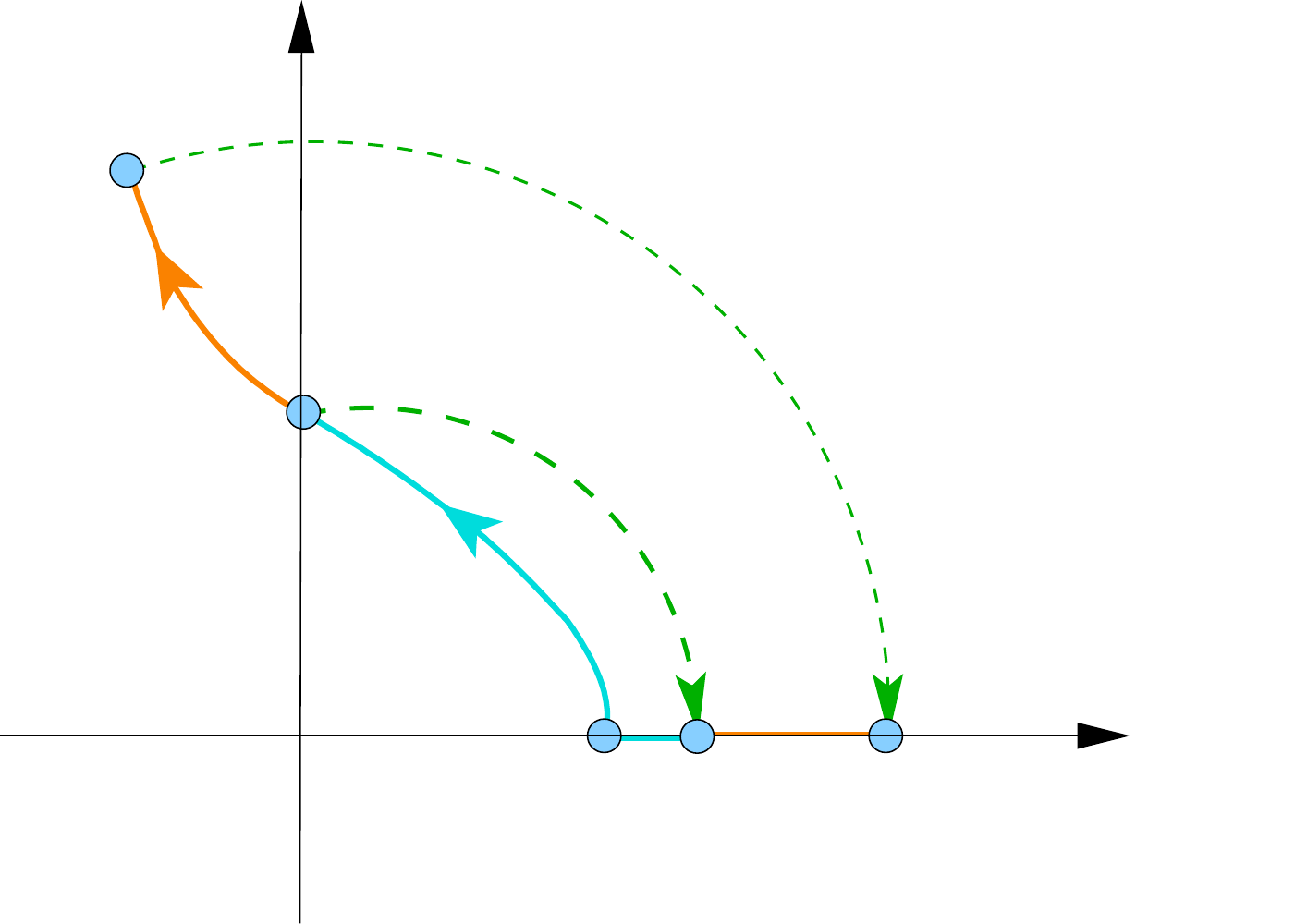}}%
	\put(-0.05,0.0){\color[rgb]{0,0,0}\makebox(0,0)[lb]{\smash{(a)}}}%
	\put(0.02886098,0.53139364){\color[rgb]{0,0,0}\makebox(0,0)[lb]{\smash{\textit{$\ssp_2$}}}}%
	\put(0.147,0.3493757){\color[rgb]{0,0,0}\makebox(0,0)[lb]{\smash{\textit{$\ssp_1$}}}}%
	\put(0.31992421,0.3974244){\color[rgb]{0,0,0}\makebox(0,0)[lb]{\smash{$\slicePhase_1$}}}%
	\put(0.51265306,0.07){\color[rgb]{0,0,0}\makebox(0,0)[lb]{\smash{\textit{$\sspRed_1$}}}}%
	\put(0.67142423,0.07){\color[rgb]{0,0,0}\makebox(0,0)[lb]{\smash{\textit{$\sspRed_2$}}}}%
	\put(0.26218268,0.65283237){\color[rgb]{0,0,0}\makebox(0,0)[lb]{\smash{\textit{$v_y$}}}}%
	\put(0.45386461,0.5442842){\color[rgb]{0,0,0}\makebox(0,0)[lb]{\smash{$\slicePhase_2$}}}%
	\put(0.39684577,0.07){\color[rgb]{0,0,0}\makebox(0,0)[lb]{\smash{\textit{$\ssp_0$}}}}%
	\put(0,0){\includegraphics[width=\unitlength,page=2]{NBBunrot.pdf}}%
	\put(0.84248574,0.09832379){\color[rgb]{0,0,0}\makebox(0,0)[lb]{\smash{\textit{$v_x$}}}}%
	\end{picture} 
    \begin{picture}(1,0.69433901)%
    \put(0,0){\includegraphics[width=\unitlength,page=1]{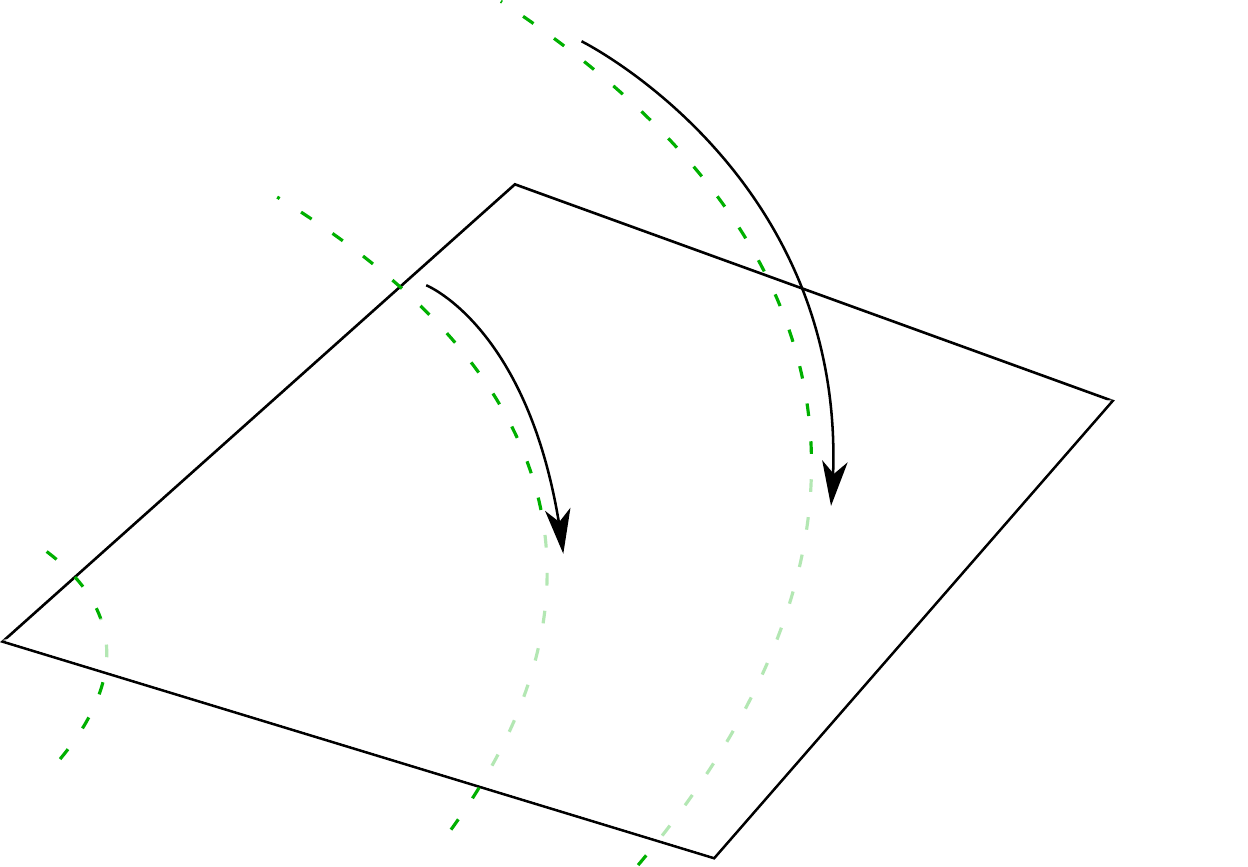}}%
    \put(-0.05,0.0){\color[rgb]{0,0,0}\makebox(0,0)[lb]{\smash{(b)}}}%
    \put(0.42248715,0.24){\color[rgb]{0,0,0}\makebox(0,0)[lt]{\begin{minipage}{0.32953426\unitlength}\raggedright $\sspRed_1$\end{minipage}}}%
    \put(0.4,0.5){\color[rgb]{0,0,0}\makebox(0,0)[lt]{\begin{minipage}{0.40874111\unitlength}\raggedright $\LieEl_z(\slicePhase_1)$\end{minipage}}}%
    \put(0.085,0.22){\color[rgb]{0,0,0}\makebox(0,0)[lt]{\begin{minipage}{0.18913151\unitlength}\raggedright $\slicep$\end{minipage}}}%
    \put(0.62,0.6){\color[rgb]{0,0,0}\makebox(0,0)[lt]{\begin{minipage}{0.40874111\unitlength}\raggedright $\LieEl_z(\slicePhase_2)$\end{minipage}}}%
    \put(0.64829519,0.29397554){\color[rgb]{0,0,0}\makebox(0,0)[lt]{\begin{minipage}{0.32953426\unitlength}\raggedright $\sspRed_2$\end{minipage}}}%
    \put(0,0){\includegraphics[width=\unitlength,page=2]{sliceHplane.pdf}}%
    \put(0.1857004,0.295){\color[rgb]{0,0,0}\makebox(0,0)[lt]{\begin{minipage}{0.32953426\unitlength}\raggedright $\ssp_0$\end{minipage}}}%
    \put(0.25,0.5){\color[rgb]{0,0,0}\makebox(0,0)[lt]{\begin{minipage}{0.32953426\unitlength}\raggedright $\ssp_1$\end{minipage}}}%
    \put(0.46,0.69570511){\color[rgb]{0,0,0}\makebox(0,0)[lt]{\begin{minipage}{0.32953426\unitlength}\raggedright $\ssp_2$\end{minipage}}}%
    \put(0,0){\includegraphics[width=\unitlength,page=3]{sliceHplane.pdf}}%
    \put(0.025,0.42){\color[rgb]{0,0,0}\makebox(0,0)[lt]{\begin{minipage}{0.18913151\unitlength}\raggedright $\sliceTan$\end{minipage}}}%
    \end{picture}
	\caption{
		(a) Schematic illustration of the symmetry reduction as a 
		polar coordinate transformation on the 
		$(v_x, v_y)$-plane. 
		(b) Schematic illustration of the symmetry reduction by a 
		slice hyperplane. 
		\label{f-slice}}
\end{figure}

Symmetry reduction is a state space coordinate transformation 
$\ssp \rightarrow \sspRed$ such that each symmetry-related 
(physically equivalent) set of states
$\{\LieEl(\phi) \ssp^* \,|\, \phi \in [0, 2 \pi)\}$ 
is represented by a 
single solution $\sspRed^*$ in the symmetry-reduced state space.
An obvious choice of such coordinates for the droplet system with the 
rotation symmetry is the time-dependent polar coordinate 
transformation that fixes the polar-angle in $(x,y)$ plane to a 
certain arbitrary value. This angle, however, is not defined when 
the droplet is at the origin, and would experience very rapid phase 
fluctuations if the droplet is close to the origin. Observing that
the droplet's horizontal velocity is always non-zero\rf{CPFB2005}, we 
define the following operation
\beq
	\sspRed(\zeit) = \LieEl(- \slicePhase(\zeit)) \ssp(\zeit) \label{e-velPolar}  
\eeq
where
\beq
	\slicePhase = \mbox{Arg}(v_x(\zeit) + i v_y(\zeit)) \, . 
\eeq
This transformation fixes the polar angle in the $(v_x, v_y)$ plane
to $0$, hence maps all symmetry-equivalent solutions to the one with
$v_y = 0$ as illustrated in \reffig{f-slice}(a), 
as long as $||{\bf v}|| \neq 0$. Thus, the 
symmetry-reduced state space coordinates \sspRed\ have 
$\hat{v}_y$ coordinate identically equal to $0$, \ie
\beq
	\sspRed = 
	\left(\hat{x}, \hat{y}, \hat{v}_x, 0, 
		  \hat{h}(r, \theta; t) \right) . \label{e-sspRed}
\eeq
Budanur \etal \rf{BudCvi14} showed that a polar coordinate transformation similar 
to \refeq{e-velPolar}, when defined for a partial differential equation
with translation symmetry, can be interpreted as a projection onto a 
codimension-1 hyperplane, named a ``slice'', in the state space. 
We are now going to take analogous steps and redefine the 
transformation \refeq{e-velPolar} in this terminology in order to 
exploit the well-established tools of this method. 

Let $\Lg$ be the generator of infinitesimal rotations satisfying 
$\LieEl (\phi) = e^{\phi \Lg}$. Its action on the state space 
coordinates can be expressed as
\bea
	\groupTan (\ssp) &=& \Lg \ssp \,, \nonumber \\
					 &=& \left(T {\bf r}(t), 
							   T {\bf v}(t), 
							   (- \partial / \partial \theta) h(r, \theta; t)
						 \right) \,,
	\label{e-InfRot}
\eea
where $T$ is the generator of the $2 \times 2$ rotations 
\refeq{e-RotMat}
\beq
	T = \begin{pmatrix}
		0 & - 1 \\
		1 &   0
		\end{pmatrix}
\eeq
and $\groupTan (\ssp)$ is called the ``group tangent'' evaluated at 
$\ssp$, since it gives the direction of an infinitesimal rotation 
at $\ssp$. 
Let us now define the ``slice template'' as the state vector
\beq
	\slicep = (\hat{x}=0, 
			   \hat{y}=0, 
			   \hat{v}_x=1, 
			   \hat{v}_y=0, 
			   \hat{h}(r, \theta; \zeit)=0) \,, \label{e-slicep}
\eeq
whose only nonzero component is $\hat{v}_x = 1$. 
It is straightforward to verify that the symmetry-reduced state space 
coordinates \refeq{e-velPolar} satisfy the half-hyperplane condition
\beq
	\inprod{\sspRed - \slicep}{\sliceTan} = 0 \quad \mbox{and} 
	\quad \inprod{\groupTan (\sspRed)}{\sliceTan} > 0 \,, 
	\label{e-sliceHplane}
\eeq 
where $\sliceTan = \groupTan (\slicep)$ and $\inprod{.}{.}$ is 
the inner product defined as
\bea
	\inprod{\ssp_m}{\ssp_n} &=& x_mx_n + y_my_n 
							+ v_{x,m}v_{x,n} + v_{y,m}v_{y,n} \continue
							&+& \iint_{\mathcal{D}}  w(r)
							h_m (r, \theta) h_n (r, \theta)\,
							r dr d\theta  \, ,
	\label{e-inprod}
\eea 
where w(r) is an appropriately chosen weight function
and the integral is computed over the whole domain $\mathcal{D}$.
Projection of the full state space dynamics onto this codimension-1 
half-hyperplane is illustrated in \reffig{f-slice} (b). 
In the slicing method, one looks for the slice phases $\slicePhase (\zeit)$, 
such that $\sspRed(\zeit) = \LieEl(- \slicePhase (\zeit)) \ssp(\zeit)$ 
satisfies \refeq{e-sliceHplane}. The main advantage of slicing
framework is that it allows for a straightforward symmetry-reduction 
of the tangent space. Let $\delta \ssp$ be a small perturbation 
(in the full state space) to $\sspRed$; it is projected onto
the slice as 
\beq
	\delta \sspRed = H(\sspRed) \delta \ssp \,,
\eeq
where
\beq
	H(\sspRed) = {\rm I} - \frac{\groupTan(\sspRed) \otimes \sliceTan}{
						   \inprod{\groupTan(\sspRed)}{\sliceTan}} , 
\eeq
${\rm I}$ is the identity matrix and $\otimes$ denotes the outer product. 
For a derivation, see the appendix of Budanur \& Hof (2017)\rf{BudHof17}.
In particular, we can obtain the symmetry reduced state space velocity
$\dot{\sspRed} = \velRed (\sspRed)$ as 
\bea
	\velRed (\sspRed) &=& H(\sspRed) \vel(\sspRed) \,, \continue
	 				  &=& \vel (\sspRed) - 	
	\frac{\inprod{\vel(\sspRed)}{\sliceTan}}{
							   \inprod{\groupTan(\sspRed)}{\sliceTan}} 
						 \groupTan(\sspRed) . \label{e-velRed1}
\eea
By taking the time derivative of 
$\sspRed = \LieEl(- \slicePhase(\zeit))$ and requiring the slice 
hyperplane condition \refeq{e-sliceHplane} to be satisfied, one can 
show that the multiplier of the group tangent in \refeq{e-velRed1} is
in fact the slice phase velocity. In other words, we can write
\bea
	\velRed (\sspRed)          &=& 
		\vel (\sspRed) - \dot{\slicePhase} \, \groupTan(\sspRed) \,,
		\label{e-velRed} \\
	\dot{\slicePhase}(\sspRed) &=& \frac{\inprod{\vel(\sspRed)}{\sliceTan}}{
									     \inprod{\groupTan(\sspRed)}{\sliceTan}} \,;
	\label{e-phaseVel}
\eea
and integrate the slice phase $\slicePhase$ along with the 
symmetry-reduced evolution. 
Rowley and Marsden\rf{rowley_reconstruction_2000}
referred to 
\refeq{e-velRed} and
\refeq{e-phaseVel} as ``reconstruction equations'' since by
integrating them simultaneously, one can reconstruct the full state 
space evolution. 
In other words, symmetry reduction does not lose any information, 
it rather
separates the time-evolution into the 
``physical'' (in-slice) and 
``redundant'' (phase) parts. Another observation to make in 
\refeq{e-phaseVel} is that it diverges when 
$\inprod{\groupTan(\sspRed)}{\sliceTan}=0$. This, in general, implies that 
a hyperplane slice that is constructed around an arbitrary template 
$\slicep$ is ``local'', and applicable in a finite neighborhood where the 
denominator of \refeq{e-phaseVel} is nonzero. 
Notice, however, that for our particular
choice of the slice template \refeq{e-slicep}, this condition becomes
$\inprod{\groupTan(\sspRed)}{\sliceTan}=\hat{v}_x = \sqrt{v_x^2 + v_y^2}$; 
thus, our 
slice is applicable as long as the droplet's velocity does not vanish. 

In the symmetry-reduced state space, a relative equilibrium 
\refeq{e-reqv} becomes an equilibrium, which satisfies
\beq
	\velRed (\sspRed_{\REQ{}}) = 0 \, . \label{e-reqRed}
\eeq
Linear stability of this solution is determined by the eigenvalue 
equation
\beq
	\StabMatRed (\sspRed_{\REQ{}}) \hat{e}_i = \lambda_i \hat{e}_i \,,\quad
	\label{e-reStability}
\eeq
where
$\StabMatRed_{ij} (\sspRed^*) = \partial \velRed_i (\sspRed) 
/ \partial \sspRed_j |_{\sspRed = \sspRed^*}$.
By straightforward algebra, this matrix is obtained as
\beq
	\StabMatRed(\sspRed) = \StabMat (\sspRed)
						 - \frac{\groupTan (\sspRed) \otimes \StabMat (\sspRed)^T \sliceTan}{
						  		 \inprod{\groupTan(\sspRed)}{\sliceTan}}
					  	 + \dot{\slicePhase} \, \frac{\groupTan (\sspRed) \otimes \Lg^T \sliceTan}{
					  	 					       \inprod{\groupTan(\sspRed)}{\sliceTan}}
				  	 	 - \dot{\slicePhase} \, \Lg \,, \label{e-StabMatRed}
\eeq
where $\StabMat_{ij} (\ssp^*) = \partial \vel_i (\ssp) 
						      / \partial \ssp_j |_{\ssp = \ssp^*}$ .
We are going to refer to $\StabMat (\ssp)$ and 
$\StabMatRed (\sspRed)$ respectively as the stability matrix and
the reduced stability matrix. $\lambda_i$ are the linear stability 
eigenvalues and $\hat{e}_i$ are the corresponding stability 
eigenvectors within the slice. 
$\Re \lambda_i < 0$ and 
$\Re \lambda_i > 0$ respectively correspond to the stable and 
unstable directions associated with the relative equilibrium and 
a bifurcation takes place when $\Re \lambda_i = 0$. 

Similar to the relative equilibria, relative periodic orbits \refeq{e-rpo}	
satisfy the periodicity condition
\beq
	\sspRed_{\RPO{}} = \flowRed{T_{\RPO{}}}{\sspRed_{\RPO{}}} \,, 
	\label{e-rpoRed}
\eeq						      
where $\sspRed(\zeit) = \flowRed{\zeit}{\sspRed(0)}$ is the 
symmetry-reduced finite-time flow induced by the dynamics.
Consequently, the linear stability of a relative periodic orbit
is determined by the eigenvalue equation
\beq
	\hat{J}^{T_{\RPO{}}}(\sspRed_{\RPO{}}) \hat{V}_i = \Lambda_i \hat{V_i}
	\, . \label{e-rpoStability}
\eeq
where
$\hat{J}^{\zeit}_{ij}(\sspRed^*) = \partial \hat{f}_i^{\zeit} (\sspRed) 
/ \partial \sspRed_j |_{\sspRed = \sspRed^*}$.
Numerically, the reduced Jacobian $\hat{J}^{\zeit}_{ij}(\sspRed^*)$ 
can be obtained as the time-ordered integral of the reduced stability 
matrix $\StabMatRed(\sspRed)$ along a symmetry-reduced state space 
trajectory. $\Lambda_i$ are known as the Floquet multipliers of the 
relative periodic orbit and $\hat{V}_i$ are the corresponding Floquet
vectors in the symmetry-reduced state space. $|\Lambda_i| < 1$ and 
$|\Lambda_i| > 1$  correspond to the stable and unstable directions 
respectively and $|\Lambda_i| = 1$ corresponds to a marginal direction. 
Note that
the symmetry-reduced state space velocity 
$\velRed (\sspRed_{\RPO{}})$ is an eigenvector of 
$\hat{J}^{T_{\RPO{}}}(\sspRed_{\RPO{}})$ with the eigenvalue $1$. 

This concludes our general introduction of the symmetry-reduction, 
which can -- in principle -- be applied to experimental, as 
well as numerical data, independent of the specifics of system in
consideration, as long as the rotation symmetry is 
present. In the next section, we are going to demonstrate the utility
of our transformation by applying it to the simulations of a 
pilot-wave model with central harmonic force.

\section{A pilot-wave model with central harmonic force}
\label{s-harmonic}

Following the efforts of Mol{\'a}{\v{c}}ek and Bush\rf{MB2013a,MB2013a}, 
Oza \emph{et al.}\rf{ORB2013} derived a trajectory equation for the 
bouncing droplets. With the 
addition of a central harmonic force $-k \pos$, this equation reads
\beq
	m \, \ddot{\pos} + D \, \dot{\pos} = - m g \, \nabla h (\pos, \zeit) - k \pos\, ,
	\label{e-TrajectoryOza}
\eeq
where $\pos = (x, y)$ is the position of the droplet, $m$ is 
its mass, $D$ is the viscous damping coefficient, $g$ is the
acceleration due to gravity, $k$ is the effective spring constant of 
the harmonic force, and $h$ is the wave field on the bath 
surface determined by the previous bounces of the droplet as
\beq
	h (\pos, \zeit) = A \sum_{n=-\infty}^{\lfloor \zeit/T_F \rfloor} 
				      J_0 (k_F | \pos - \pos(n T_F)|)
				      e^{- (\zeit - nT_F)/T_F \memory} \, , \label{e-Field}
\eeq
where $A$ is the initial amplitude of the waves generated, $J_0$ is the 
Bessel function of the first kind, $T_F$ is the time between the bounces, 
known as the `Faraday time', $k_F = 2 \pi / \lambda_F$
is the wave number corresponding to the least stable Faraday wave,
and $\lambda_F$ is the corresponding wavelength. 
The tunable parameter $\memory$ is called ``memory'' and 
it determines the damping speed of the waves generated in previous bounces. 
One of the
approximations that were made by Oza \etal \rf{ORB2013} is already 
apparent 
from \refeq{e-Field}, where each bounce instantly generates a Bessel wave, 
with no time needed for propagation. 

With the wave field given as in \refeq{e-Field}, the trajectory 
equation \refeq{e-TrajectoryOza} is a delay-differential system, which 
is very hard to study both analytically and numerically. Therefore, 
Oza \etal \rf{ORB2013} and several subsequent 
studies\rf{LOPB2016,THORB2016,KOB2017} approximated 
the sum in \refeq{e-Field} by the integral
\beq
	h (\pos, \zeit) = \frac{A}{T_F} \int_{-\infty}^{\zeit}
					J_0 (k_F | \pos - \pos(\zeit')|)
					e^{- (\zeit - \zeit')/T_F \memory} d\zeit' \, . \label{e-FieldInt}	
\eeq
This approximation is justified since the time scale of the horizontal 
motion is much greater than the time scale of bouncing\rf{ORB2013}. At this
stage, \refeq{e-TrajectoryOza} is a four-dimensional 
integro-differential system. Labousse\rf{labousse2014thesis,PerrardLabousse2018} 
showed 
that this system can further be manipulated into an 
infinite-dimensional ODEs and this is the
form we are going to use in the following. It follows from Graf's 
addition theorem\rf{watson1944} that \refeq{e-FieldInt} can be written 
as\small{
\bea
h (\pos, \zeit) = A \sum_{n=0}^{\infty} && \bigg[(2\,-\,\delta_{n,0}) J_n (k_F \rpos) 
\continue
&& \times  (C_n (\zeit) \cos (n \theta)
				+ S_n (\zeit) \sin (n \theta)) \bigg]
\,  \label{e-GrafSum}	
\eea}
where
\bea
C_n (\zeit) &=& \int_{-\infty}^\zeit \frac{d \tau'}{T_F} J_n (k_F \rpos ) 
e^{- (\zeit - \tau') / T_F \memory} \cos (n \theta (\tau')) 
\label{e-Cn} \, , \\
S_n (\zeit) &=& \int_{-\infty}^\zeit \frac{d \tau'}{T_F} J_n (k_F \rpos ) 
e^{- (\zeit - \tau') / T_F \memory} \sin (n \theta (\tau'))	
\label{e-Sn} \, ,
\eea
We can obtain the time derivatives of the mode amplitudes $C_n$ and
$S_n$ as
\begin{eqnarray}
	\dot{C}_n &=& - \frac{C_n}{T_F \memory} + \frac{1}{T_F} J_n (k_F \rpos) \, \cos n \theta   \, , \\
	\dot{S}_n &=& - \frac{S_n}{T_F \memory} + \frac{1}{T_F} J_n (k_F \rpos) \, \sin n \theta \, .	
\end{eqnarray}

Rearranging \refeq{e-TrajectoryOza},
the acceleration of the walker in $x$ and $y$ directions can be 
written explicitly as
\begin{eqnarray}
	\ddot{x} &=& - (D / m)\, \dot{x} - (k / m)\, x 
			   - g \, \partial_x h \, , \label{e-ddotx} \\
	\ddot{y} &=& - (D / m)\, \dot{y} - (k / m)\, y 
			   - g \, \partial_y h \, , \label{e-ddoty} 
\end{eqnarray}
where the partial derivatives are
\begin{eqnarray}
	\partial_x h &=& (\cos \theta \, \partial_r 
				    - r^{-1} \, \sin \theta \, \partial_\theta) \, h \,, \\
				 &=& A \sum_{n=0}^{\infty} \bigg[ (2 - \delta_{n, 0} )  \nonumber \\
				 	 && \times \bigg(
				 	 k_F \, J_{n-1}(k_F r) \, K_n \cos \theta \,
				 	 - \frac{n \, J_n (k_F r)}{r} L_n \bigg)
				 	 \bigg] \,, \\
	\partial_y h &=& (\sin \theta \, \partial_r 
				    + r^{-1} \, \cos \theta \, \partial_\theta) \, h \,, \\
				 &=& A \sum_{n=0}^{\infty} \bigg[ (2 - \delta_{n, 0} )  \nonumber \\
				 	 && \times \bigg(
				 	 k_F \, J_{n-1}(k_F r) \, K_n \sin \theta \,
				 	 + \frac{n \, J_n (k_F r)}{r} M_n \bigg) \bigg] \,.			 	 				 	 
\end{eqnarray}
with $K_n, L_n$ and $M_n$ defined as
\bea
	K_n &=& C_n \cos (n \theta) + S_n \sin (n \theta) \, , \\
	L_n &=& C_n \cos [(n + 1) \theta] +  S_n \sin [(n + 1) \theta] \, , \\
	M_n &=& S_n \cos [(n + 1) \theta] - C_n \sin [(n + 1) \theta] \, .
\eea

Finally, we nondimensionalize trajectory equations \refeq{e-ddotx} and 
\refeq{e-ddoty} by scaling time as $\zeit / T_F \rightarrow \zeit$ and lengths as 
$(k_F x ,k_F y) \rightarrow (x, y)$ and rewrite everything compactly as
\begin{eqnarray}
	\ddot{x}\, &&= - \eta \, \dot{x} - \chi \, x \nonumber \\
			 && - \mu \,  \sum_{n=0}^{\infty} (2 - \delta_{n, 0}) 
			     \left[
			     J_{n-1}(r) \, K_n \cos \theta - \frac{n J_n(r)}{r} L_n
			     \right], \label{e-ddotxNonDim} \\
	\ddot{y}\, &&= - \eta \, \dot{y} - \chi \, y \nonumber \\ 
			 && - \mu \,  \sum_{n=0}^{\infty} (2 - \delta_{n, 0}) \left[
			     J_{n-1}(r) \, K_n \sin \theta + \frac{n J_n(r)}{r} M_n
			     \right], \label{e-ddotyNonDim}		\\
	\dot{C}_n\, &&= - \memory^{-1} C_n + J_n (\rpos) \, \cos (n \theta )\, ,  
				  \label{e-CnNonDim} \\
	\dot{S}_n\, &&= - \memory^{-1} S_n + J_n (\rpos) \, \sin (n \theta )\, , 
				  \label{e-SnNonDim} 
\end{eqnarray}
where,
\beq
	\eta = \frac{D T_F}{m} \, , \quad 
		  	 \chi = \frac{k T_F^2}{m} \, , \quad 
			 \mu = g A k_F^2 T_F^2  \, .  \label{e-parameters} 
\eeq
With the addition of trivial equations
$\dot{x} = \dot{x}$ and 
$\dot{y} = \dot{y}$,
equations \refeqs{e-ddotxNonDim}{e-SnNonDim}
yields a set of first 
order ODEs \refeq{e-ODEform} for the state space vector
\begin{equation}
\ssp = (x, \, y, \, \dot{x},  \, \dot{y}, \, C_0, 
		\, C_1, \, S_1, \, C_2, \, S_2, \, \ldots ) \, . \label{e-ssp}
\end{equation}
Formally, this state space is infinite dimensional. In the numerical 
work to follow, the Fourier-Bessel expansion (\ref{e-GrafSum}) is truncated 
at a finite number of $N+1$ 
modes (starting counting from the $0$th mode), yielding a 
$2N+5$ dimensional representation. 
We define the $L_2$ inner product in this 
representation as
\beq
	\inprod{\ssp_1}{\ssp_2} = \sum_{i = 1}^{2N + 5} \ssp_{1, i} 
													\ssp_{2, i} 
	\, .
\eeq

For the state space coordinates ordered as in \refeq{e-ssp}, we can 
write the rotation-operator \refeq{e-Rotation} explicitly as 
\begin{equation}
	\LieEl (\phi) = \begin{pmatrix}
	\cos \phi & - \sin \phi & 0 & 0 & 0 & 0 & 0 & \ldots \\
	\sin \phi &   \cos \phi & 0 & 0 & 0 & 0 & 0 & \ldots \\
	0 & 0 & \cos \phi & - \sin \phi & 0 & 0 & 0 & \ldots \\
	0 & 0 & \sin \phi &   \cos \phi & 0 & 0 & 0 & \ldots \\
	0 & 0 & 0 & 0 & 1 & 0 & 0 & \ldots \\			 					
	0 & 0 & 0 & 0 & 0 & \cos \phi & - \sin \phi & \ldots \\
	0 & 0 & 0 & 0 & 0 & \sin \phi &   \cos \phi & \ldots \\	 	
	\vdots & \vdots & \vdots & \vdots & \vdots & 
	\vdots & \vdots & \ddots 		 		
	\end{pmatrix}\,,
\label{e-LieEl}
\end{equation}
with the subsequent $2\times2$ blocks along its diagonal 
populated with the rotation matrices \refeq{e-RotMat}
$R(2 \phi), R(3 \phi), \ldots$
The corresponding generator of infinitesimal rotations then 
becomes
\begin{equation}
\Lg 		  = \begin{pmatrix}
0 & - 1 & 0 & 0 & 0 & 0 & 0 & 0 & 0 & \ldots \\
1 &   0 & 0 & 0 & 0 & 0 & 0 & 0 & 0 & \ldots \\
0 & 0 & 0 & - 1 & 0 & 0 & 0 & 0 & 0 & \ldots \\
0 & 0 & 1 &   0 & 0 & 0 & 0 & 0 & 0 & \ldots \\
0 & 0 & 0 & 0 & 0 & 0 & 0 & 0 & 0 & \ldots \\
0 & 0 & 0 & 0 & 0 & 0 & - 1 & 0 & 0 & \ldots \\
0 & 0 & 0 & 0 & 0 & 1 &   0 & 0 & 0 & \ldots \\
0 & 0 & 0 & 0 & 0 & 0 & 0 & 0 & - 2 & \ldots \\
0 & 0 & 0 & 0 & 0 & 0 & 0 & 2 &   0 & \ldots \\
\vdots & \vdots & \vdots & \vdots & \vdots & 
\vdots & \vdots & \vdots & \vdots & \ddots 		 		
\end{pmatrix} \,.
\label{e-Lg}
\end{equation}

In addition to the rotational symmetry, the pilot wave model 
with a central force is also equivariant under the reflection 
$y \rightarrow -y$, or equivalently $\theta \rightarrow -\theta$.
The action of this transformation on the state space coordinates is
represented by the diagonal matrix
\beq
	\sigma = \mathrm{diag}[1, -1, 1, -1, 1, 1, -1, 1, -1, \ldots] \,,
	\label{e-sigma}
\eeq
and the finite-time flow mapping implied by the time-evolution under
\refeqs{e-ddotxNonDim}{e-SnNonDim} satisfies 
\beq
	\flow{\zeit}{\sigma \ssp} = \sigma \flow{\zeit}{\ssp} \,. 
	\label{e-Eqvarsigma}
\eeq

The model parameters \refeq{e-parameters} are determined by the 
experimental conditions as explained by Oza \etal \rf{ORB2013} and 
Perrard \etal \rf{PLMFC2014}. In the following numerical work, we 
are going to adopt the numbers reported by Tambasco \etal \rf{THORB2016}
and set 
$m = 0.25\times10^{-6} kg$, 
$f = 80Hz$, 
$D = 2.0\times10^{-6} kg/s$, 
$A = 3.5\times10^{-6} m$, and
$k = 3.2\times10^{-6} N / m$. These choices yield the
nondimensional system parameters
\beq
	\eta =  0.2 \,,	\quad
	\chi =  0.008 \,, \quad
	\mu =  0.0375482401231 \, . \label{e-parsNum}
\eeq

In the next section, we are going to explore the bifurcations of this
model and the onset of chaos numerically for $\memory \in [10, 19]$. 
We observed for this parameter range that a truncation of 
the Fourier-Bessel expansion \refeq{e-GrafSum} at $N = 25$ ensures at least $8$
order of magnitude drop in the mode amplitudes $||C_n + S_n||$, 
hence the numerical results that we are going to report adopts this
resolution. Note that the resulting state space representation is 
$55$-dimensional. In the numerical results to follow, we simulate 
this $55$-dimensional ODE system 
\refeqs{e-ddotxNonDim}{e-SnNonDim} using 
\texttt{odeint} function
of \texttt{scipy.integrate} module\rf{scipy} of Python programming 
language. 
      Note that \texttt{odeint} adapts the time-steps in order to keep 
	  estimated numerical integration errors below $10^{-8}$. 	
	  In the results of the following sections, the trajectories are 
	  sampled with a time-step $\delta \zeit = 0.01$.
For simulations in the symmetry-reduced state space, 
we integrate \refeq{e-velRed1}, which we obtain 
explicitly by projecting \refeqs{e-ddotxNonDim}{e-SnNonDim}.

\subsection{Local bifurcations and the onset of chaos}

As we shall now demonstrate, the explicit ODE form of the dynamical
equations \refeqs{e-ddotxNonDim}{e-SnNonDim} along with the symmetry 
reduced dynamics \refeqs{e-velRed}{e-phaseVel} allow us to use
standard numerical analysis methods for a bifurcation study of the 
pilot-wave system. In this section and the rest of the paper, we are going to 
use the following naming convention for the solutions we describe.
Initial capital letters will refer to the type of the solution: 
EQ: equilibrium, 
REQ: relative equilibrium, 
PO: periodic orbit, 
RPO: relative periodic orbit, 
C: chaotic. The following integer (except for chaotic solutions)
will refer to the number of times the solution intersects the 
Poincar\'e section \refeq{e-PoincSy}, which we are going to define and 
use for our illustrations later. The following lowercase letter will refer to 
the shape of the solution on the $(x,y)$-plane: 
c: circle, 
o: oval, 
l: lemniscate, 
t: trefoil. 
The solutions that are not reflection-symmetric will appear in 
pairs of positive and negative average angular momenta. We will
refer to the ones with negative average angular momentum with an 
additional $\sigma$ in front. For example, the circular solution 
is a relative equilibrium that does not intersect the Poincar\'e 
section \refeq{e-PoincSy}; thus, the circular solution with positive
angular momentum will be $\REQ{0c}$ and its reflection copy with 
negative angular momentum will be $\sigma \REQ{0c}$. 
When needed, additional letters will be used to further specify the 
solutions and these will be explained in place. 
For reference, 
\reftab{e-InvSols} shows a summary of invariant solutions in the 
numerical bifurcation 
study to follow. In \reftab{e-InvSols} and the rest of the trajectory
plots of this paper, the lengths are scaled by $2 \pi$ so that 
one unit of distance is equal to one Faraday wavelength.

\begin{table*}
	\caption{Summary of invariant pilot-wave solutions that 
			are studied in this paper. Each solution has a reflection-copy 
			with the exception of $\PO{2l}$, which itself is symmetric under 
		    the reflection \refeq{e-sigma}. The last column shows 
	        the $\memory$ value at which the trajectories are plotted. 
        	\label{e-InvSols}}
	\begin{tabular}{m{.1\textwidth}|
					m{.2\textwidth}|
					m{.1\textwidth}|
					l|
					l|
					m{.1\textwidth}}
		Name & Type & Shape & Trajectory & Reduced trajectory & at $\memory$ \\ 
		\hline
		$\REQ{0c}$ & Relative equilibrium & Circle &
		\begin{minipage}{.16\textwidth}
			\includegraphics[width=\textwidth]{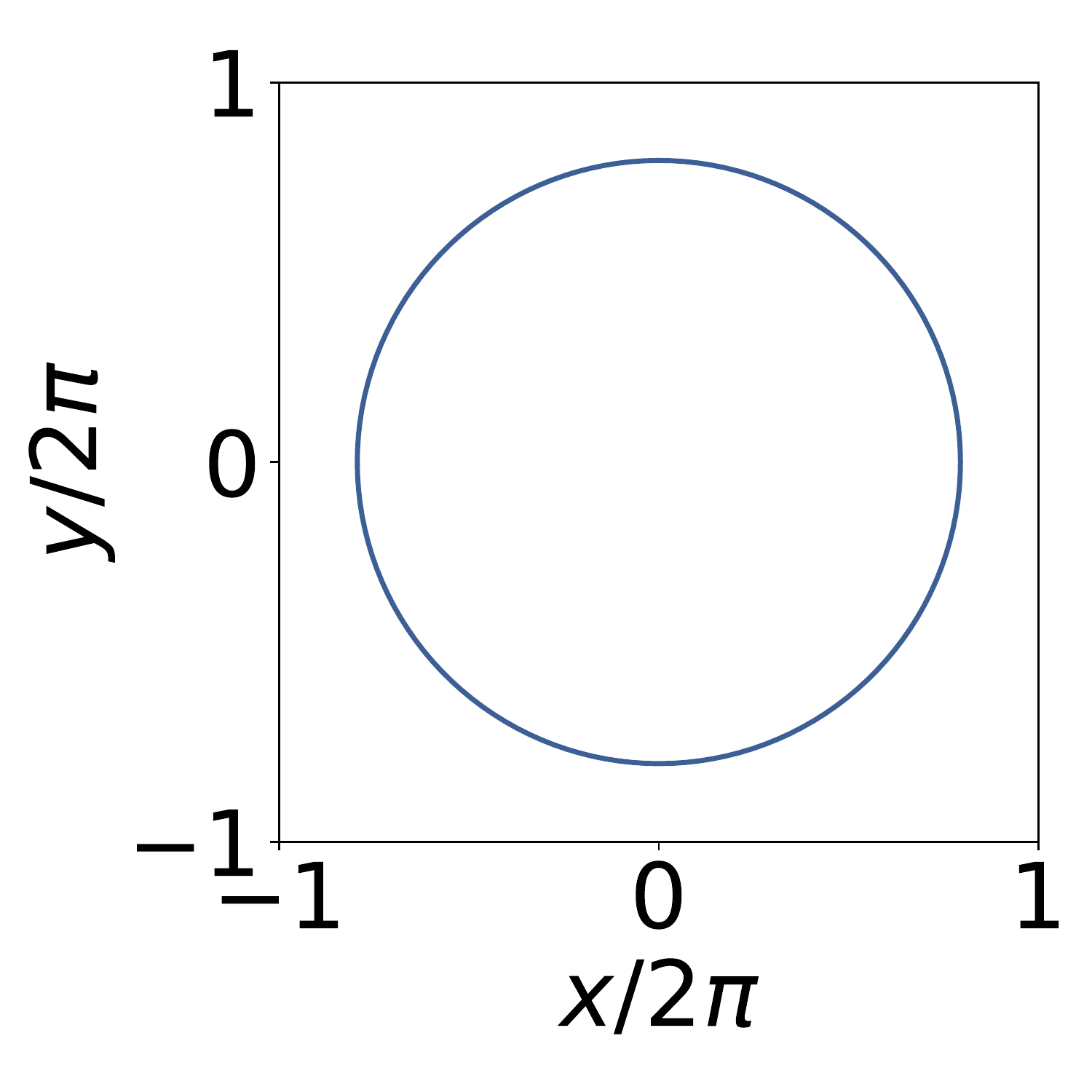}
		\end{minipage}& 
		\begin{minipage}{.16\textwidth}
			\includegraphics[width=\textwidth]{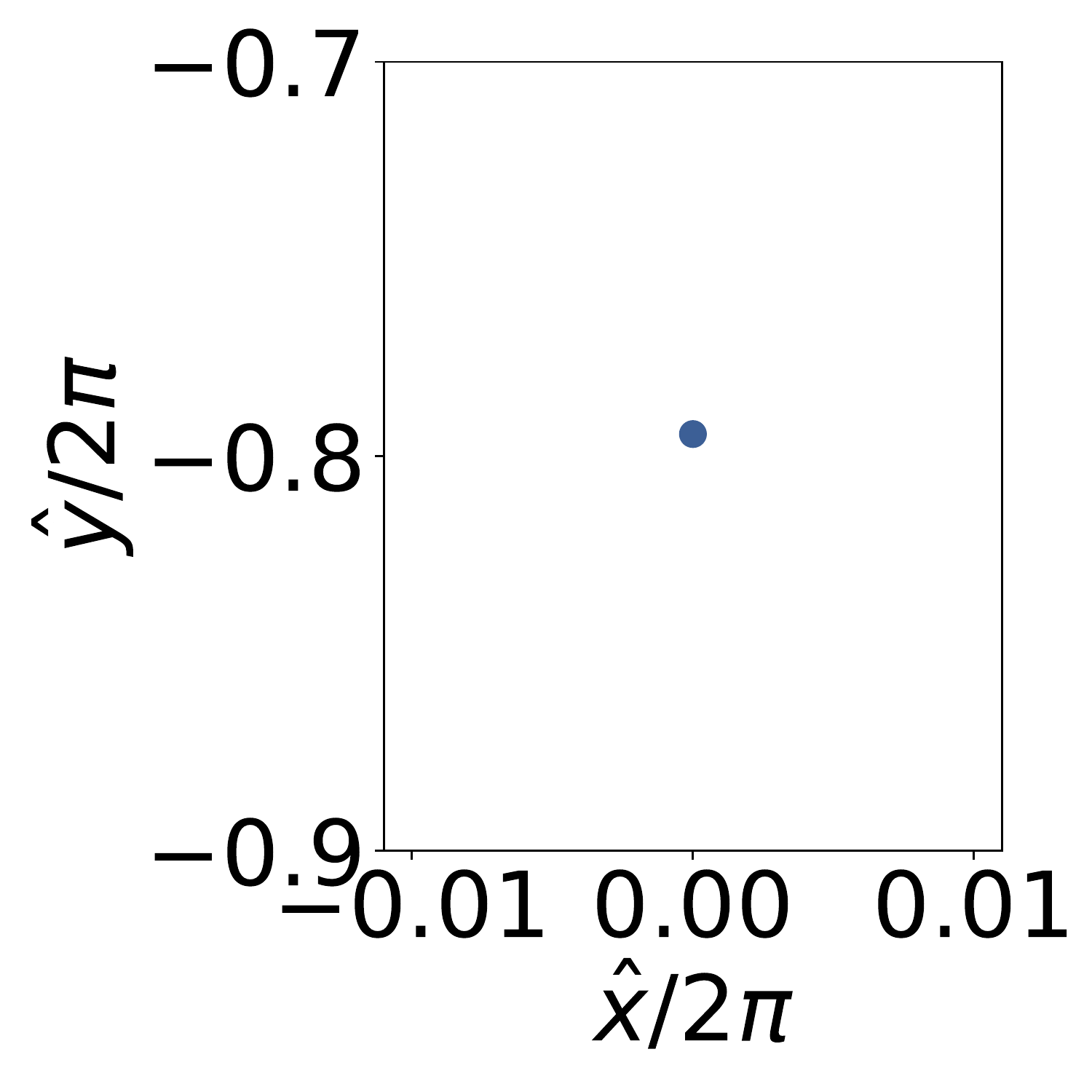} 
		\end{minipage}&  
		 $10.0$ \\ 
		\hline
		$\RPO{1o}$ & Relative periodic orbit & Oval & 
		\begin{minipage}{.16\textwidth}
			\includegraphics[width=\textwidth]{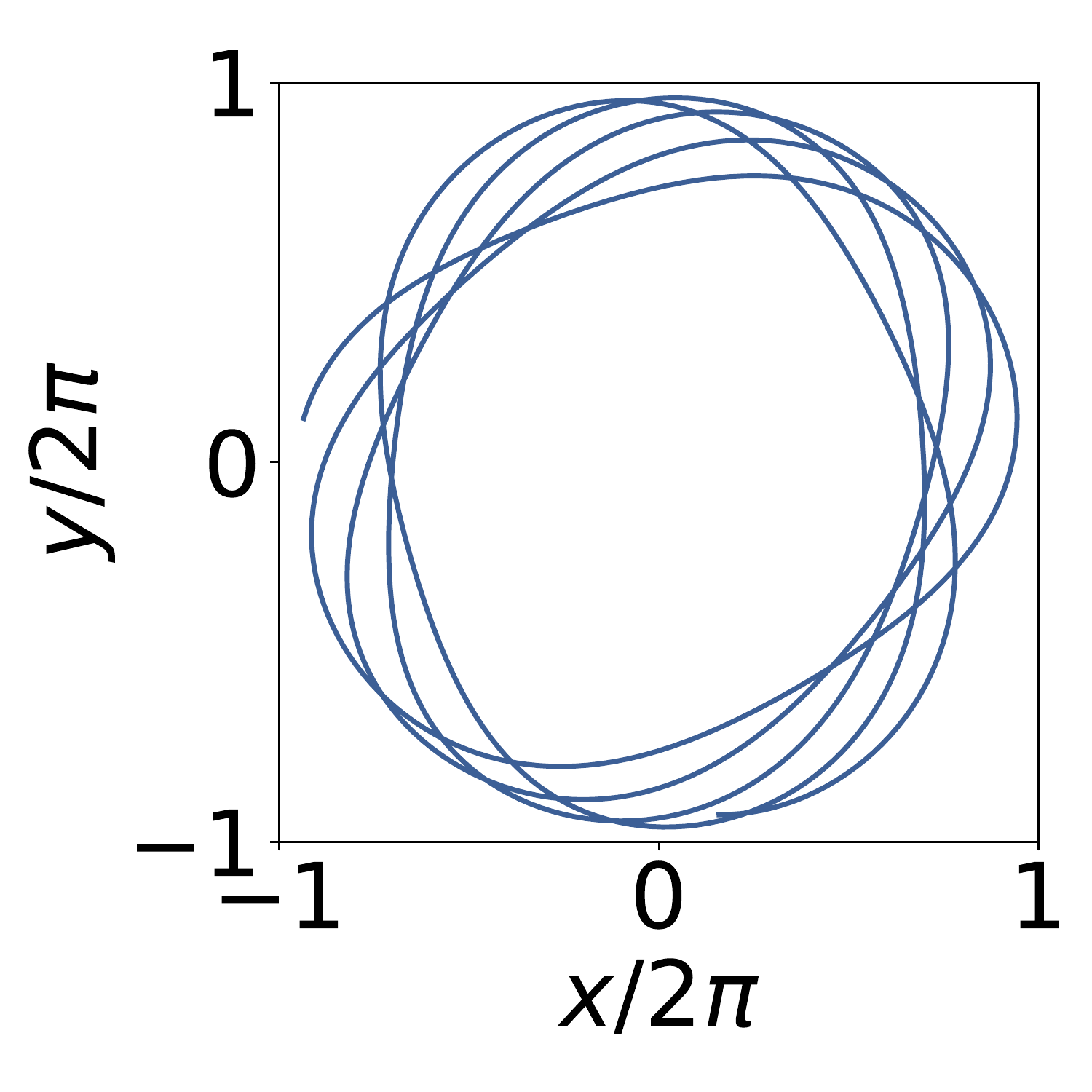}
		\end{minipage}& 
     	\begin{minipage}{.16\textwidth}
			\includegraphics[width=\textwidth]{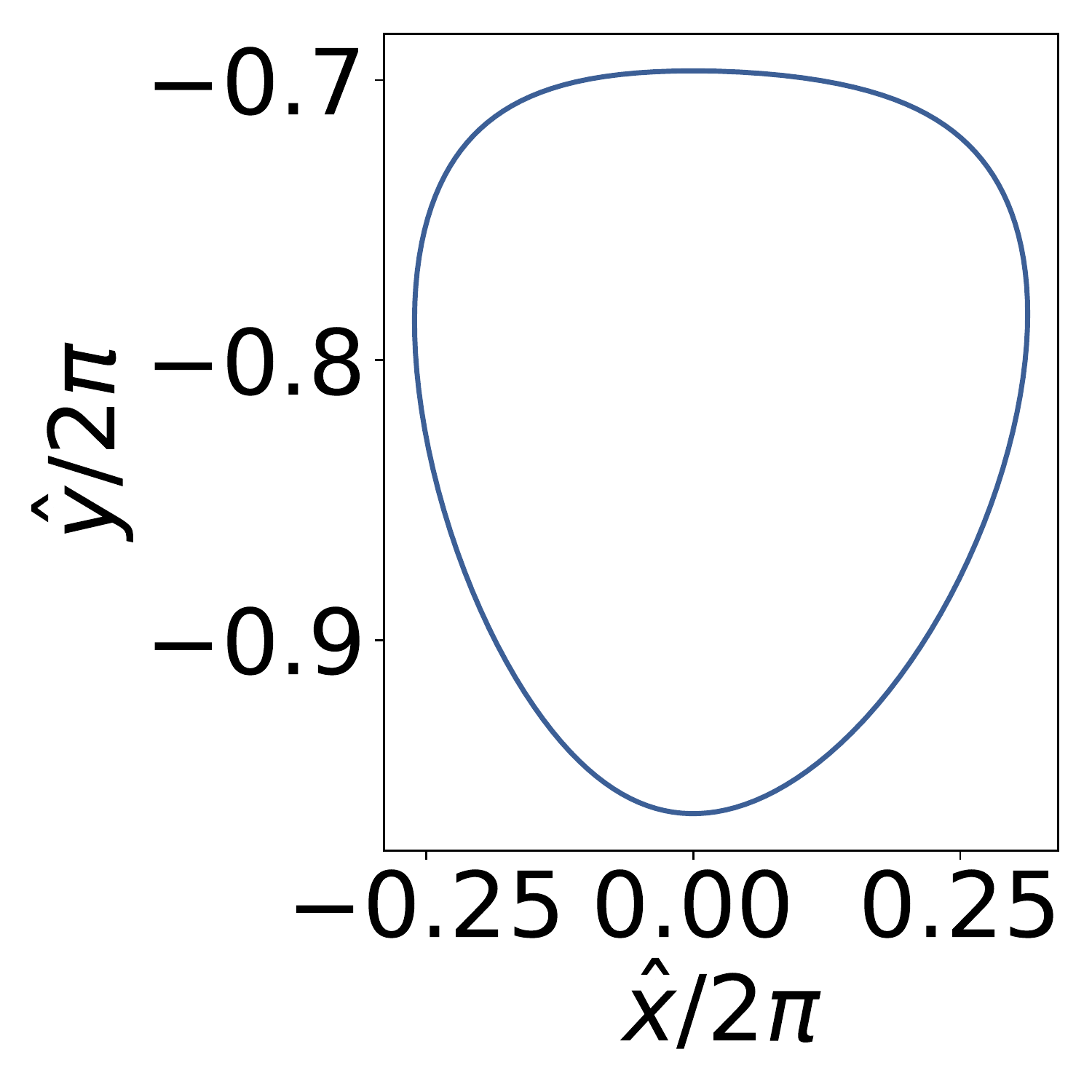} 
		\end{minipage}&  
		$15.0$ \\ 
		\hline
		$\RPO{4os}$ & Relative periodic orbit & Oval & 
		\begin{minipage}{.16\textwidth}
			\includegraphics[width=\textwidth]{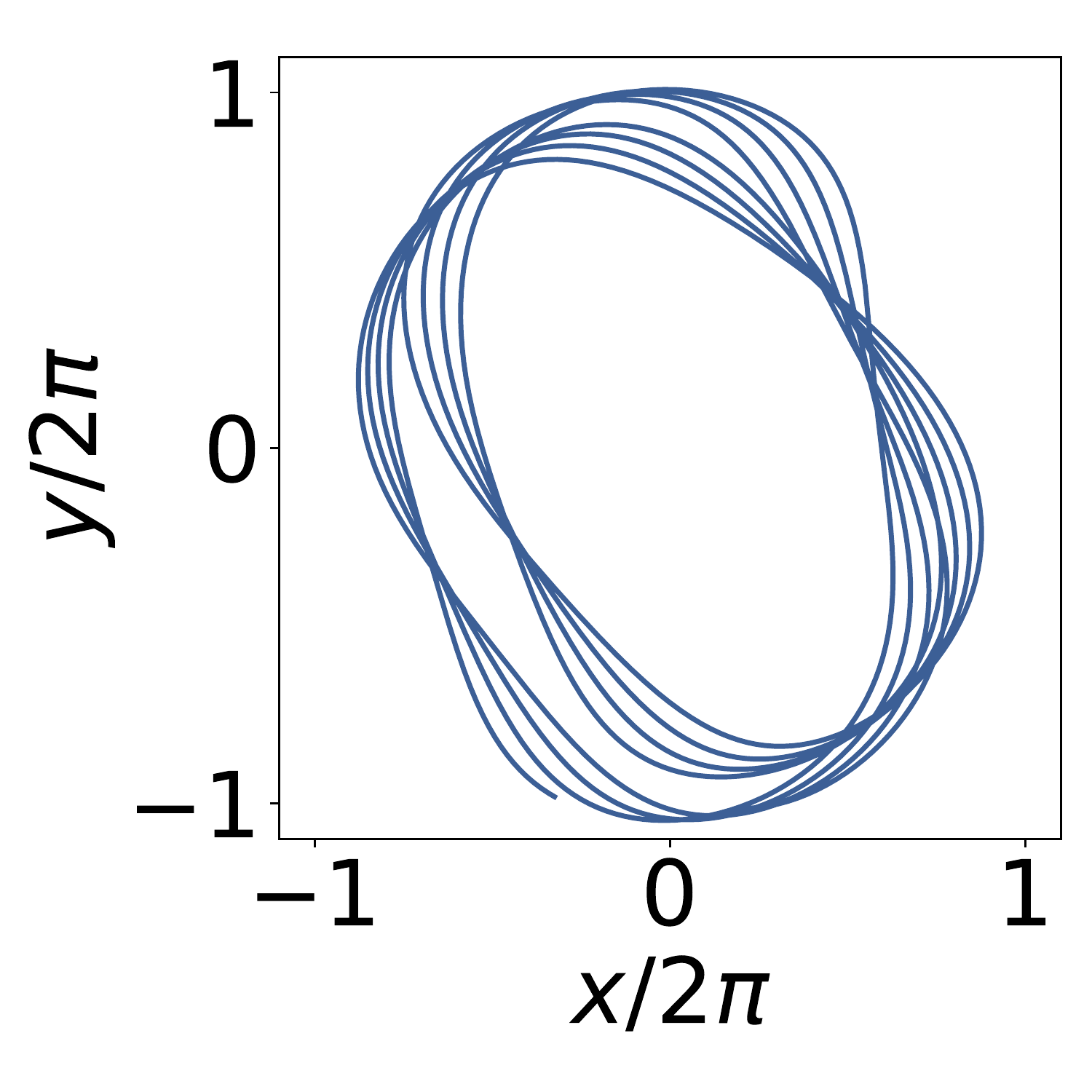}
		\end{minipage}& 
		\begin{minipage}{.16\textwidth}
			\includegraphics[width=\textwidth]{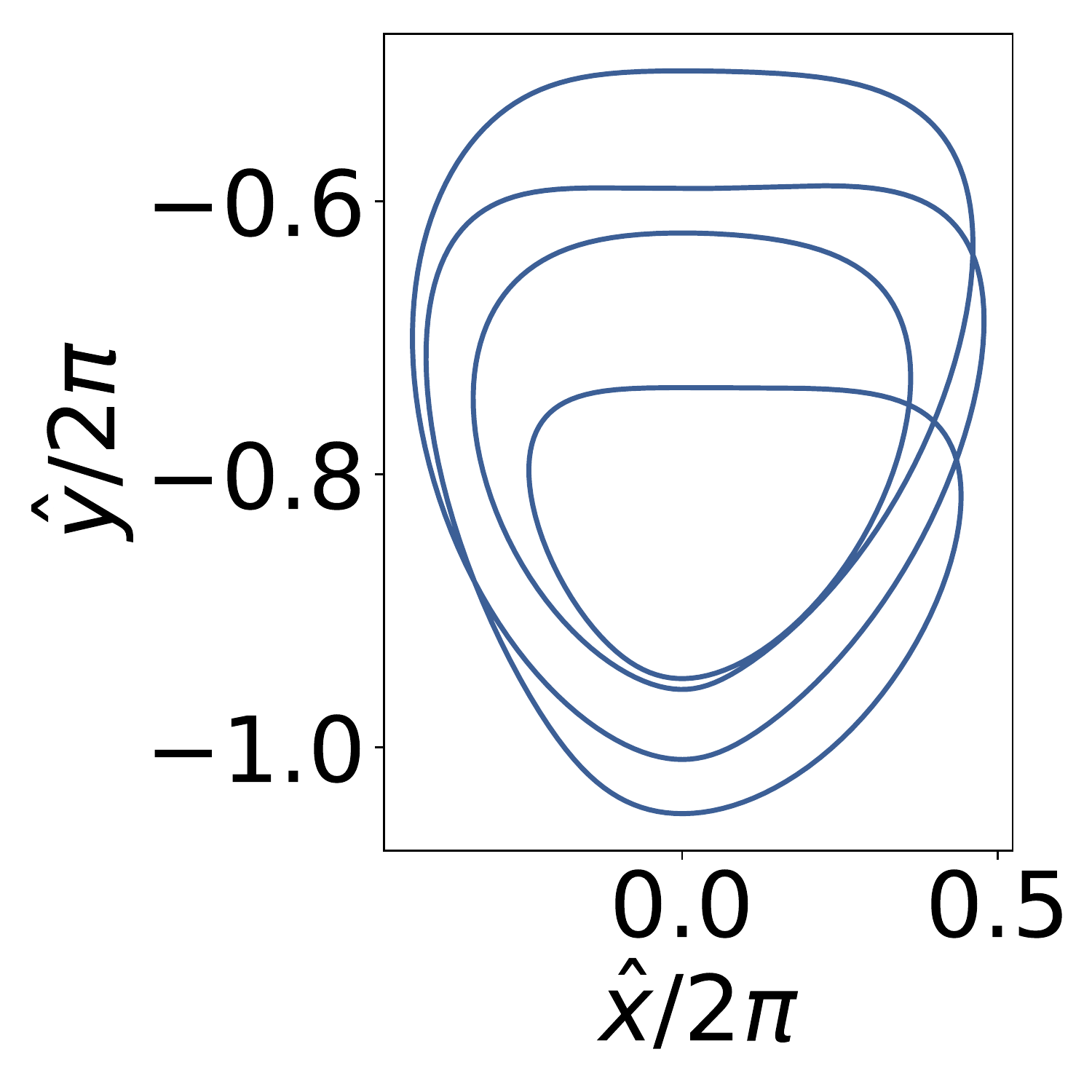} 
		\end{minipage}&  
		$18.0$ \\ 
		\hline
		$\RPO{4ou}$ & Relative periodic orbit & Oval & 
		\begin{minipage}{.16\textwidth}
			\includegraphics[width=\textwidth]{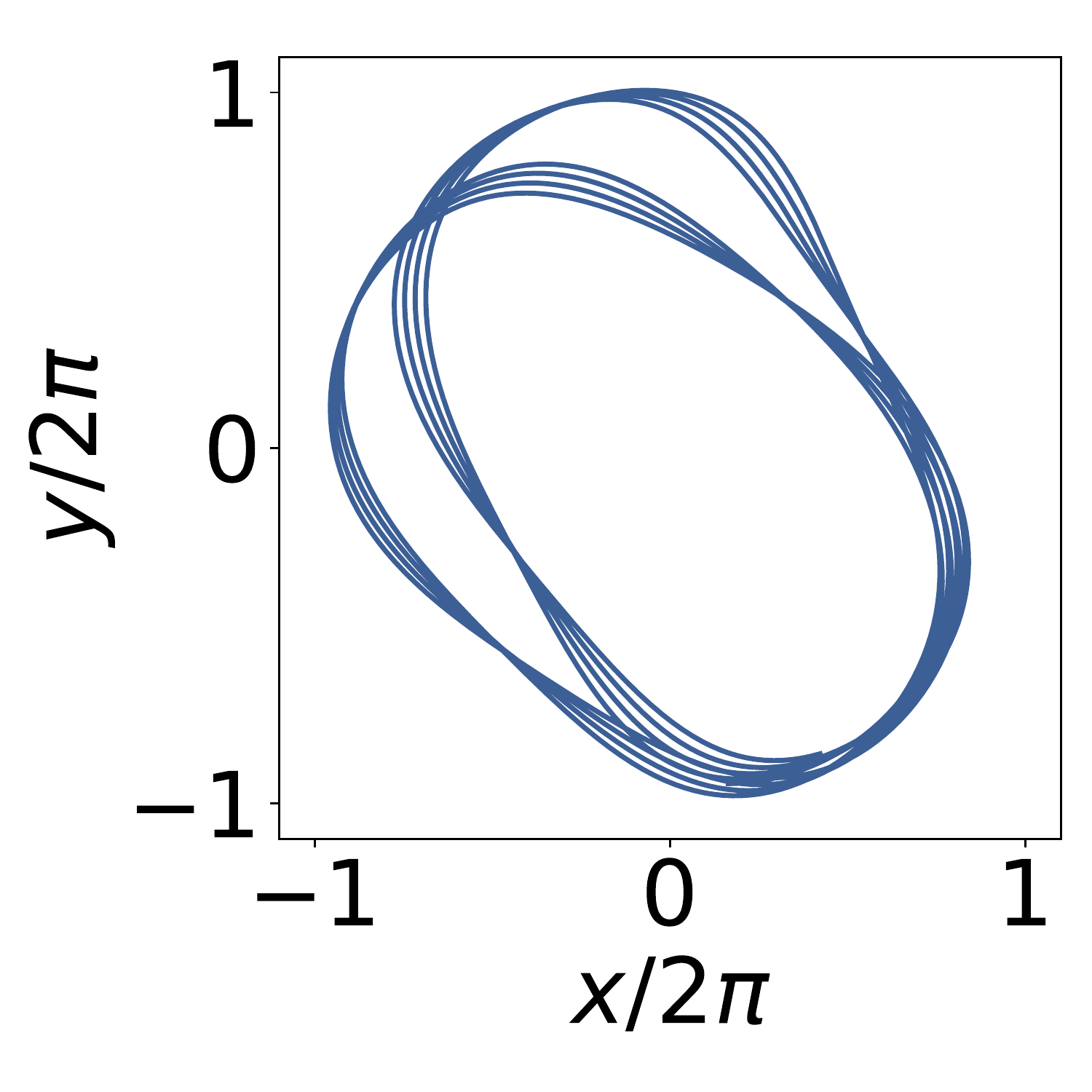}
		\end{minipage}& 
		\begin{minipage}{.16\textwidth}
			\includegraphics[width=\textwidth]{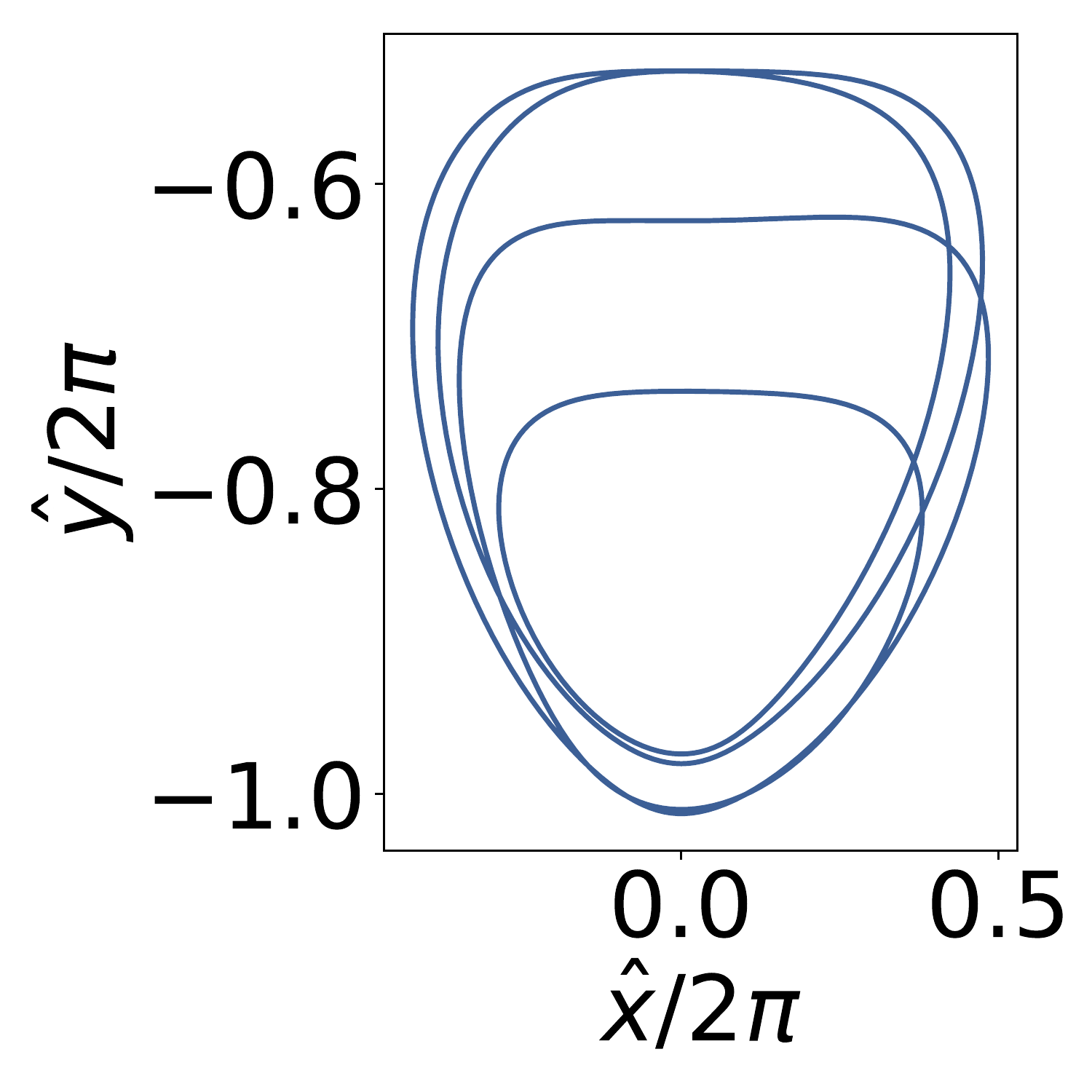} 
		\end{minipage}&  
		$18.0$\\ 
		\hline		
		$\PO{2l}$ & Periodic orbit & Lemniscate & 
		\begin{minipage}{.16\textwidth}
			\includegraphics[width=\textwidth]{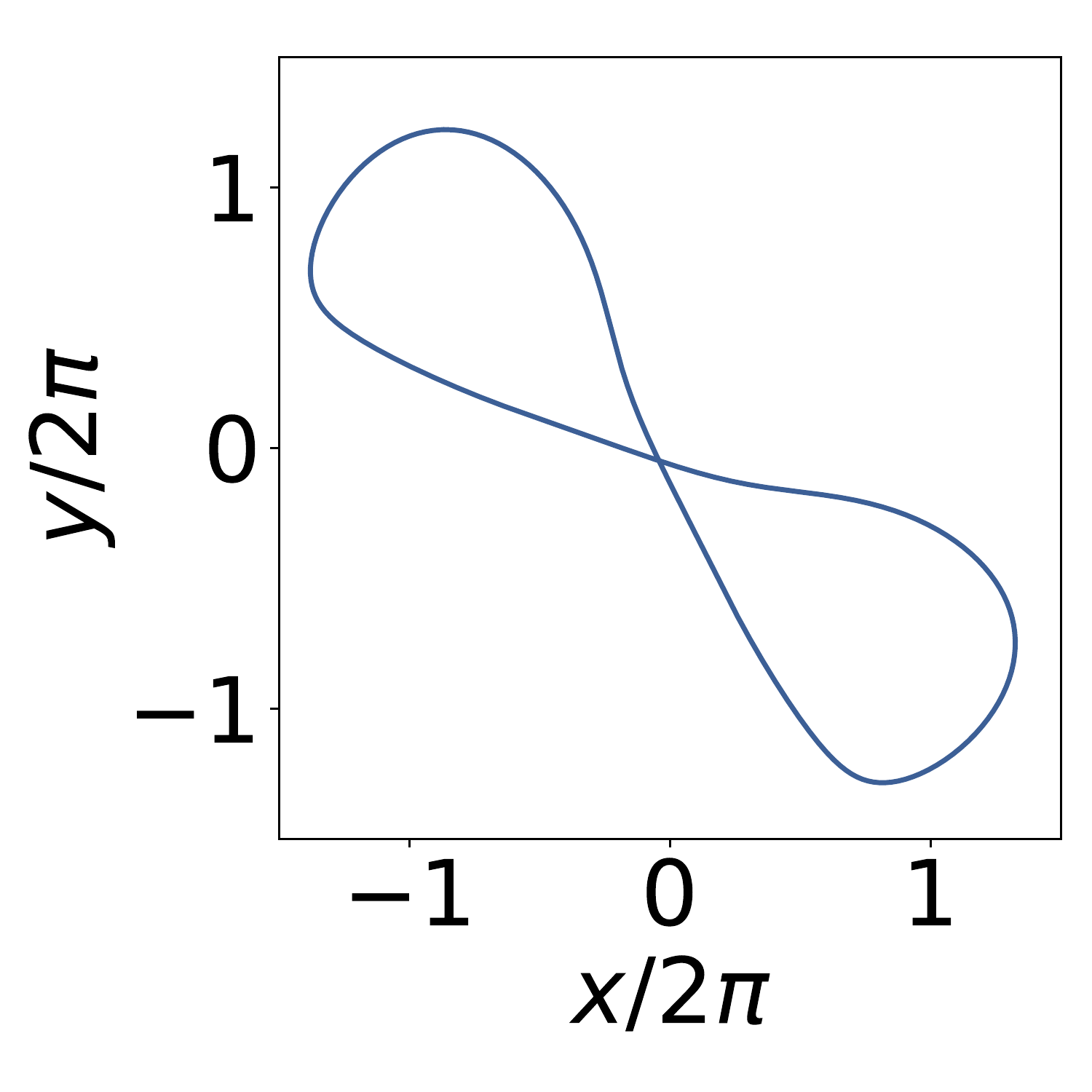}
		\end{minipage}& 
		\begin{minipage}{.16\textwidth}
			\includegraphics[width=\textwidth]{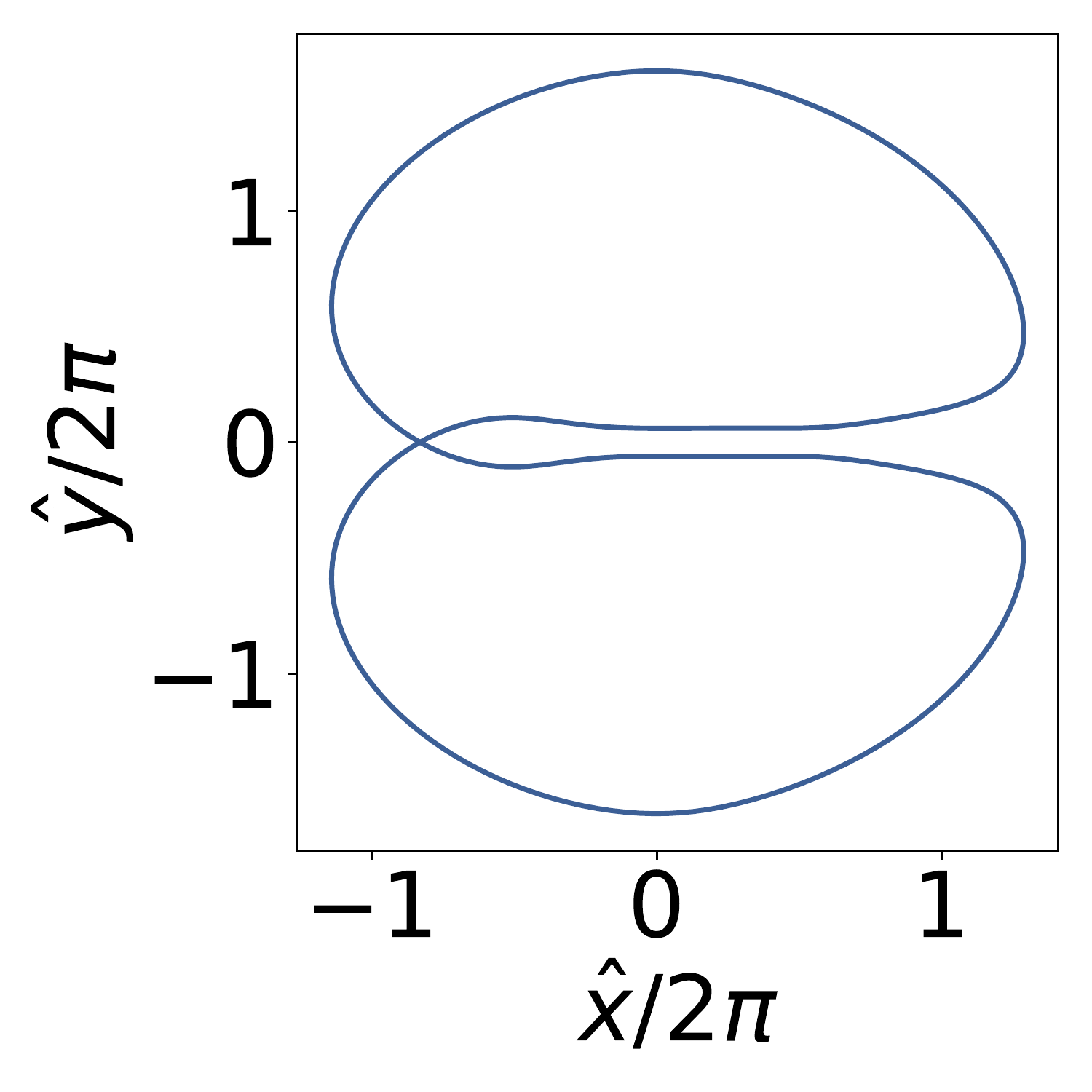} 
		\end{minipage}&  
		$14.0$ \\ 		
		\hline
		$\RPO{2l}$ & Relative periodic orbit & Lemniscate & 
		\begin{minipage}{.16\textwidth}
			\includegraphics[width=\textwidth]{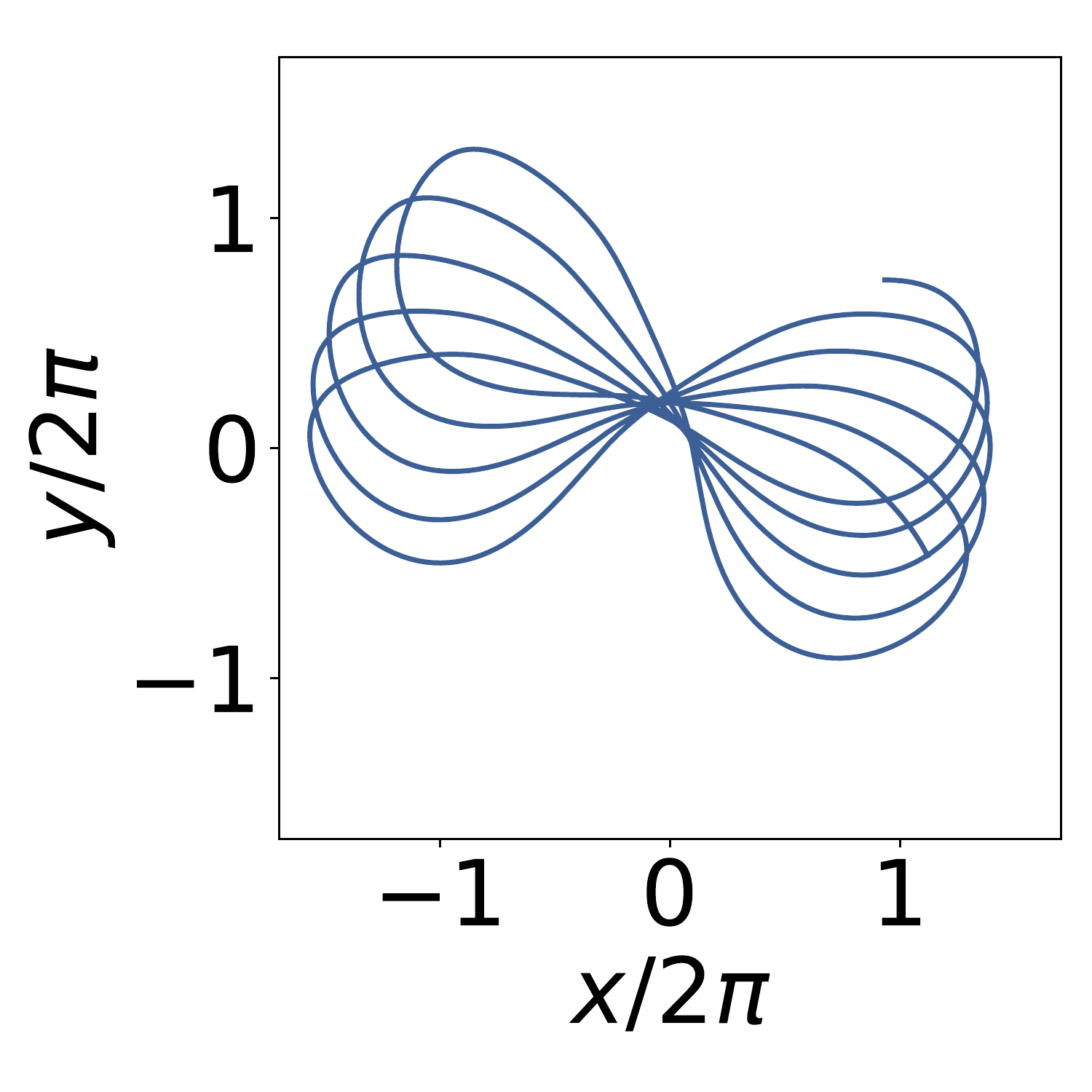}
		\end{minipage}& 
		\begin{minipage}{.16\textwidth}
			\includegraphics[width=\textwidth]{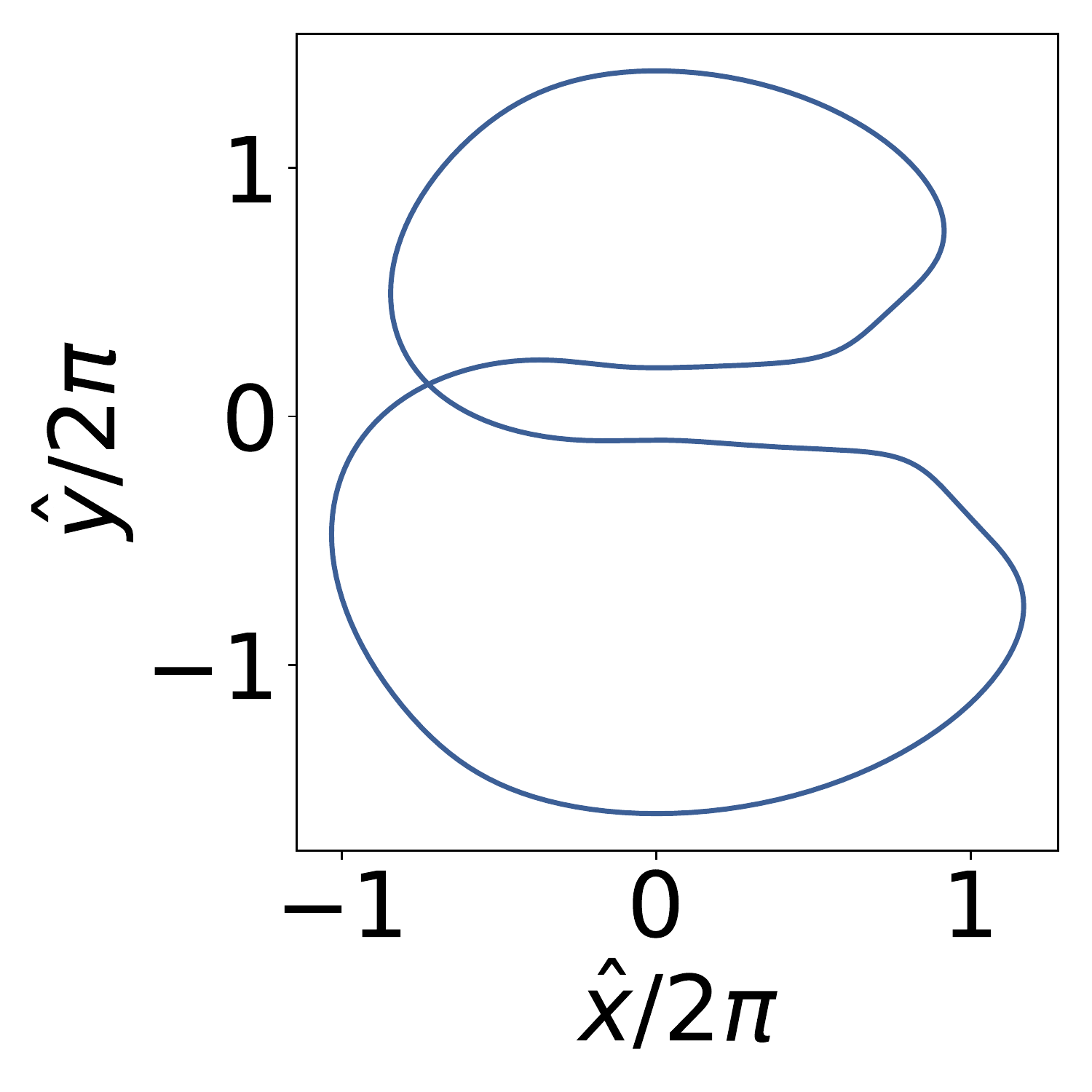} 
		\end{minipage}&  
		$16.0$ \\ 		
		\hline
		$\RPO{1t}$ & Relative periodic orbit & Trefoil & 
		\begin{minipage}{.16\textwidth}
			\includegraphics[width=\textwidth]{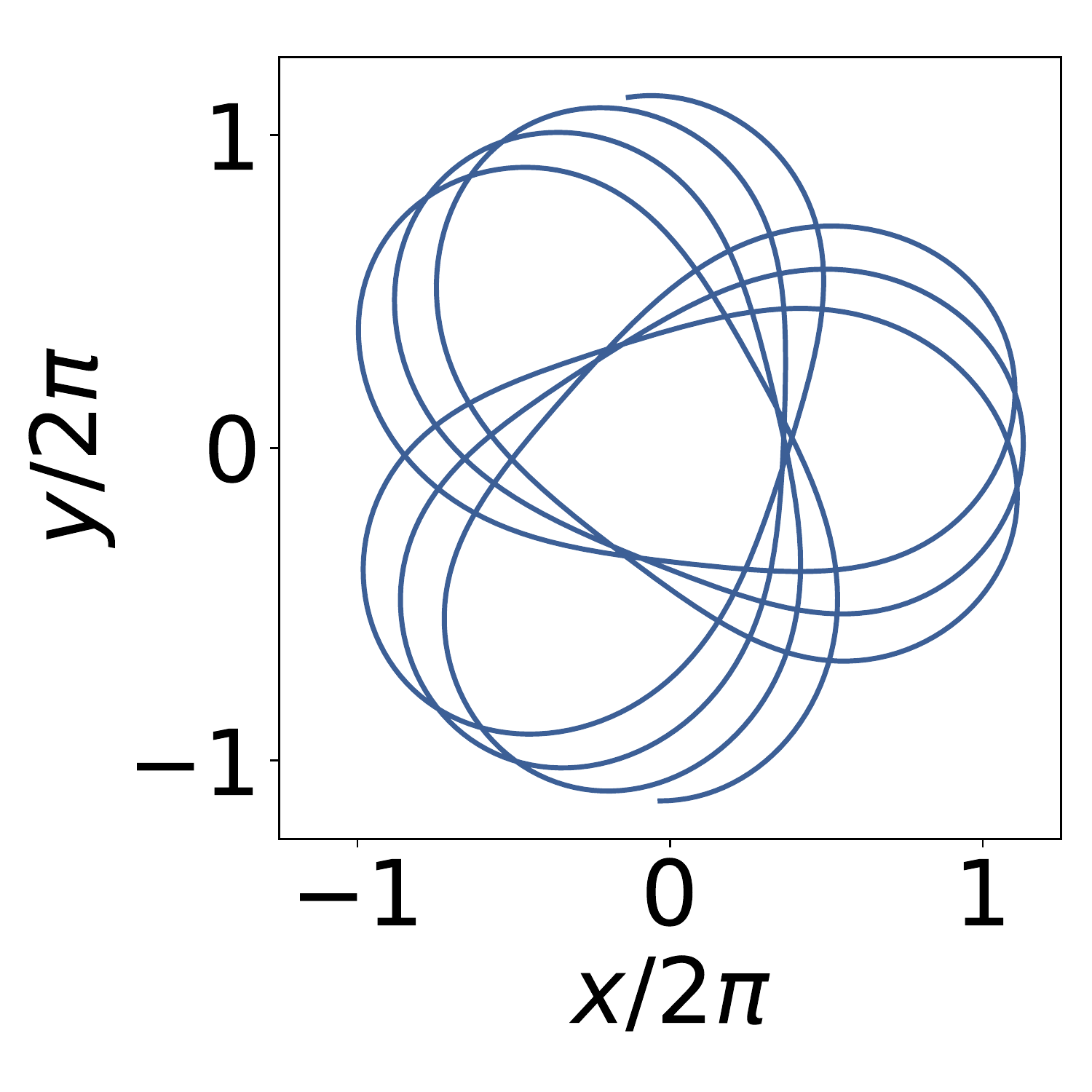}
		\end{minipage}& 
		\begin{minipage}{.16\textwidth}
			\includegraphics[width=\textwidth]{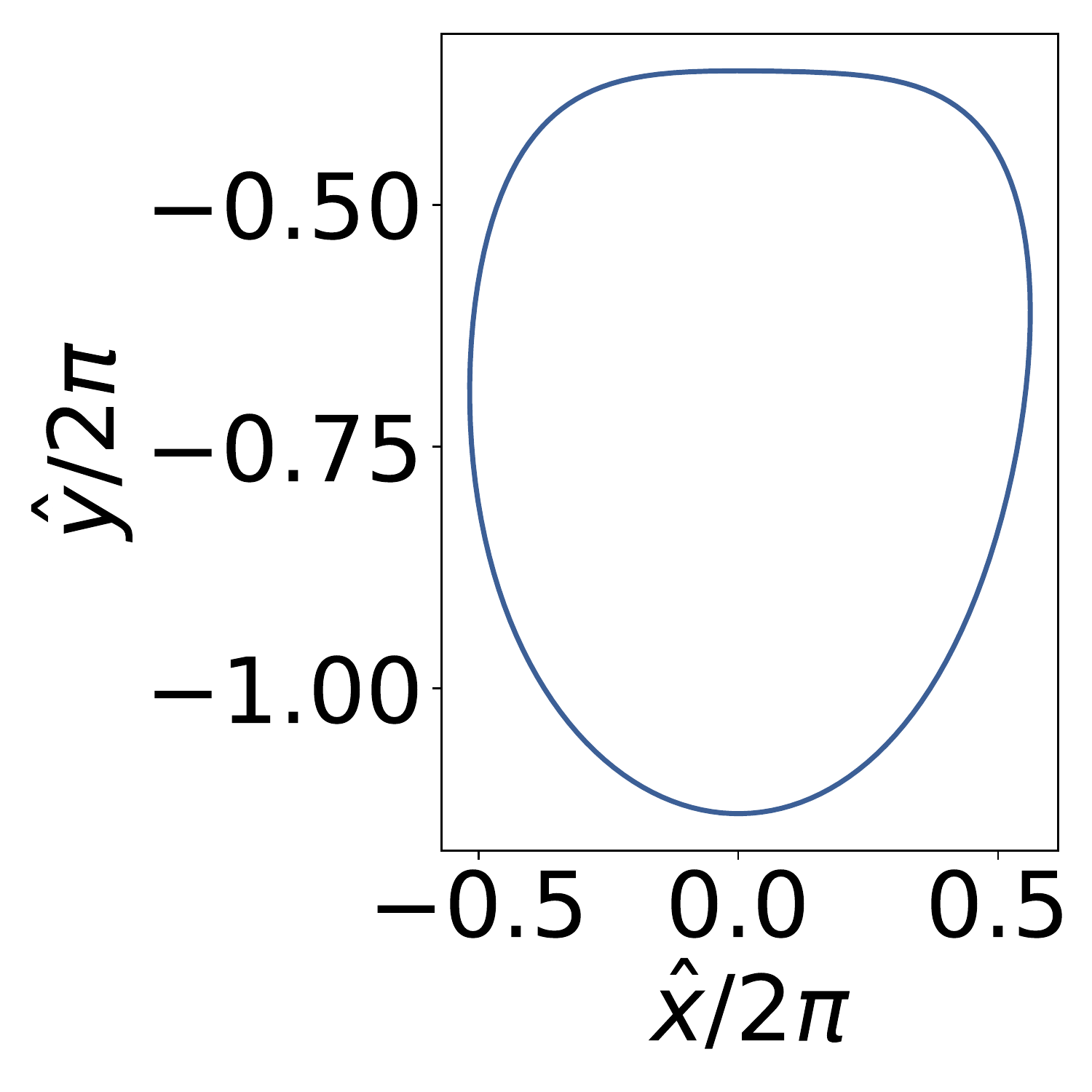} 
		\end{minipage}&  
		$14.0$	
	\end{tabular} 
\end{table*}

\begin{figure}
	\centering
	\begin{overpic}[width=0.23\textwidth]{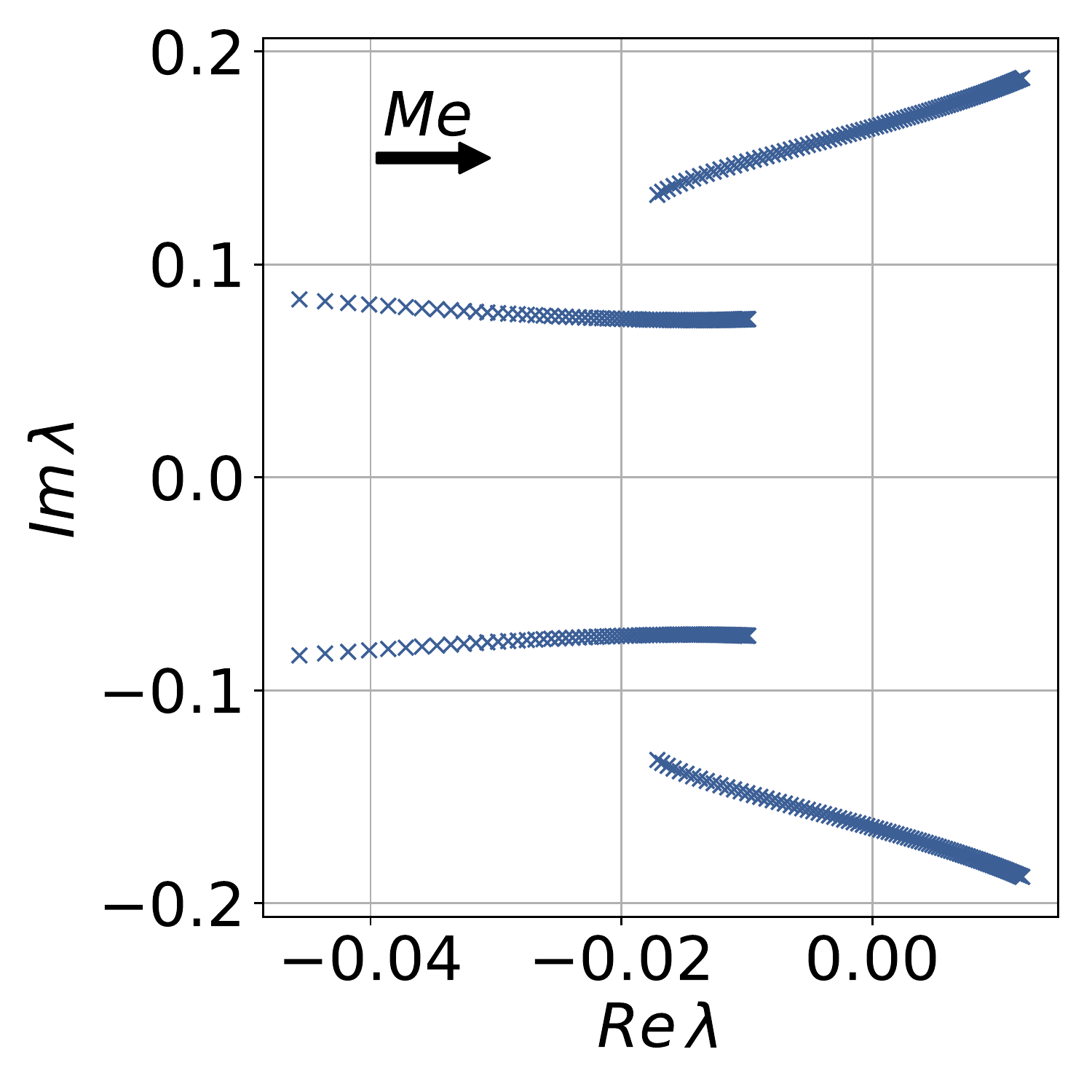}
		\put (0,0) {(a)}
	\end{overpic}                             
	\begin{overpic}[width=0.23\textwidth]{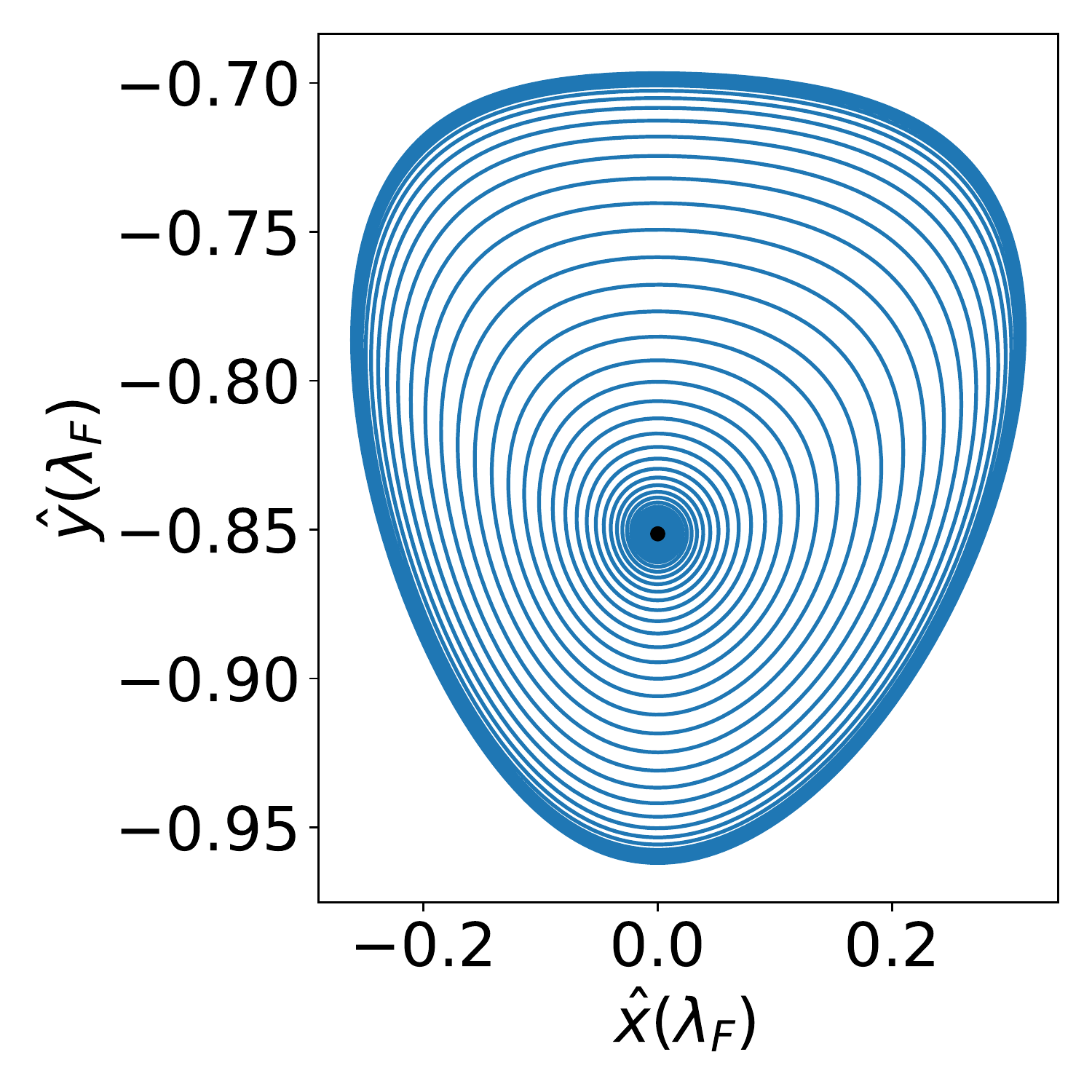}
		\put (0,0) {(b)}
	\end{overpic}
	\caption{Destabilization of the circular trajectory with 
		increasing $\memory \in [10, 19]$. 
		(a) Real and imaginary parts
		of the leading two pairs of complex 
		conjugate stability eigenvalues of the circular trajectory.
		Direction of the increasing memory is indicated by the 
		annotated arrow. (b) An orbit with the initial condition 
		\refeq{e-spiralIC} at $\memory = 15$ illustrating the 
		connection of the  unstable manifold of \REQ{0c} (marked 
		black) to \RPO{1o}. 		 
	}
	\label{f-circleDestab}
\end{figure}

\subsubsection{Bifurcations from the circular solution}
\label{s-circBif}

As shown by Labousse \etal \rf{LOPB2016} and Tambasco \etal \rf{THORB2016}, 
when $\memory = 10$,
the dynamical system described by \refeqs{e-ddotxNonDim}{e-SnNonDim}
has a stable circular orbit with radius $r \approx 0.8 \lambda_F$, 
which we have visualized in \reffig{f-circle}.
This solution is a relative equilibrium \refeq{e-reqv}, which we are 
going to refer to as $\REQ{0c}$. 
$\REQ{0c}$ intersects the slice hyperplane 
\refeq{e-sliceHplane} at a single point $\sspRed_{\REQ{0c}}$ 
corresponding to the snapshot
on \reffig{f-circle}(a). This solution satisfies the equilibrium
condition \refeq{e-reqRed} in the symmetry-reduced state space. 
Starting from the stable $\sspRed_{\REQ{0c}}$ at $\memory = 10.0$, 
we increased the $\memory$ in steps of $0.1$, find the 
root of \refeq{e-reqRed}
numerically  using
\texttt{fsolve} function from \texttt{scipy.optimize}\rf{scipy} 
and compute the spectrum 
of \refeq{e-StabMatRed}
at this point for each $\memory$ value. 
In these 
computations, we discarded the $4^{th}$ column of 
$\velRed (\sspRed)$, which corresponds to the 
$\hat{v}_y$ direction that is identically set to $0$ in 
the slice. Similarly, when computing the eigenvalues of 
\refeq{e-StabMatRed} we discard the $4^{th}$ column and row.

The leading (with the largest 
$\Re \lambda$) two pairs of linear stability eigenvalues of 
the circular trajectory computed this way are plotted in 
\reffig{f-circleDestab}(a). As shown, the leading pair of 
complex conjugate eigenvalues $\lambda_{1,2}$ cross the 
imaginary axis upon 
increasing \memory . This corresponds to the interval 
$\memory \in (13.7, 13.8)$ and at $\memory = 13.8$, the 
circular trajectory is unstable. For verification, we translated
this \memory\ value to the to the forcing acceleration 
$\gamma$ which is related to $\memory$ by 
$\memory\ T_F = T_d / (1 - \gamma / \gamma_F)$, where 
$T_d$ is the decay time of waves in the absence of forcing
and $\gamma_F$ is the Faraday instability threshold. 
With $T_F = 2 / f = 0.025 s$ 
and $T_d \approx 0.0182$\rf{LOPB2016}, $\memory = 13.8$
corresponds to $\gamma / \gamma_F \approx 0.9472$. 
This value is in agreement with Tambasco \etal \rf{THORB2016} who 
reported the destabilization of the circular orbit at 
$\gamma/\gamma_F = 0.948$.

The destabilization of the circular orbit as \memory\ is increased 
corresponds to a supercritical Hopf bifurcation as we confirm by the
appearance of a relative periodic orbit in its vicinity. This is 
illustrated in \reffig{f-circleDestab}(b), where we show an orbit
at $\memory = 15$ starting approximately on the unstable manifold 
of \REQ{0c}\ with the initial condition
\beq
	\sspRed(0) = \sspRed_{\REQ{0c}} + 10^{-4} \Re \hat{e}_1 \, .
	\label{e-spiralIC}
\eeq

\begin{figure*}
	\centering
	\begin{overpic}[width=0.25\textwidth]{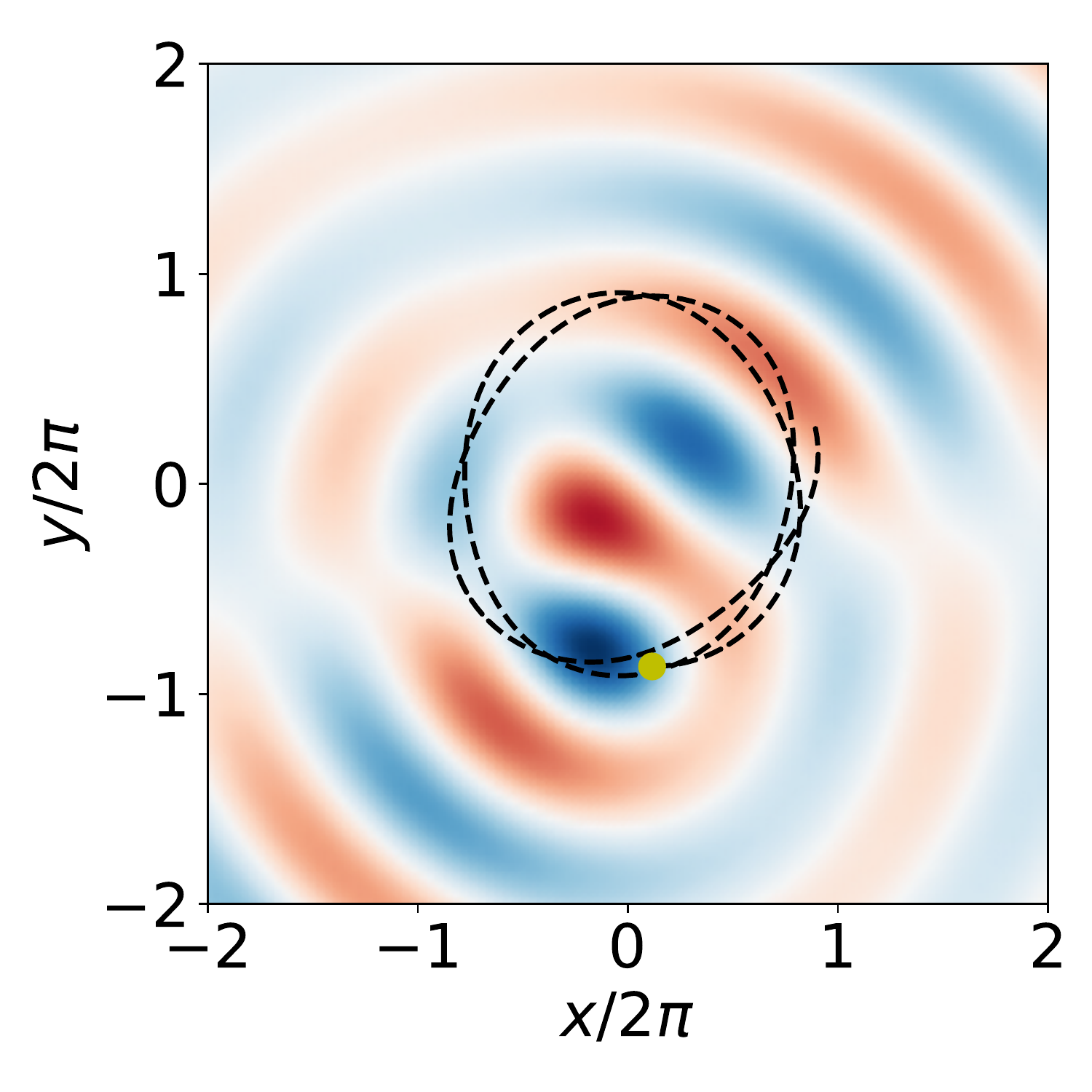}
		\put (0,2) {(a)}
	\end{overpic}                             
	\begin{overpic}[width=0.25\textwidth]{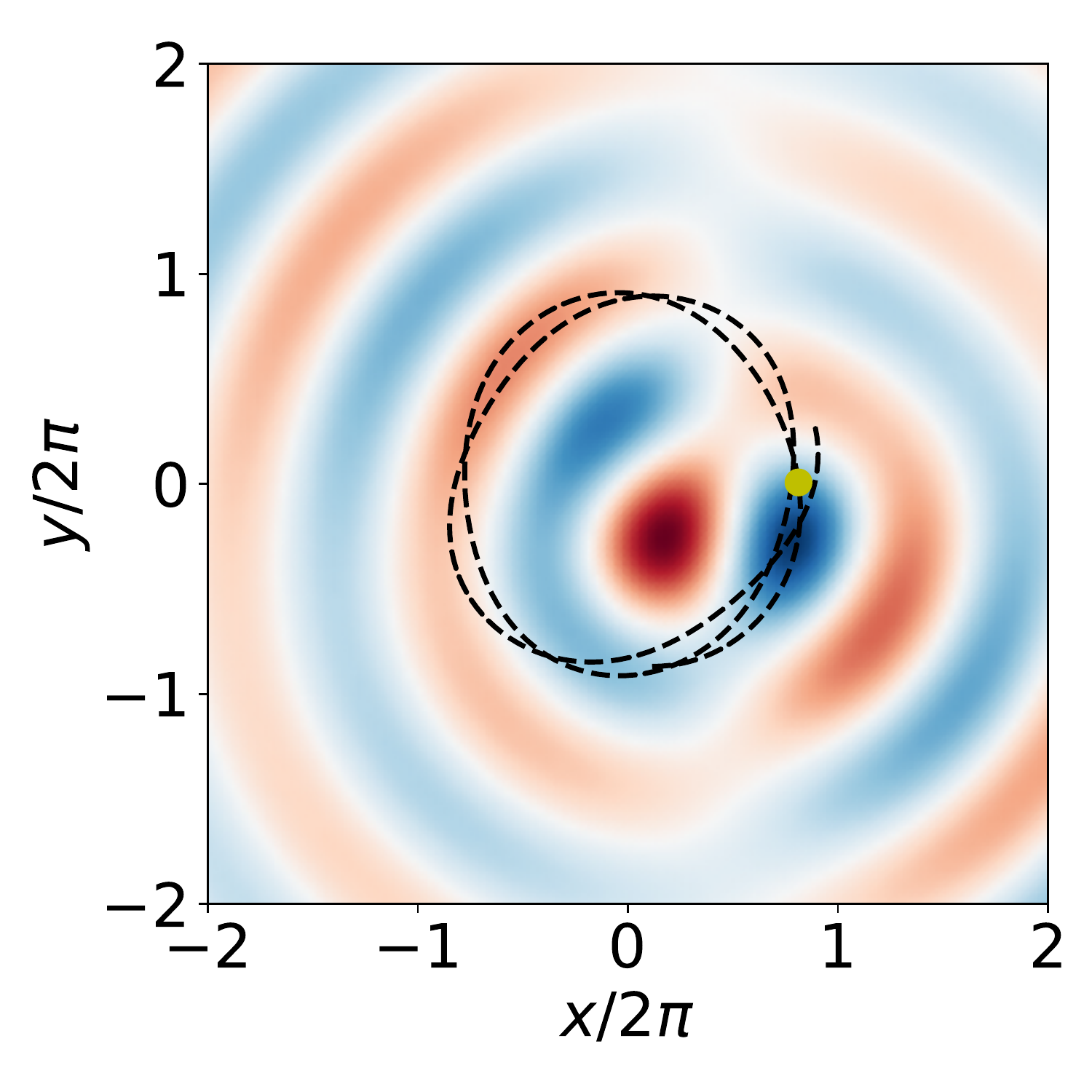}
		\put (0,2) {(b)}
	\end{overpic}
	\begin{overpic}[width=0.25\textwidth]{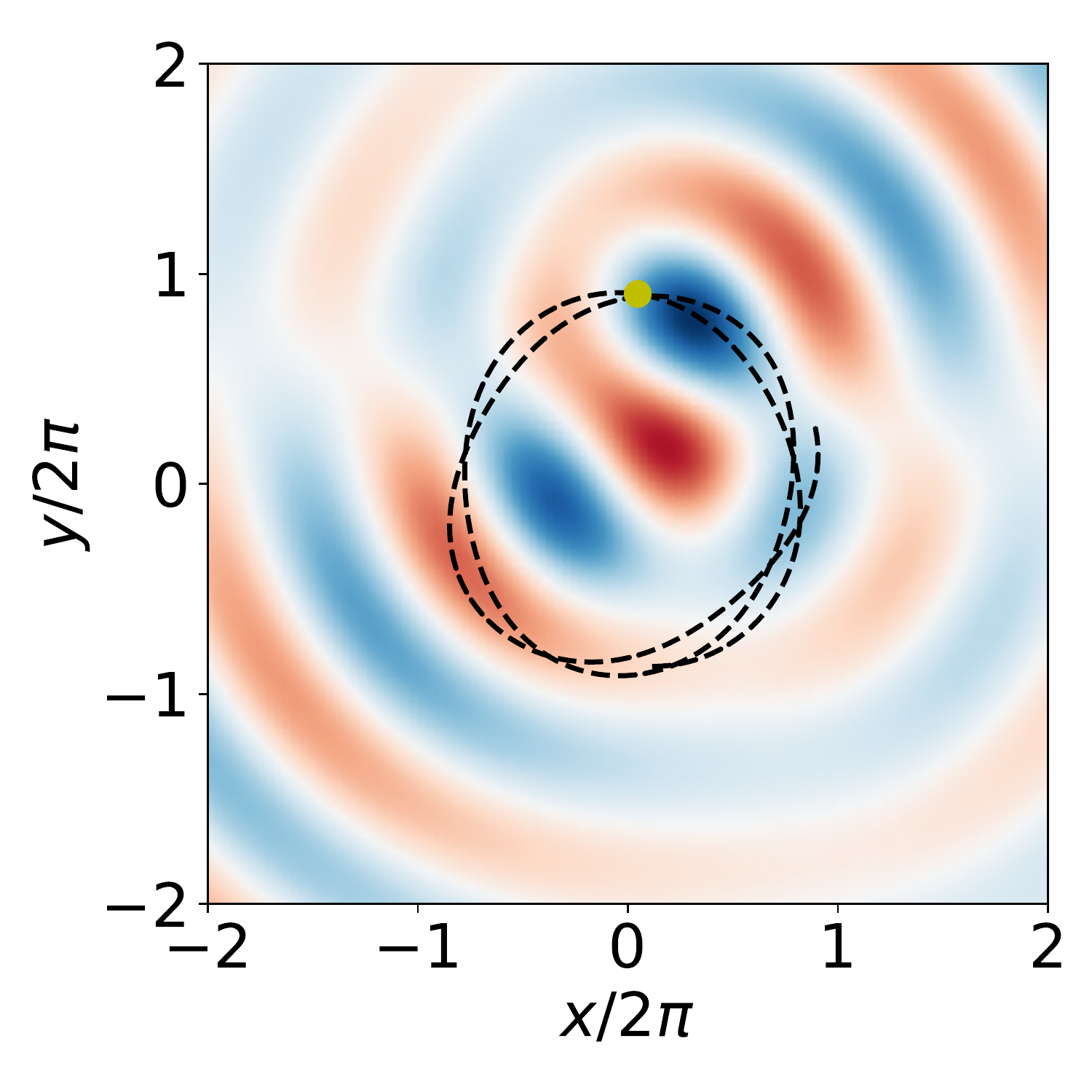}
		\put (0,2) {(c)}
	\end{overpic}
	\includegraphics[height=0.25\textwidth]{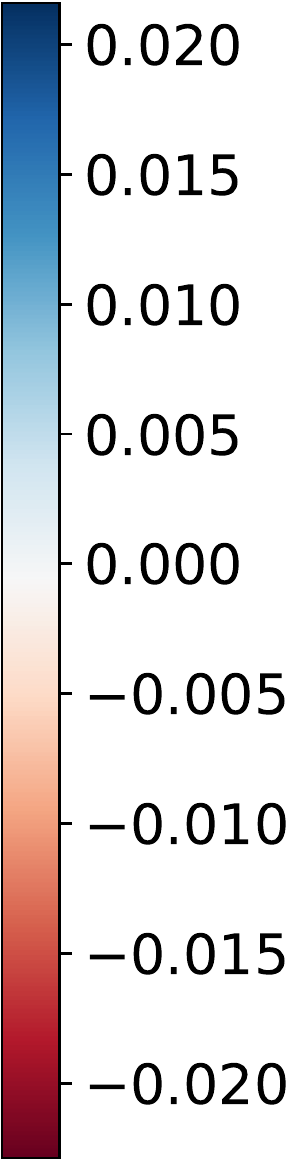}
	\caption{
		Three snapshots of a simulated trajectory of \RPO{1o}
		where the trace of the droplet for five periods
		is drawn as a dashed curve
		and its instantaneous position at the respective snapshot is 
		indicated with a yellow dot in each figure. 
		The magnitude of the wave field color coded in each snapshot. 
		(a) $\zeit = 0$, (b) $\zeit = 20$, (c) $\zeit = 40$. 
	}
	\label{f-rpoZero}
\end{figure*}

In order to find this relative periodic orbit to the numerical precision, 
we implemented Newton's method  (\refAppe{s-Newton})
for periodic orbits to find the 
roots of \refeq{e-rpoRed}. Once again, discarding the $4$-th element of
the equation, which is set to $0$ by the symmetry reduction. 
We visualized the relative periodic orbit's trajectory on 
$(x,y)$ plane along with the corresponding wave field at $\memory=15$ 
as three snapshots in \reffig{f-rpoZero}. Perrard \etal\ 
\rf{PLMFC2014} referred to the orbits with similar trajectories as 
``ovals'', which we will also adopt for the family of solutions to follow. 
At $\memory = 15$, 
\RPO{1o} is stable and its period
is $T_{\RPO{1o}} \approx 40.04$, thus, snapshots in 
\reffig{f-rpoZero}(a) and (c)
are approximately one period apart. Notice that the wave fields in 
\reffig{f-rpoZero} (a) and (c) are almost the same up to a rotation
of the coordinates.

\begin{figure}
	\centering
	\begin{overpic}[width=0.23\textwidth]{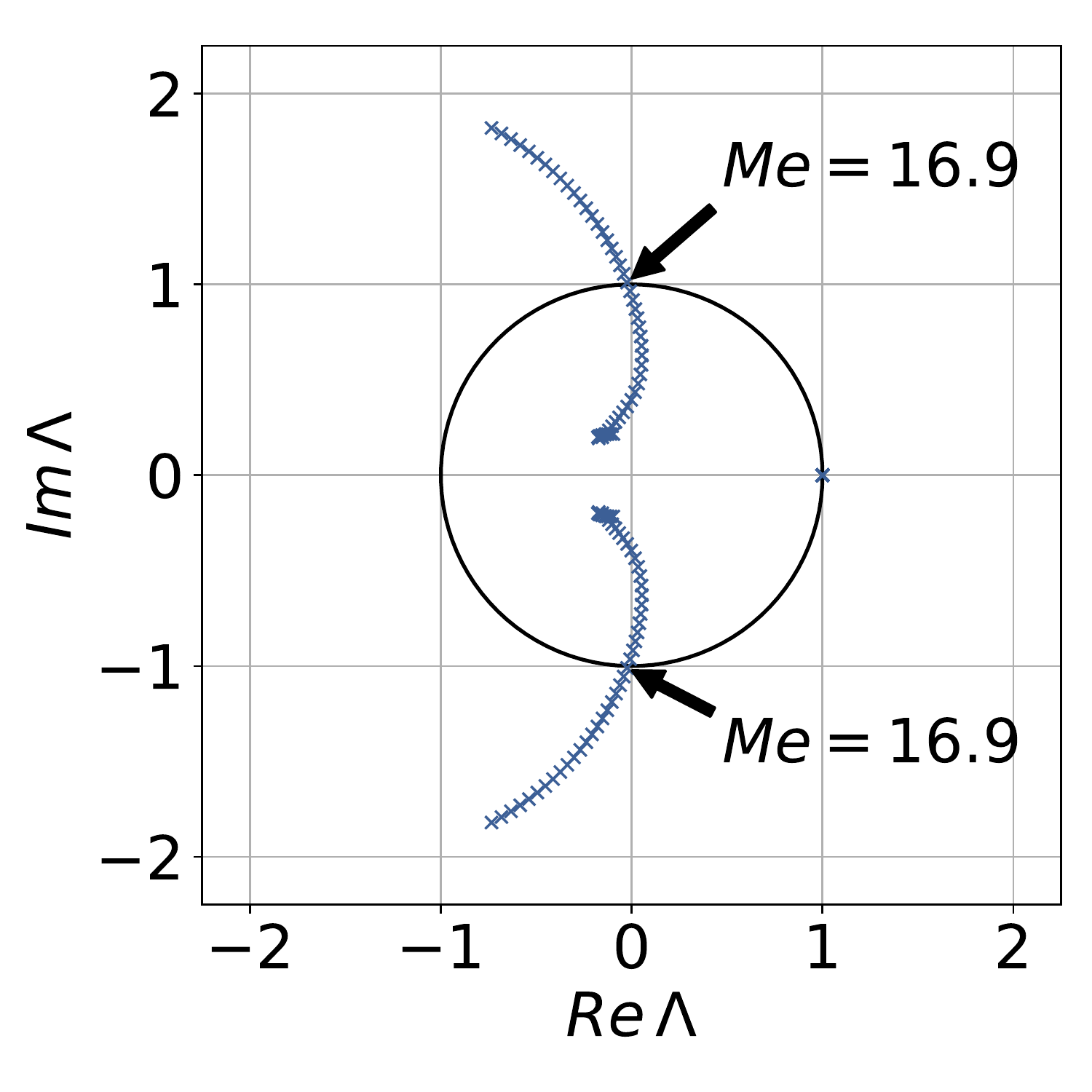}
		\put (0,0) {(a)}
	\end{overpic}                             
	\begin{overpic}[width=0.23\textwidth]{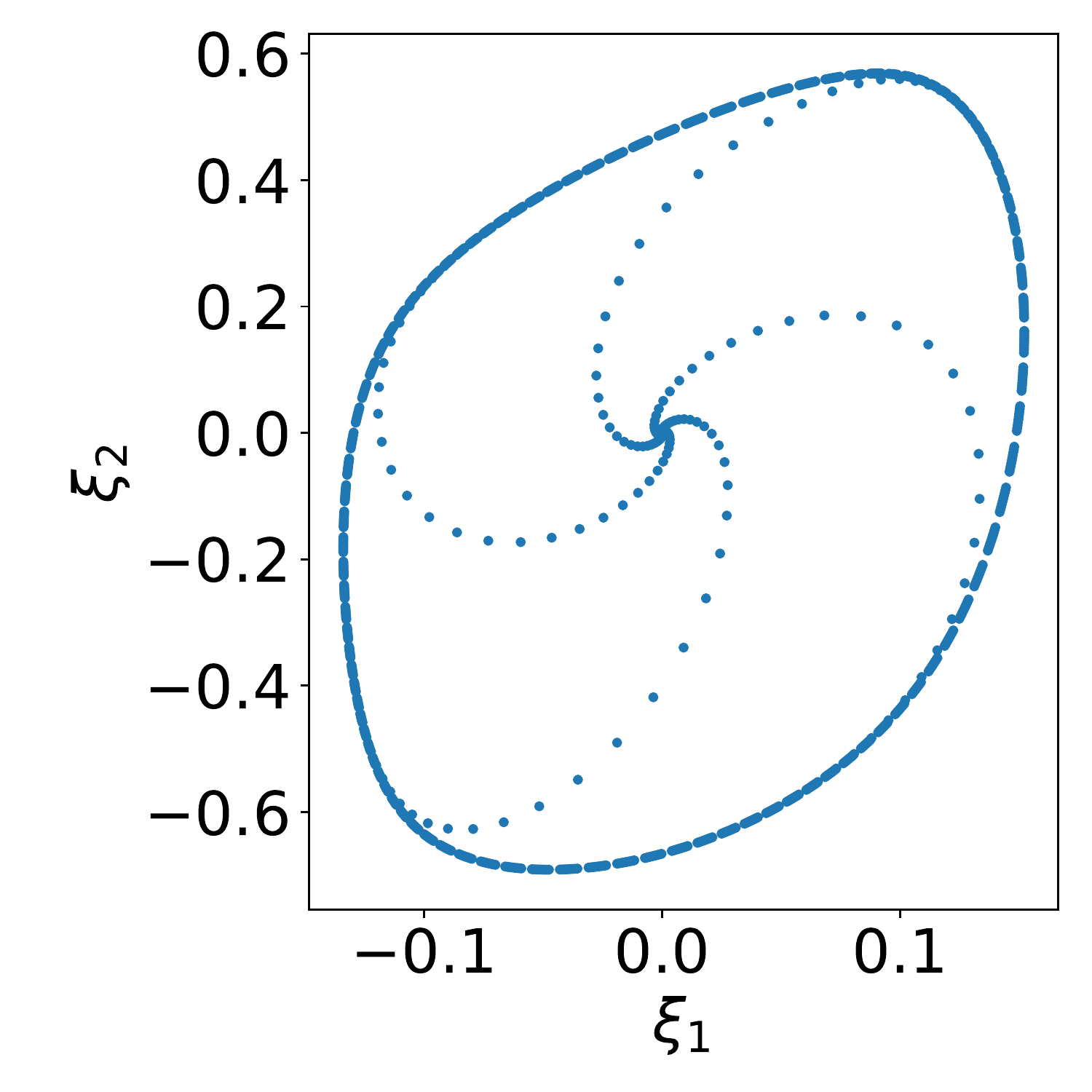}
		\put (0,0) {(b)}
	\end{overpic}
	\caption{Destabilization of the \RPO{1o} with 
		increasing $\memory \in [14, 19]$. 
		(a) Real and imaginary parts
		of the leading three Floquet multipliers of \RPO{1o}.
		$\memory = 16.9$ at which the leading complex conjugate 
		Floquet multipliers cross the unit circle is annotated. 
		(b) An orbit with the initial condition 
		\refeq{e-rpo0sipralIC} at $\memory = 17$ illustrating 
		the connection of \RPO{1o}'s unstable manifold to the 
		surrounding 2-torus as a projection from the Poincar\'e 
		section \refeq{e-PoincS}. 		 
	}
	\label{f-rpoZeroDestab}
\end{figure}

As we did with the circular solution, we varied 
$\memory \in [14, 19]$ with steps of $0.1$ and numerically solved
\refeq{e-rpoRed} using Newton's method
and obtained the Floquet 
multipliers of \RPO{1o} by computing the eigenvalues of 
\refeq{e-rpoStability} at each step. We plotted the leading 
(with largest $|\Lambda_i|$) three Floquet multipliers of 
\RPO{1o} for $\memory \in [14,19]$ 
on \reffig{f-rpoZeroDestab}(a). For each \memory\ value, there 
is a marginal Floquet multiplier with $\Lambda = 1$ corresponding
to the perturbations along the periodic orbit, as expected. Leading
non-marginal pair of complex conjugate Floquet multipliers cross 
the unit circle at $\memory \in (16.8, 16.9)$ (annotated in 
\reffig{f-rpoZeroDestab}(a)) and after this point, \RPO{1o} is 
unstable.

In order to study the dynamics on the $(\Re V_1, \Im V_1)$-plane
in the vicinity of $\sspRed_{\RPO{1o}}$ after it becomes unstable,
we define a Poincar\'e section as the half-hyperplane
\beq
	\inprod{\sspRed_\PoincS - \sspRed_{\RPO{1o}}}{
			\velRed(\sspRed_{\RPO{1o}})} = 0 \,,\quad
	\inprod{\velRed(\sspRed_\PoincS)}{
			\velRed(\sspRed_{\RPO{1o}}) } > 0, \label{e-PoincS}
\eeq
where $\sspRed_{\RPO{1o}}$ is an arbitrary point on $\RPO{1o}$ and
$\sspRed_\PoincS$ denotes the state space points on the Poincar\'e 
section. We started a simulation with the initial condition
\beq
	\sspRed(0) = \sspRed_{\RPO{1o}} + 10^{-4} \Re V_1 
	\label{e-rpo0sipralIC}
\eeq
and tracked its intersections with the Poincar\'e section 
\refeq{e-PoincS}. In order to visualize the dynamics on this 
Poincar\'e section, we projected it onto the orthonormal bases 
$(\hat{\xi}_1, \hat{\xi}_2)$ that span the $(\Re V_1, \Im V_1)$-plane
in \reffig{f-rpoZeroDestab}(b) at $\memory = 17$. As suggested 
by the appearance of an invariant 2-torus in the vicinity of \RPO{1o},
its leading complex conjugate Floquet multipliers' crossing of the 
unit circle corresponds to a supercritical Neimark-Sacker 
bifurcation\rf{Kuznetsov2010}. 

\begin{figure}
	\centering
	\begin{overpic}[height=0.23\textwidth]{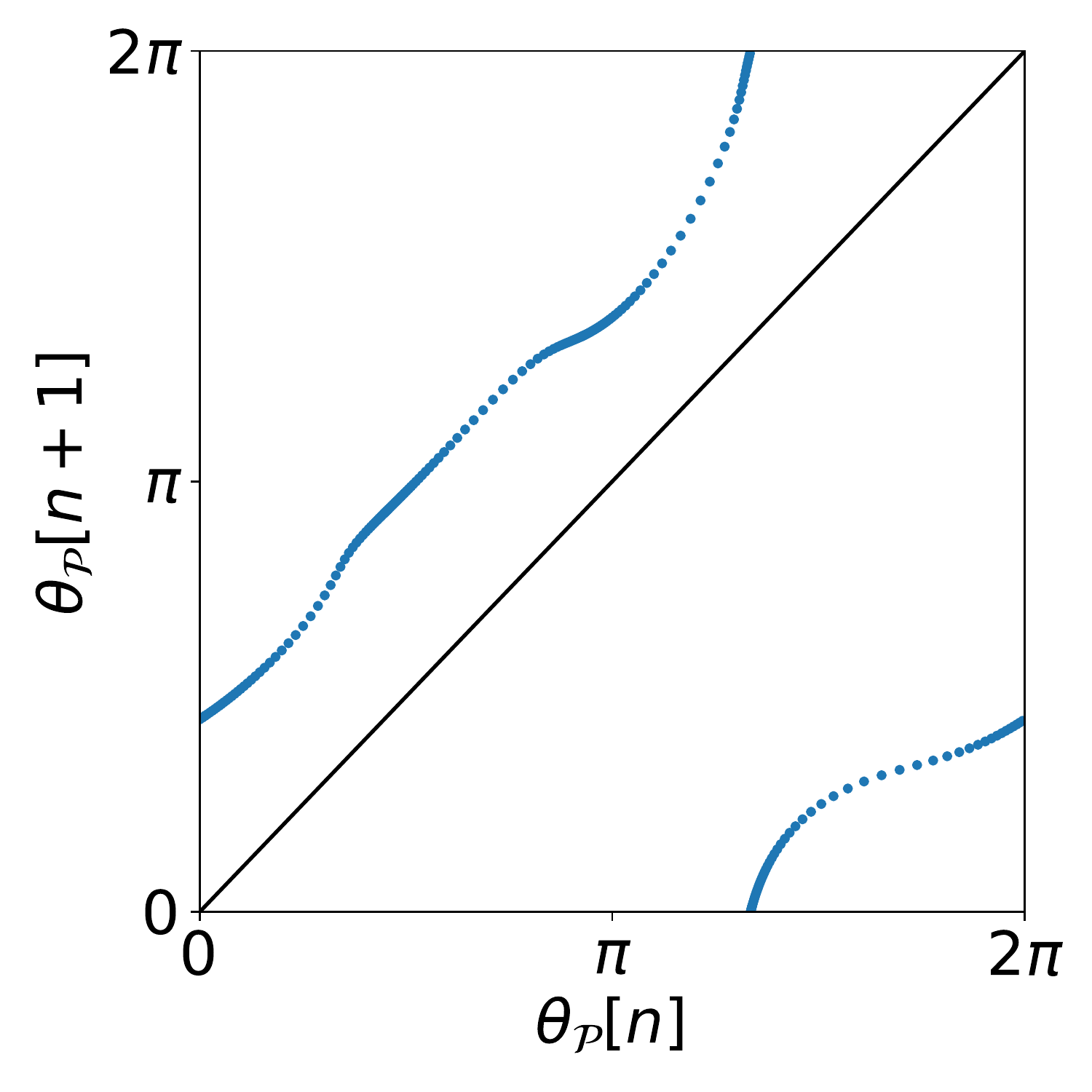}
		\put (0,0) {(a)}
	\end{overpic}                             
	\begin{overpic}[height=0.23\textwidth]{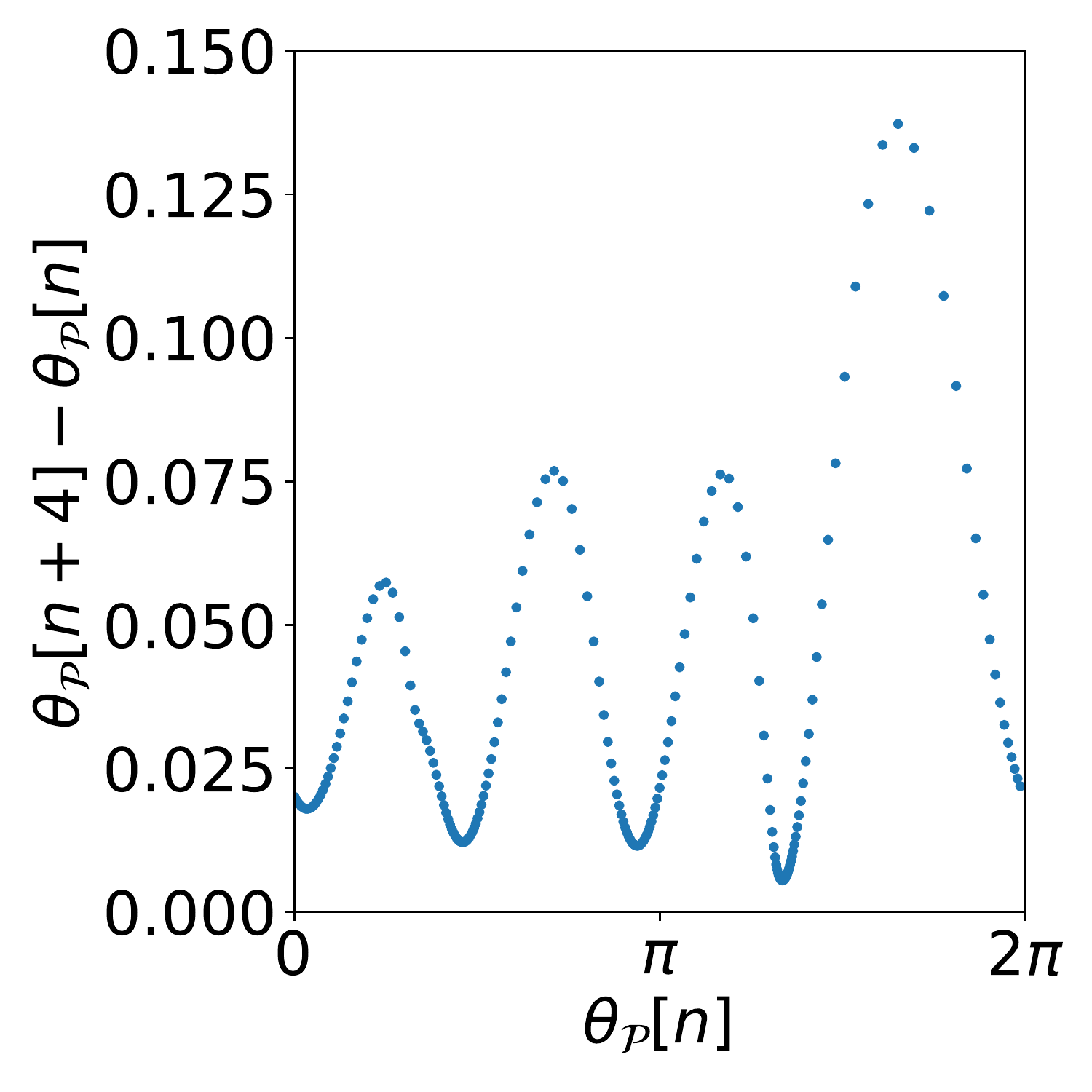}
		\put (0,0) {(b)}
	\end{overpic}
	\caption{(a)
		Poincar\'e return map of the dynamics on the 2-torus 
		surrounding \RPO{1o} at $\memory = 17.1$. 
		(b)
		Intersection of the $4$-th iterate of the return map 
		in (a) and the identity map, illustrating the map's 
		progress towards a period-4 resonance.
	\label{f-circleMap}
	}
\end{figure}

\begin{figure}
	\centering
	\begin{overpic}[width=0.35\textwidth]{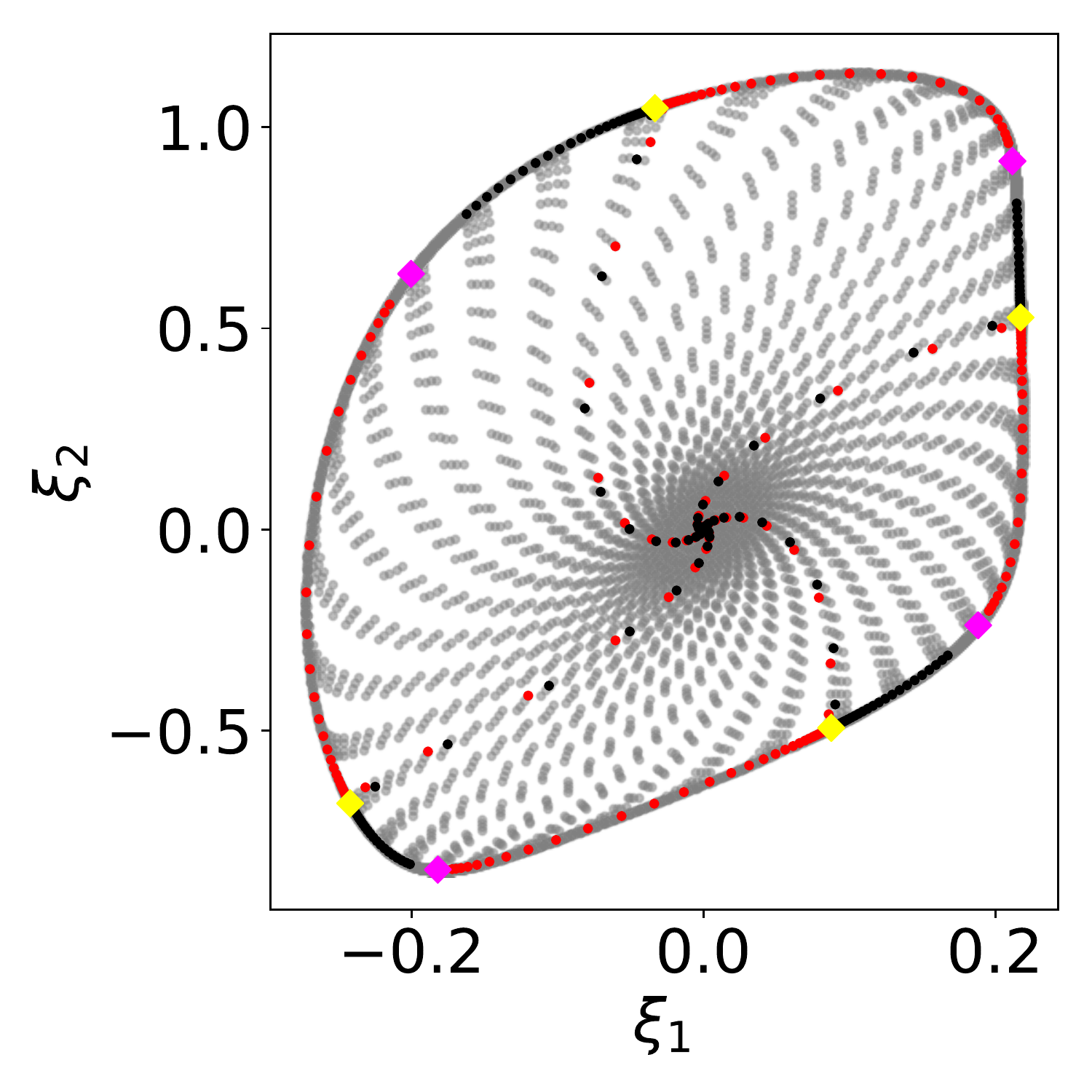}
		\put (0,0) {}
	\end{overpic}                             
	\caption{
		The unstable manifold of \RPO{1o} (at the center) at $\memory = 17.2$
		visualized as the intersections (gray dots) of the 
		time-forward dynamics of initial conditions
		\refeq{e-UnstManICs} with the Poincar\'e section \refeq{e-PoincS}.
		Stable (\RPO{4os}) and unstable (\RPO{4ou}) pair of period-$4$ 
		orbits are marked respectively with magenta and yellow diamonds. 
		Two orbits that lie on different sides of \RPO{4ou}'s stable 
		manifold and then follow its unstable manifold in opposite
		directions marked red and black. 
		\label{f-period4}
	}
\end{figure}

The invariant 2-torus surrounding \RPO{1o} disappears at 
$\memory \in (17.1, 17.2)$ and leaves its place to a stable and 
unstable pair of period-4 (on the Poincar\'e section) orbits. 
In order to elucidate this process, we analyzed the system
before ($\memory = 17.1$) and after ($\memory = 17.2$) the 
bifurcation. At $\memory = 17.1$, we parameterized the invariant 
torus surrounding \RPO{1o}\ with $\theta_{\PoincS} \in [0, 2 \pi)$ and 
generated a return map 
$\theta_{\PoincS}[n+1] = f_{\PoincS}(\theta_{\PoincS}[n])$
as shown in \reffig{f-circleMap}(a). Upon our investigation of the 
fourth iterate of this map, we have found that at $4$ locations 
it comes close to being tangent to the identity map as plotted in
\reffig{f-circleMap}(b). This is a typical situation as the system 
proceeds towards a saddle-node bifurcation that would give rise to 
a stable-unstable pair of period-$4$ orbits. In order to elucidate 
the dynamics after this bifurcation, we approximated \RPO{1o}'s 
unstable manifold by forward integrating the initial conditions
\beq
	\sspRed (\delta, \phi) = 
	\sspRed_{\RPO{1o}} + \epsilon \Lambda_1^{\delta} 
					    (\Re V_1 \cos \phi + \Im V_1 \sin \phi), 
	\label{e-UnstManICs}
\eeq
where we used $4$-equidistant values for $\delta \in [0, 1)$, 
$36$-equidistant values for $\phi \in [0, 2\pi)$ and set 
$\epsilon = 10^{-4}$. Time-forward dynamics
of the initial conditions \refeq{e-UnstManICs} approximately cover
the linearized unstable manifold of \RPO{1o}, thus 
we expect their further evolution to take the shape of the 
nonlinear unstable manifold\rf{BudCvi15}. We visualized these orbits
at their intersections with the Poincar\'e section \refeq{e-PoincS}
as projections onto $(\Re V_1, \Im V_1)$-plane on \reffig{f-period4}.
We found that all of these trajectories converged to a new stable
periodic orbit (marked magenta on \reffig{f-period4}) 
that intersects the Poincar\'e section 4 times. 
We will refer to this orbit as \RPO{4os} with the additional ``s'' 
signifying that this solution appears as the stable one of a 
saddle-node pair. 
We have also observed
that this manifold was separated in two parts at 4 different 
locations, where nearby trajectories went in opposite directions 
when they arrived at the edge of the manifold as illustrated by 
red and black orbits on \reffig{f-period4}. By utilizing a Newton
search, we confirmed that these separation points corresponded to 
an unstable period-4 orbit \RPO{4ou} (``u'' standing for ``unstable''),
which we marked yellow on 
\reffig{f-period4}. These observations together confirm that a 
saddle-node bifurcation on the invariant torus surrounding 
\RPO{1o} took place at $\memory \in (17.1, 17.2)$.

\begin{figure}
	\centering
	\begin{overpic}[width=0.35\textwidth]{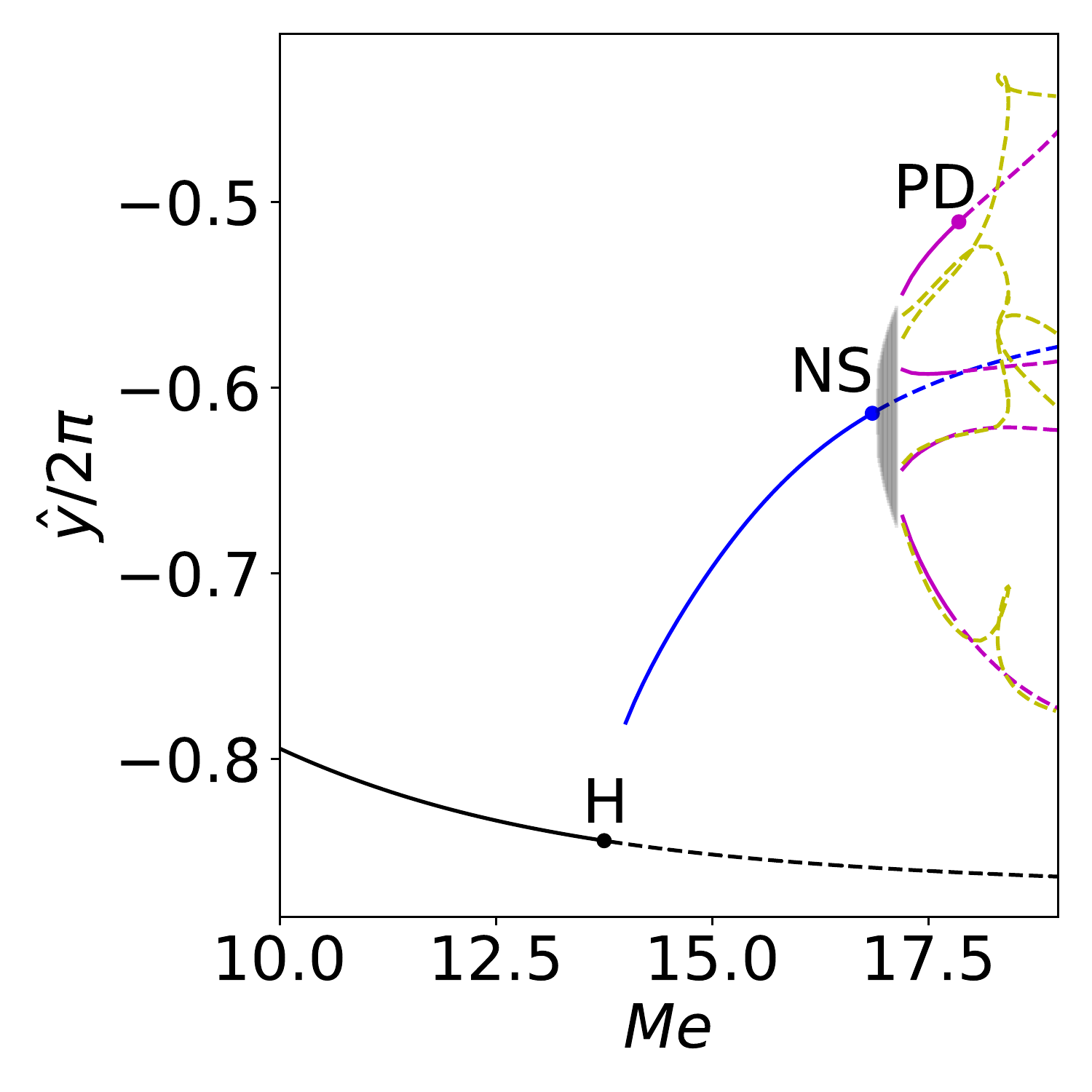}
		\put (0,0) {}
	\end{overpic}                             
	\caption{
		Diagram summarizing bifurcations starting from the circular 
		solution, where stable solutions 
		are drawn as solid curves and unstable ones are dashed. 
		Approximate bifurcation points are annotated with letters 
		indicating the type of bifurcation; 
		H: Hopf, 
		NS: Neimark-Sacker,
		PD: Period doubling. 
		Different colors correspond to different solutions; 
		black: \REQ{0c},
		blue: \RPO{1o},
		gray(transparent): invariant torus,
		magenta: \RPO{4os},
		yellow: \RPO{4ou}.		
		\label{f-bifDiag}}
\end{figure}

We then continued both \RPO{4ou} and \RPO{4os} up to $\memory = 19.0$. 
For this computation, we needed to implemented
pseudoarclength continuation\rf{DKK1991} since \RPO{4ou} underwent
several fold bifurcations and became more and more unstable. 
\RPO{4os}, on the other hand, 
underwent a period doubling bifurcation 
at $\memory \in (17.8, 17.9)$, marking the beginning of the
cascade which leads to chaos. 

      Until the period doubling point, the bifurcation sequence 
	  we described here qualitatively\footnote{The exact 
	  	bifurcation points are slightly 
	  	different, which could be due to differences 
	  	in the numerical methods.}
	  agrees with the observations of Tambasco
	  \etal\rf{THORB2016}, who studied the same system with a 
	  different numerical method.
	  Notably, they also observed that the 
  	  relative two-torus is replaced by a period-4 relative 
  	  periodic orbit, which they refer to as 
  	  ``a frequency-locked wobbling orbit''.
      Upon further increase of $\memory$, 
      Tambasco \etal\rf{THORB2016} observes a new 
      incommensurate frequency in the Fourier spectrum of the 
      wobbling orbit, which does not agree with the period 
      doubling bifurcation we find here. Consequently 
      Tambasco \etal 's description to transition to chaos 
      disagrees with ours, which we will present
      in \refsect{s-chaos}. In \refsect{s-conclusion}, 
      we discuss the differences between two
      transition scenarios in more detail.
  	  
We summarized the 
bifurcations that we analyzed up to this point in \reffig{f-bifDiag}. 
The vertical axis of \reffig{f-bifDiag} shows
the $\hat{y}$ coordinate of $\sspRed_{\REQ{0c}}$ and 
for the rest of the solutions, 
the $\hat{y}$ coordinate of their intersections with the 
Poincar\'e section defined as the half-hyperplane
\beq
	\inprod{\sspRed_{\PoincS_y} - \hat{t}_{\PoincS_y}}{\hat{n}_{\PoincS_y}} = 0 
	\,, \quad
	\inprod{\velRed(\sspRed_{\PoincS_y})}{\hat{n}_{\PoincS_y}} > 0	
	\label{e-PoincSy}
\eeq
where $\hat{t}_{\PoincS_y} = (0, 1, 0, \ldots)$ and 
$\hat{n}_{\PoincS_y} = (1, 0, 0, \ldots)$. The Poincar\'e section
\refeq{e-PoincSy} is equivalent to marking the reduced state space 
trajectories' intersections with the $\hat{y}$-axis in $+\hat{x}$ 
direction. 

At the period doubling bifurcation of \RPO{4os}, annotated with
PD in \reffig{f-bifDiag}, a period-8 cycle appears in its vicinity, 
which itself undergoes another period doubling upon further increase in
\memory. We postpone the analysis of the chaotic attractor, which
forms following this cascade of period doublings to 
\refsect{s-chaos}, and turn our focus to another set of solutions 
that coexist with the ones we presented so far. 

\subsubsection{The lemniscate solution and the symmetry breaking}

In the previous section, we followed primary and subsequent 
bifurcations from the counterclockwise rotating circular solution. 
All the solutions we presented so far had 
positive angular momentum 
$L = x v_y - y v_x > 0$ at all times. 
Since the system is equivariant under the reflection symmetry 
\refeq{e-sigma}, all of these solutions have reflection-symmetric
counterparts with $L < 0$. A different type of stable periodic 
solution, which we will refer to as \PO{2l} coexist in the 
state space at $\memory = 14.0$. This solution can be found by 
initiating a simulation with a random initial condition at $\memory = 14.0$:
We observed at this regime that an initial condition $\ssp(0)$ 
that is populated by standard normally distributed numbers 
eventually lands on $\PO{2l}$, or $\RPO{4}$ 
(or its reflection copy $\sigma \RPO{4}$). 
Both of these orbits 
are stable at $\memory = 14.0$.

\begin{figure*}
	\centering
	\begin{overpic}[width=0.25\textwidth]{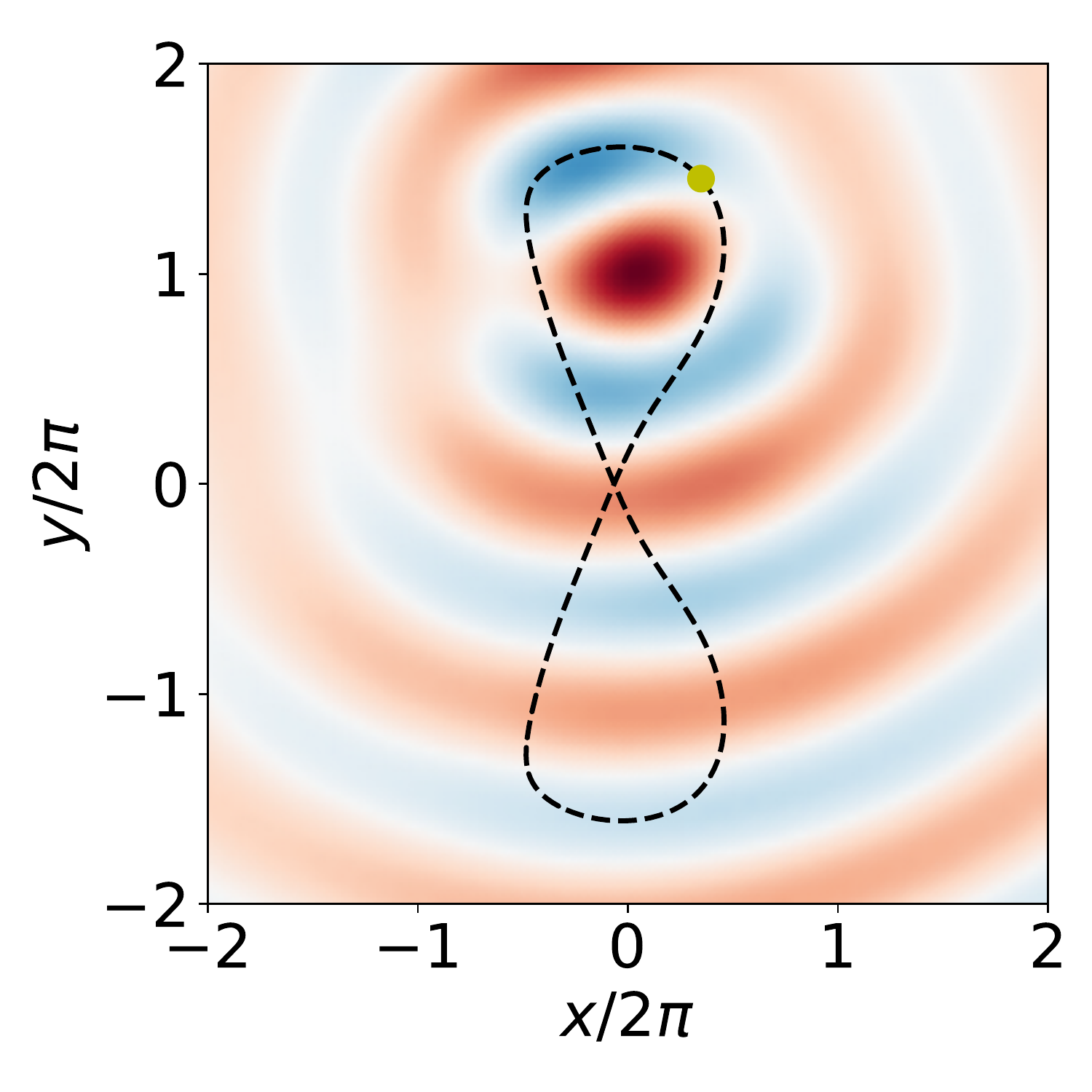}
		\put (0,2) {(a)}
	\end{overpic}                             
	\begin{overpic}[width=0.25\textwidth]{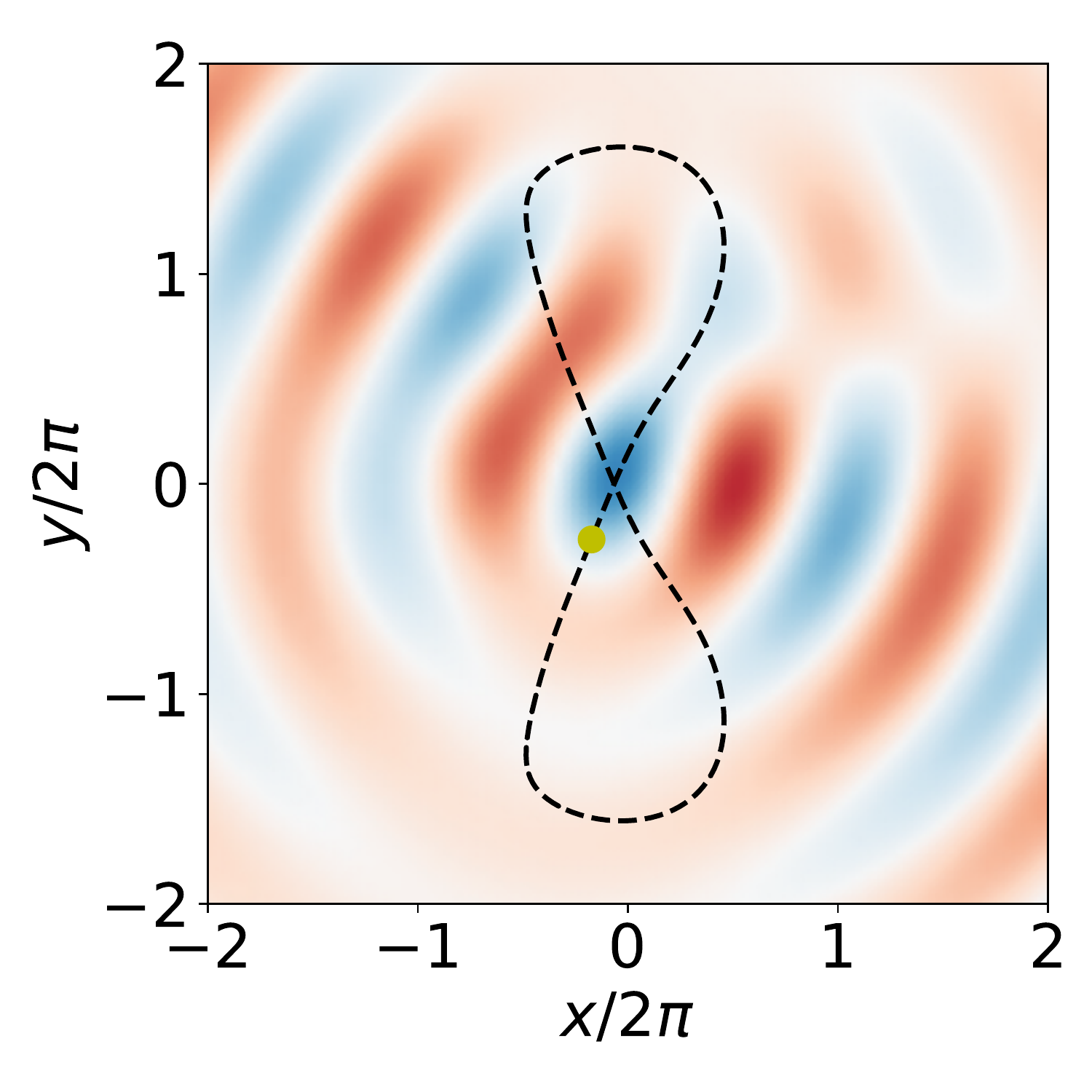}
		\put (0,2) {(b)}
	\end{overpic}
	\begin{overpic}[width=0.25\textwidth]{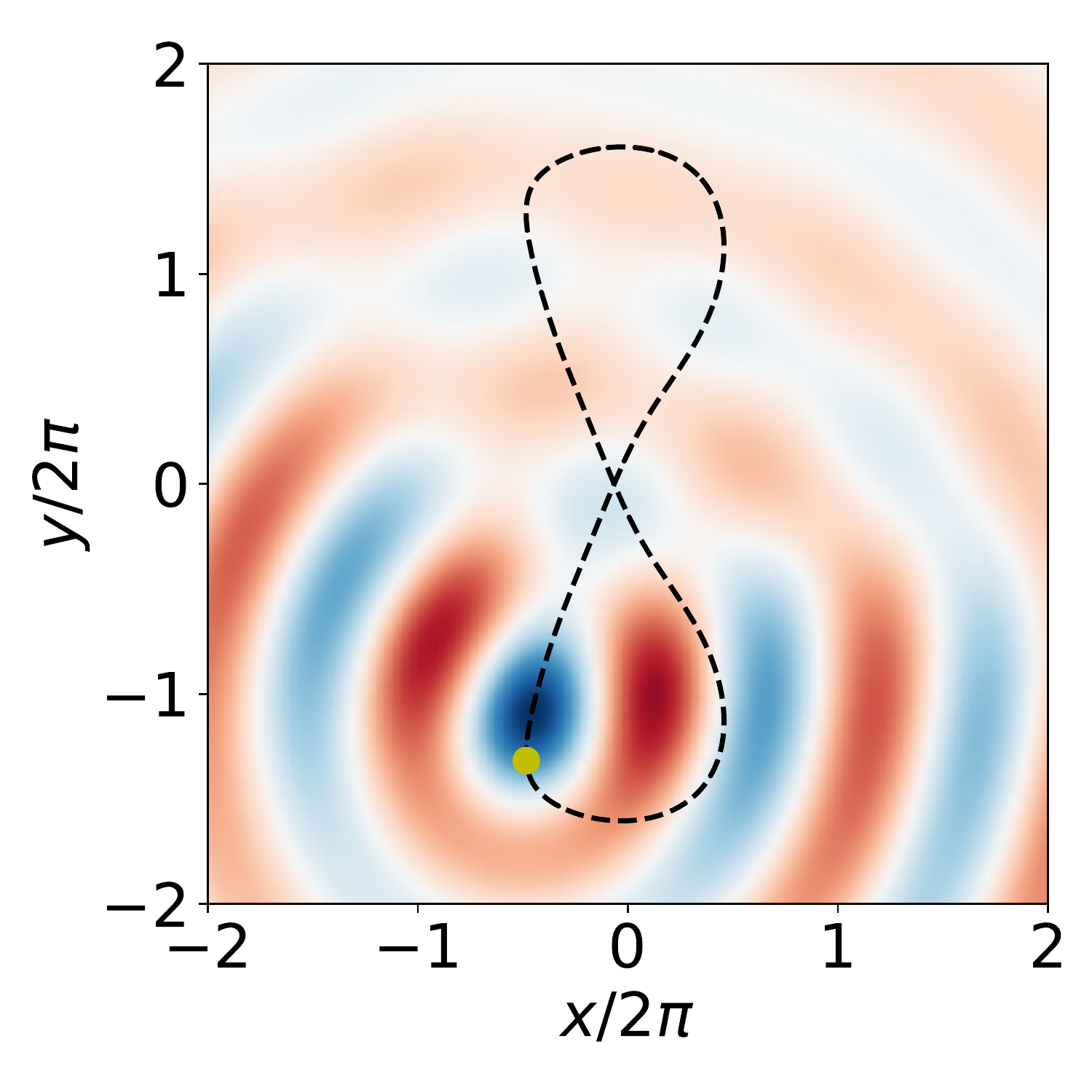}
		\put (0,2) {(c)}
	\end{overpic}
    \includegraphics[height=0.25\textwidth]{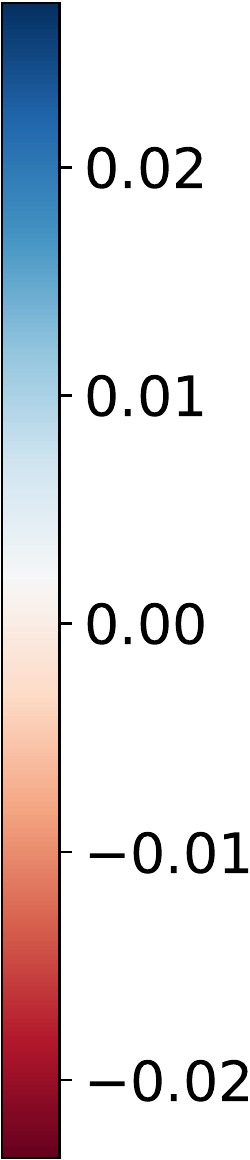}
	\caption{
		Three snapshots of a simulated trajectory of \PO{2l}
		at $\memory = 14.0$. The trace of the droplet for one period
		is drawn as a dashed curve
		and its instantaneous position at the respective snapshot is 
		indicated with a yellow dot in each figure. 
		The magnitude of the wave field color coded in each snapshot
		(a) $\zeit = 0$, (b) $\zeit = 25$, (c) $\zeit = 50$. 
	}
	\label{f-poTwo}
\end{figure*}

We visualized three snapshots from one full period of 
$\PO{2l}$ in \reffig{f-poTwo}. Perrard \etal \rf{PLMFC2014} called 
similar solutions ``lemniscates'', referring to their figure-8 shape. 
From \reffig{f-poTwo}, it can be seen that the droplet's angular 
momentum is reversed as it traverses through one full period. 
In fact, the angular momentum along $\PO{2l}$ is exactly zero as the
second half of the orbit is the reflection symmetry of the first 
half. Similar solutions of the 
Kuramoto-Sivashinsky system are 
sometimes referred to as ``pre-periodic''\rf{SCD07,BudCvi15} since
they can be treated as relative periodic orbits satisfying
\beq
	\ssp_{\PO{}} = \sigma
					\flow{T_{\PO{}} / 2}{\ssp_{\PO{}}}\,. 
	\label{e-ppo}
\eeq
In other words, after half period $T_{\PO{}} / 2$, the periodic orbit
comes to the reflection of its initial position. Such solutions generically 
do not undergo period doubling bifurcations \rf{SwiWie84} and when 
the Jacobian of \refeq{e-ppo} has a negative real eigenvalue outside the 
unit circle, this corresponds to the symmetry-breaking. 

\begin{figure}
	\centering
	\begin{overpic}[width=0.23\textwidth]{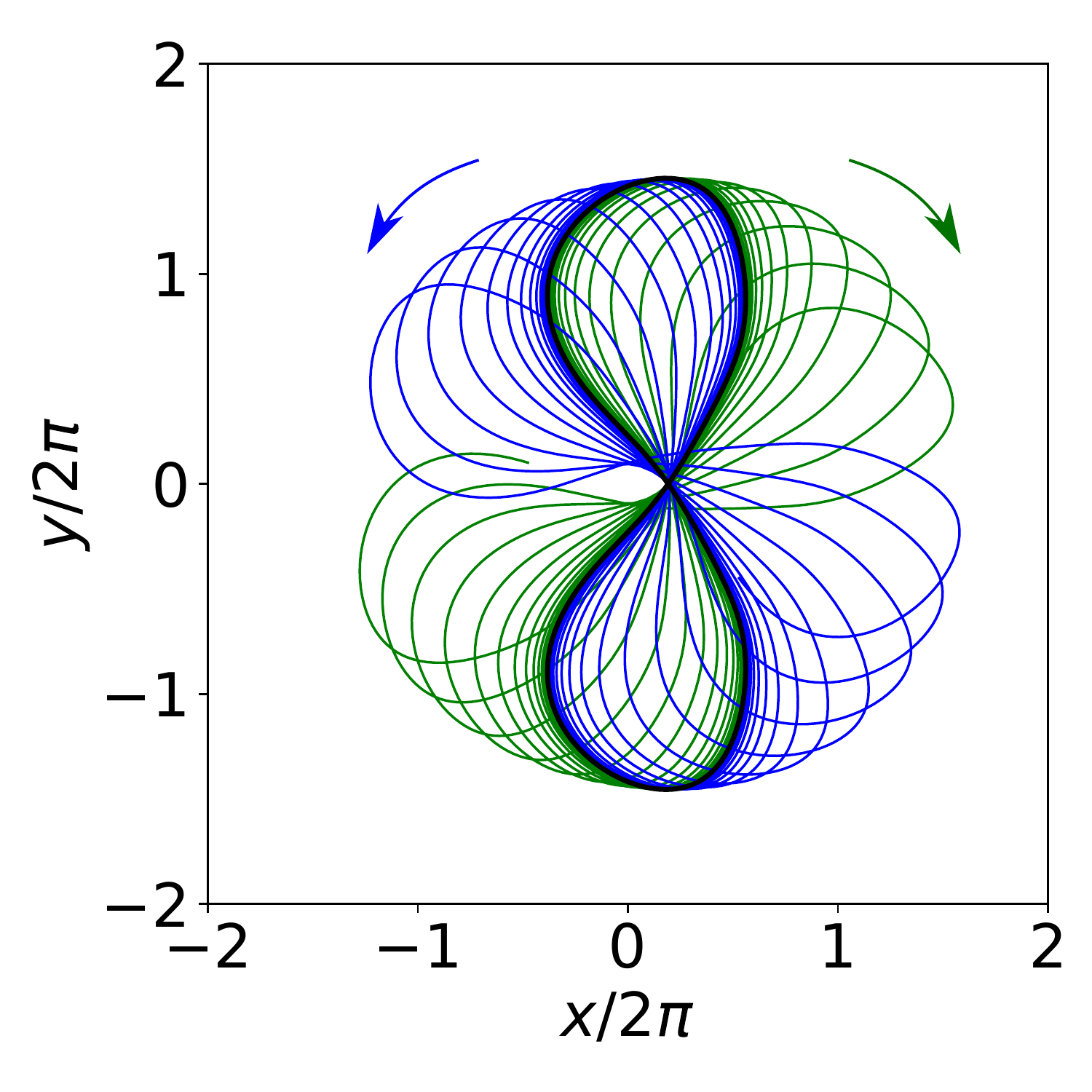}
		\put (0,0) {(a)}
	\end{overpic}                             
	\begin{overpic}[width=0.23\textwidth]{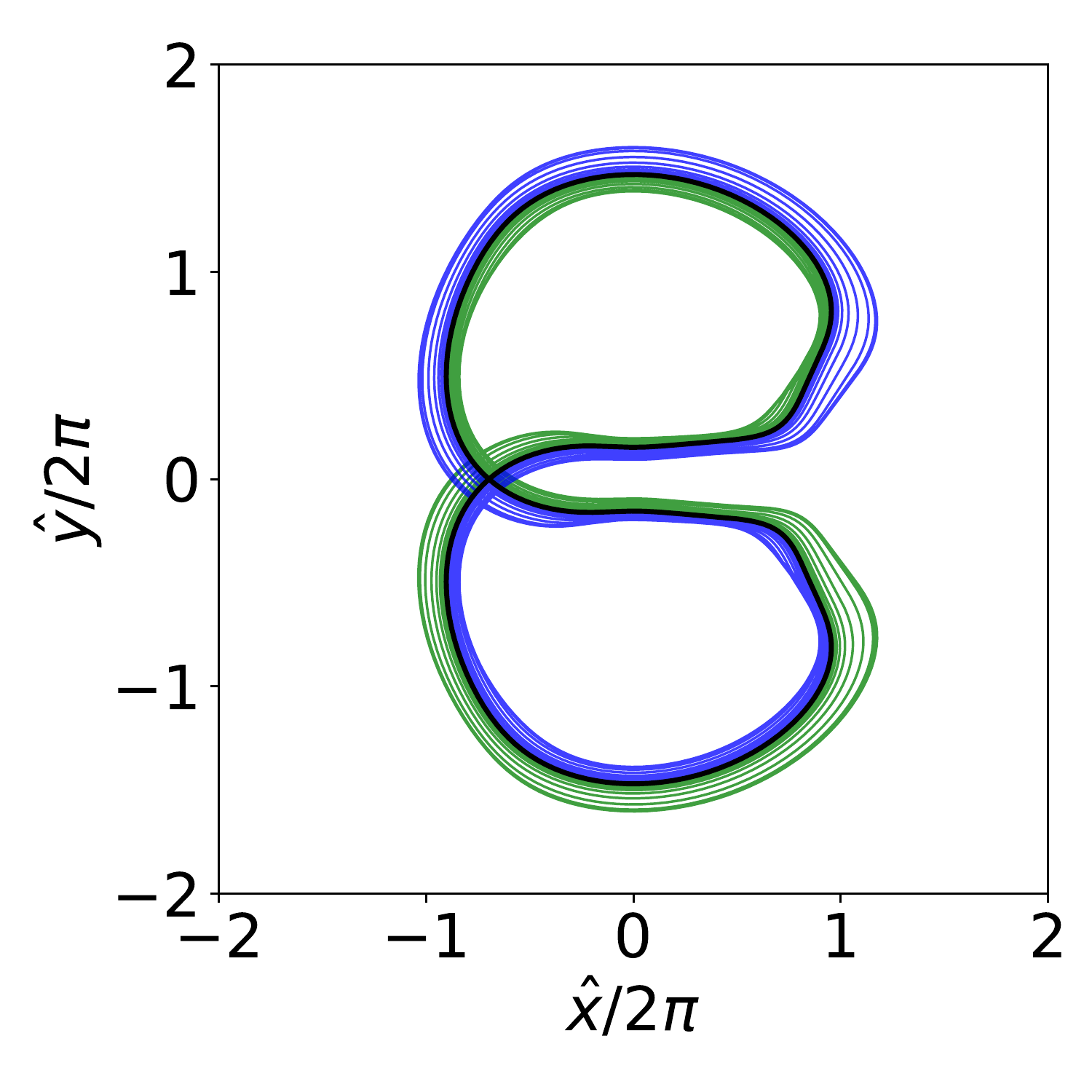}
		\put (0,0) {(b)}
	\end{overpic}
	\caption{\PO{2l} (black) at $\memory = 16.0$ and symmetry-broken 
		dynamics on its unstable manifold illustrated as the 
		trajectories of perturbations to \PO{2l} in $\pm V_1$ 
		(green/blue) directions;
		(a) in full state space, 
		(b) after continuous symmetry reduction. 
              Azimuthal drift directions of blue and green orbits 
			  are indicated with arrows in (a).
	}
	\label{f-symBreak}
\end{figure}

A symmetry-breaking bifurcation of \PO{2l} takes place at 
$\memory \in (15.6, 15.7)$ and there appears 
drifting lemniscate solutions in its vicinity. We visualized
these dynamics in \reffig{f-symBreak} by showing the trajectories 
of small perturbations $\pm 10^{-2} V_1$ to \PO{2l}, where 
$V_1$ is the leading (unstable) Floquet vector. 
\refFig{f-symBreak} (a) shows the symmetry-broken dynamics in 
full state space along with the unstable \PO{2l}. 
In \reffig{f-symBreak} (b), when the symmetry is reduced, 
it becomes apparent that two new relative periodic orbits appear 
in $\PO{2l}$'s vicinity. We are going to refer to these drifting 
lemniscates as $\RPO{2l}$ and $\sigma \RPO{2l}$. 

At $\memory \in (16.5, 16.6)$, \RPO{2l} undergoes a period doubling
bifurcation, which leads to a cascade and a chaotic attractor 
formation. Before examining these dynamics in \refsect{s-chaos}, we
must present another type of solution, which coexist with the ones
we studied so far. 

\subsubsection{The trefoil}

In the previous section, we have stated that at $\memory = 14.0$, 
the pilot-wave system has two stable solutions, $\RPO{1o}$ and 
$\PO{2l}$, and a random initial condition eventually lands on one or 
the other. A natural question to ask at this stage is the following. 
What separates the basins of attraction of these two solutions? 
Experience from previous studies\rf{GOY1982,GOY1983,ASY1997ch10} of 
low dimensional systems suggests that such disconnected regions 
might be separated by the stable manifold of another invariant 
solution. 
One of the most prominent examples of this kind of state 
space structure is manifested in the subcritical transition to 
turbulence in shear flows, where laminar and turbulent
flows are coexisting attractors. 
In numerical simulations of
channel flow, Itano and Toh\rf{IT01} were first to find a traveling 
wave solution (a relative equilibrium) 
by a shooting method which bisects between trajectories
whose energy fluctuations eventually reach turbulent levels and 
those that vanishes.  
This suggested that the turbulent and laminar 
regions of the state space were separated by the stable manifold of
this solution.  
Following numerical 
studies\rf{TI03,SchEckYor07,SGLDE08,MMSE09,ScMaEc10,ZamEck14a,KKSDEH2016}
of shear flows found similar 
solutions\footnote{The exact type of the invariant solution 
	depends on the system under study and the parameters.} with shooting 
methods akin to the one used by Itano and Toh \rf{IT01}. 
Such separating 
solutions in subcritical transition literature are now generally 
referred to as ``edge states'', following the terminology introduced 
by Schneider \etal \rf{SchEckYor07} 
and the bisection methods for locating them 
is usually called ``edge tracking''. We are going to adopt this 
terminology in what follows. 

In order to investigate the edge between the initial conditions that 
eventually land on \RPO{1o} and those that go to \PO{2l}, we formulated 
the following edge tracking method. 
For tracking the laminar-turbulent boundary
in shear flows, the 
kinetic energy of turbulent fluctuations is usually used as a proxy 
to determine whether the solution eventually becomes 
turbulent. In our case, we found the time-averaged angular momentum 
of the final state to be an appropriate quantity and formualted
the edge tracking algorithm in \refappe{s-EdgeTrack}. 

\begin{figure}
	\centering
	\begin{overpic}[width=0.35\textwidth]{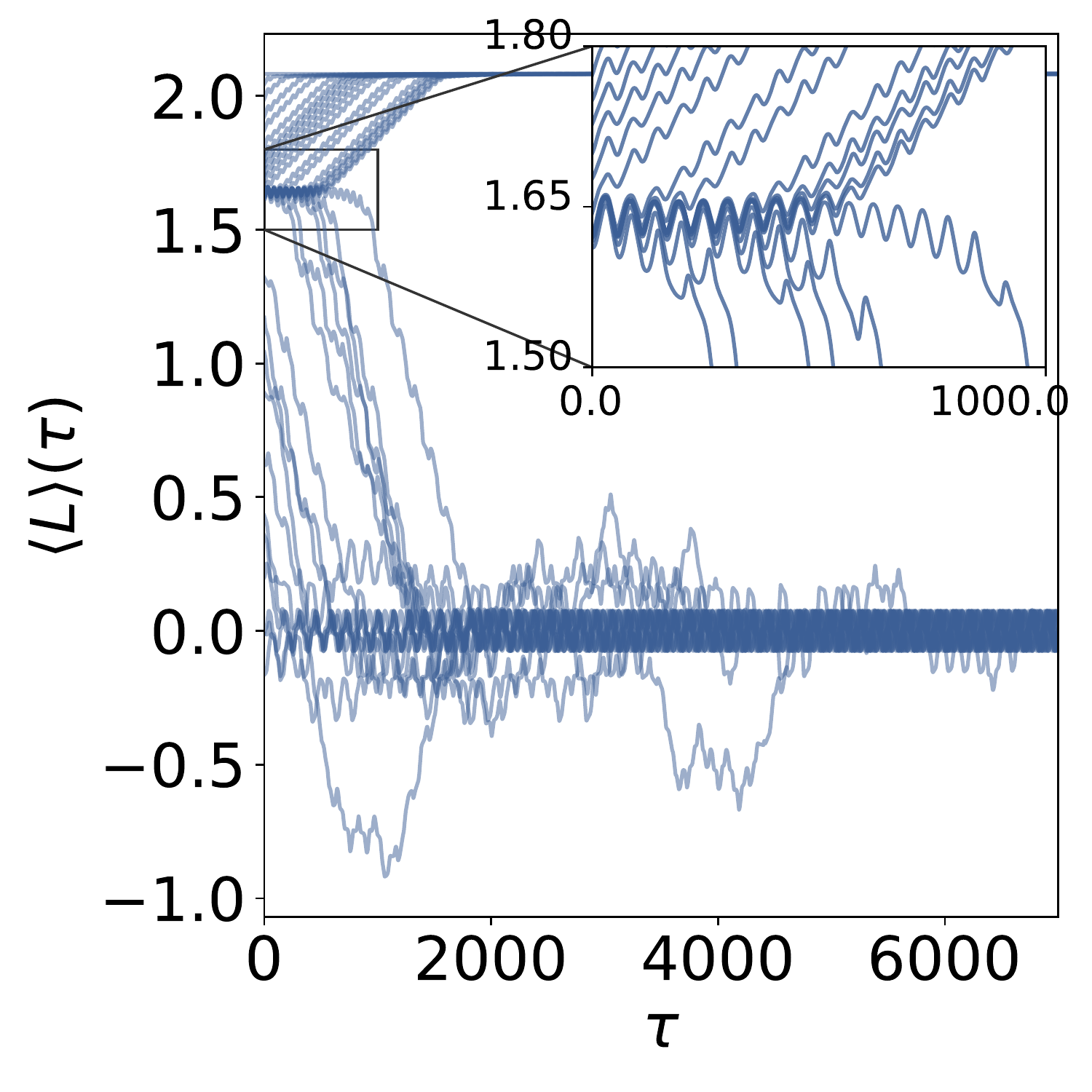}
		\put (0,0) {}
	\end{overpic}    
	\caption{
		Time-series of the window-averaged angular momentum of bisection
		trajectories. Inset: Zoom-in showing nearly-periodic 
		temporal dynamics.
		\label{f-LavgtInset}
	}
\end{figure}

    Let $\sspRed_{\PO{2l}}$ and $\sspRed_{\RPO{1o}}$ be
	random points on \RPO{1o} and \PO{2l}.  
	In order to search for the basin boundary between 
	\PO{2l} and \RPO{1o}, we run edge tracking 
	algorithm (\refappe{s-EdgeTrack}) with 	
	$\epsilon = 10^{-9}\,,\, 
	\zeit_f = 10000\,,\, 
	\zeit_w = 1000, 
	L_{th, 1} = 1.8\,,\,
	L_{th, 2}=0.1$.
We plotted the time-series of the window-averaged
angular momentum in \reffig{f-LavgtInset}. As expected, the 
average angular momentum eventually settles around 
$\langle L \rangle \approx 0$ (\PO{2l}) or 
$\langle L \rangle \approx 2.1$ (\RPO{1o}). Notice, however,
that before landing on these final states, the average angular
momentum of the bisection trajectories oscillates around 
$\langle L \rangle \approx 1.65$. We observed approximately 
periodic symmetry-reduced dynamics in these episodes; 
and upon a Newton search, found a new relative periodic orbit, 
which we will refer to as $\RPO{1t}$.

\begin{figure*}
	\centering
	\begin{overpic}[width=0.25\textwidth]{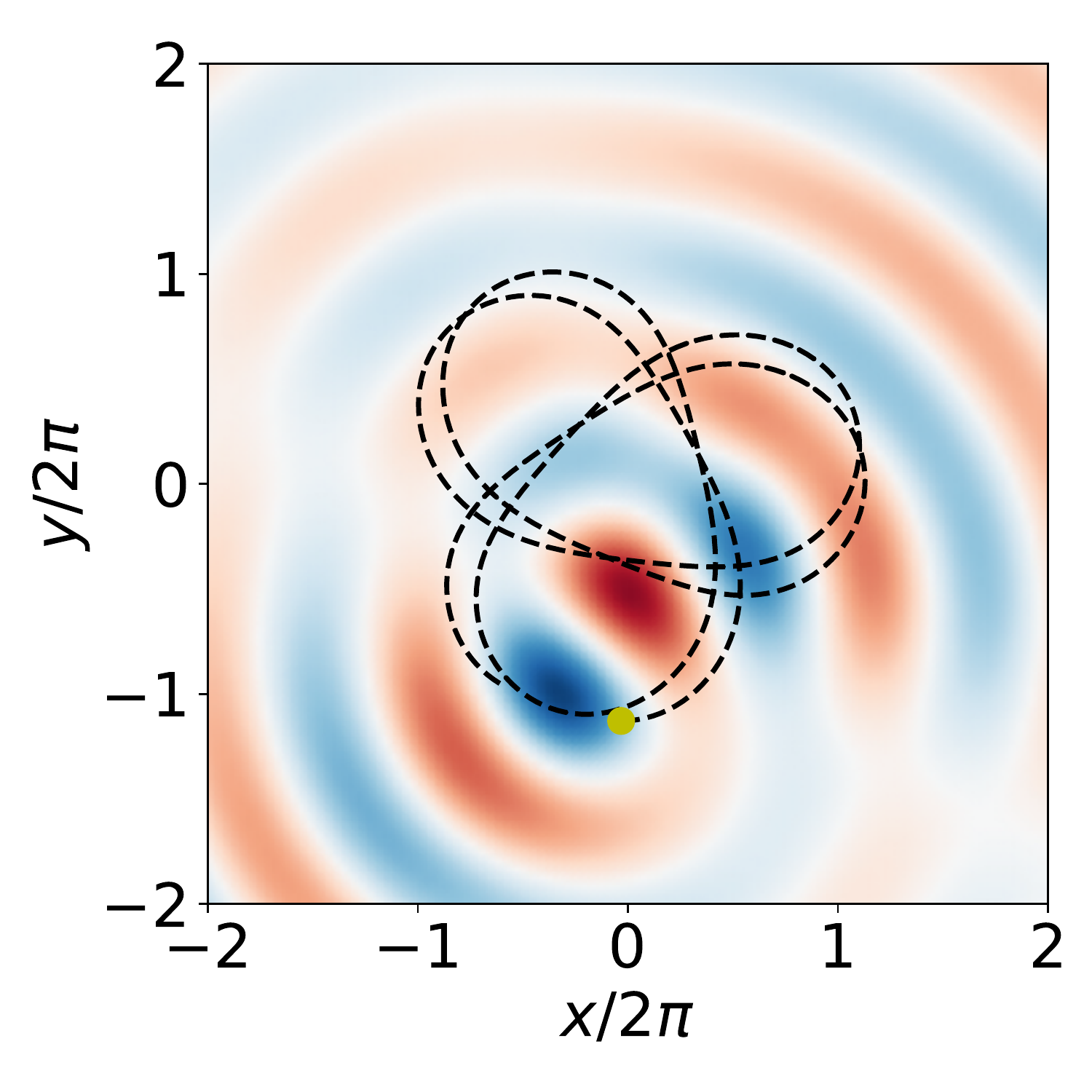}
		\put (0,2) {(a)}
	\end{overpic}                             
	\begin{overpic}[width=0.25\textwidth]{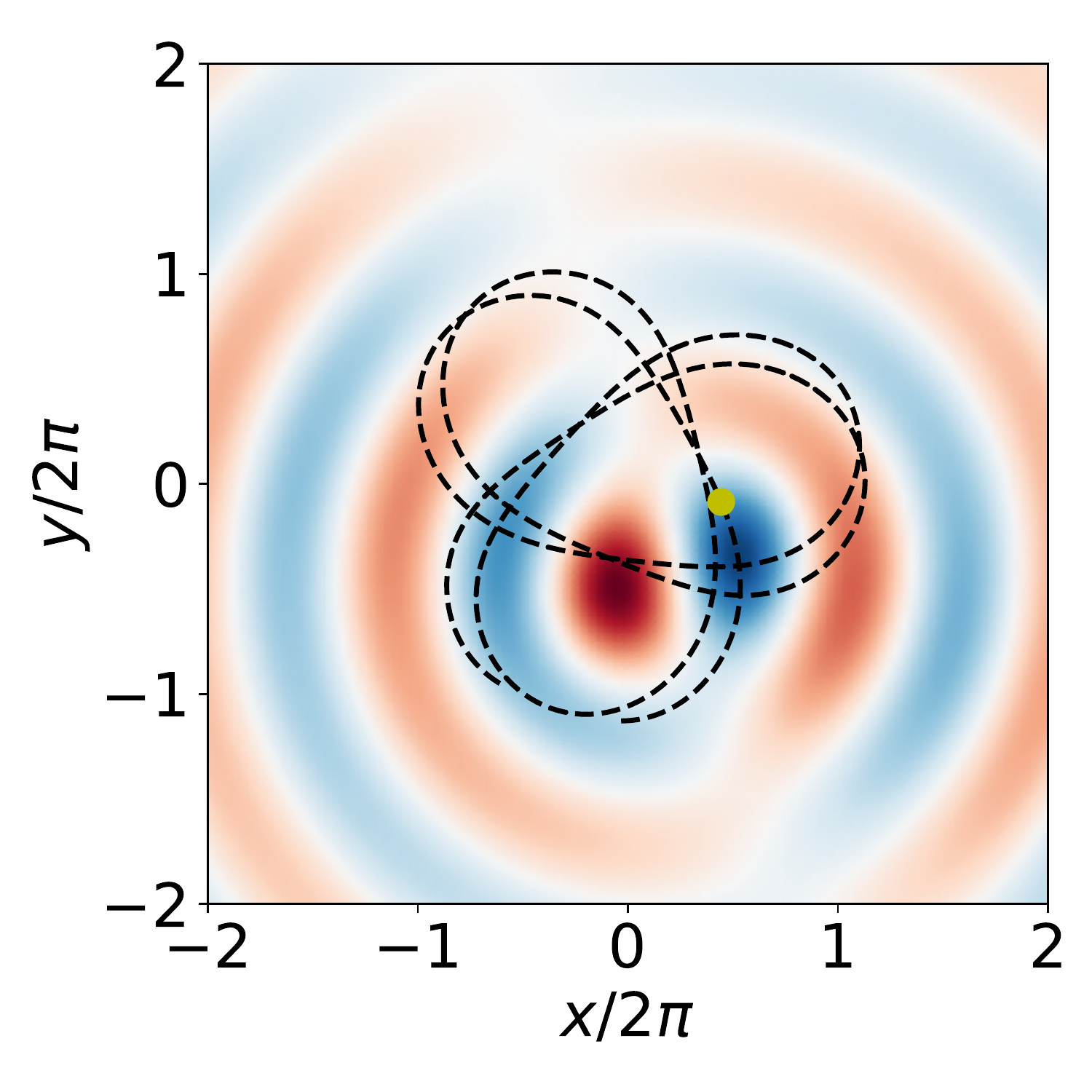}
		\put (0,2) {(b)}
	\end{overpic}
	\begin{overpic}[width=0.25\textwidth]{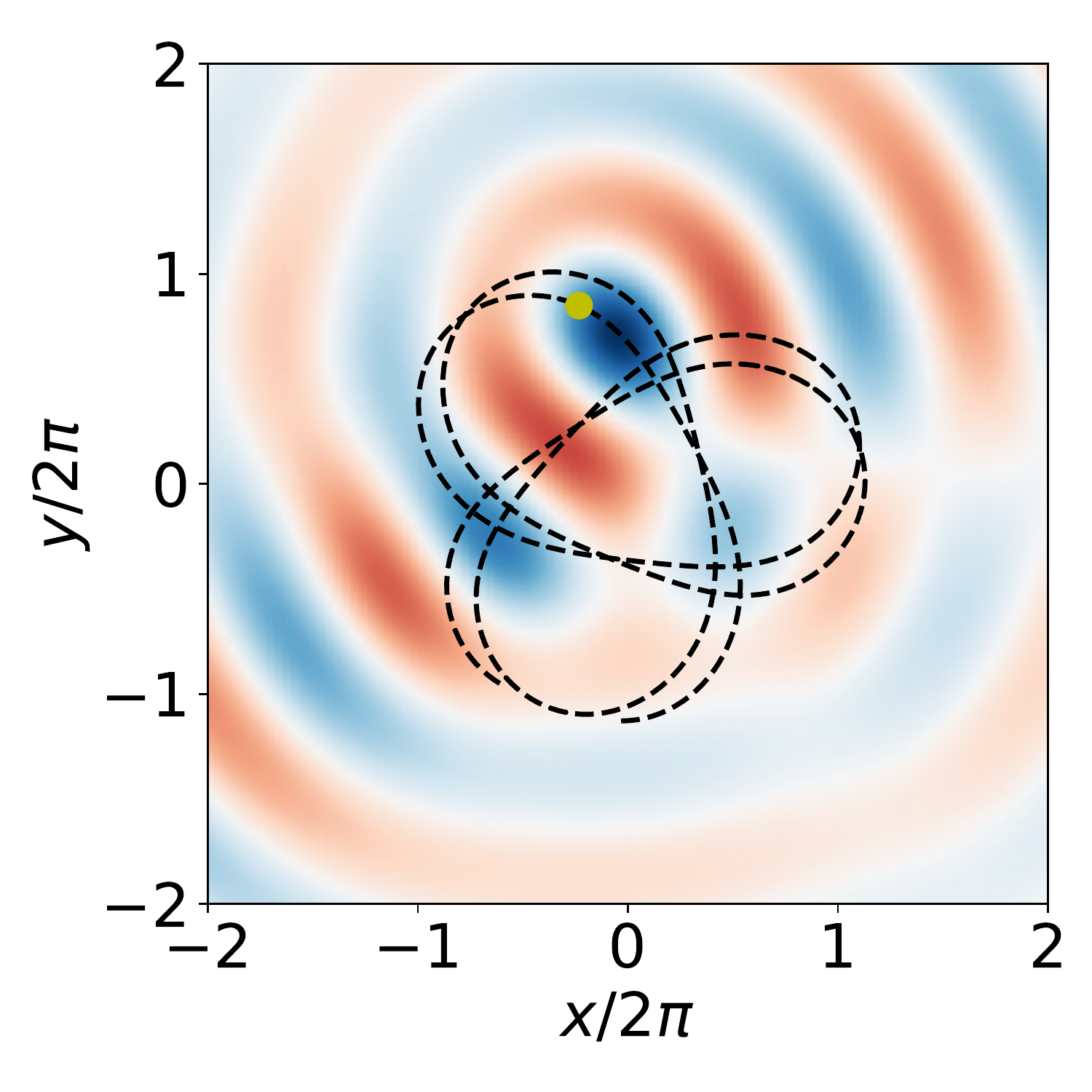}
		\put (0,2) {(c)}
	\end{overpic}
	\includegraphics[height=0.25\textwidth]{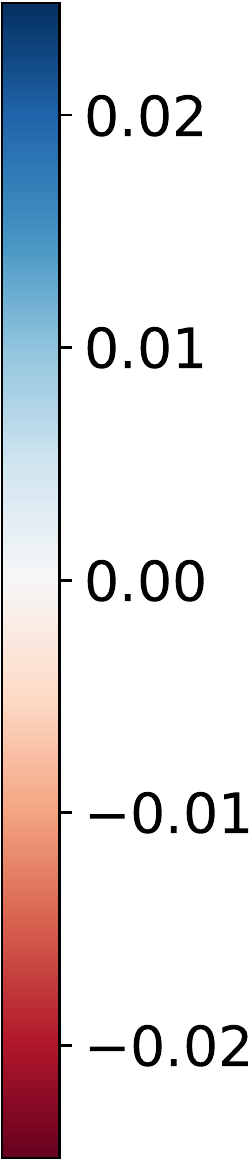}
	\caption{
		Three snapshots of a simulated trajectory of \RPO{1t}
		at $\memory = 14.0$. The trace of the droplet for six periods
		is drawn as a dashed curve
		and its positions at the instance of 
		snapshots are indicated with a yellow dot at each figure. 
		The magnitude of the wave field at the instance of 
		snapshot is color coded. 
		(a) $\zeit = 0$, (b) $\zeit = 20$, (c) $\zeit = 40$. 
	}
	\label{f-rpo1t}
\end{figure*}

In \refFig{f-rpo1t}, we visualize three snapshots from 
$\RPO{1t}$'s trajectory simulated in the full state space. Experimentally 
observed solutions with similar shapes were named ``trefoils'' by
Perrard \etal \rf{PLMFC2014}. 
By numerical continuation, we confirmed that 
when $\memory \in [12.91, 19.0]$, \RPO{1t} has one 
positive Floquet multiplier outside the unit circle, which renders 
\RPO{1t}'s stable manifold codimension-1. This is in agreement with our 
initial intuition since a codimension-1 manifold divides the state space 
into two distinct regions.  Thus, we conclude that at $\memory = 14.0$, 
the edge between the basins of attraction of \RPO{1o} and \PO{2l} is
set by the stable manifold of the trefoil solution \RPO{1t}. 

      \RPO{1t} appears in a saddle-node bifurcation at 
	  $\memory \approx 12.91$ along with a stable branch, trajectory 
	  of which is also trefoil-shaped. We observed this stable solution 
  	  for $\memory \in [12.91, 12.95]$, however, our efforts in 
      continuing this solution to higher $\memory$ have failed.
      Trefoil-shaped trajectories have been observed in both
      laboratory\rf{PLMFC2014} and numerical\rf{KOB2017} 
      experiments, which could only be possible if the associated
      solution is stable. 
      Our observation of the stable trefoil orbit 
      following the saddle-node
      bifurcation provides one explanation for experimental findings
      by fine-tuning the system parameters and initial conditions. 
      It is conceivable that 
      other trefoil-shaped solutions might appear at 
      saddle-node bifurcations at different parameter values.

\subsection{Chaotic attractors and their global bifurcations}
\label{s-chaos}

In the previous sections, we showed that the droplet system has 
three different families of solutions: (1) the circular solution 
and the subsequent ovals, (2) the lemniscate solutions, and (3) 
the trefoil. We have already mentioned that two of these solutions, 
namely \RPO{4os} and \RPO{2l}, undergo period-doubling bifurcations.
In this section, we are going to illustrate that both of these 
period-doublings yield chaotic dynamics, which subsequently merge
and generate full complexity of the chaotic pilot-wave dynamics. 

\begin{figure}
	\centering
	\begin{overpic}[width=0.35\textwidth]{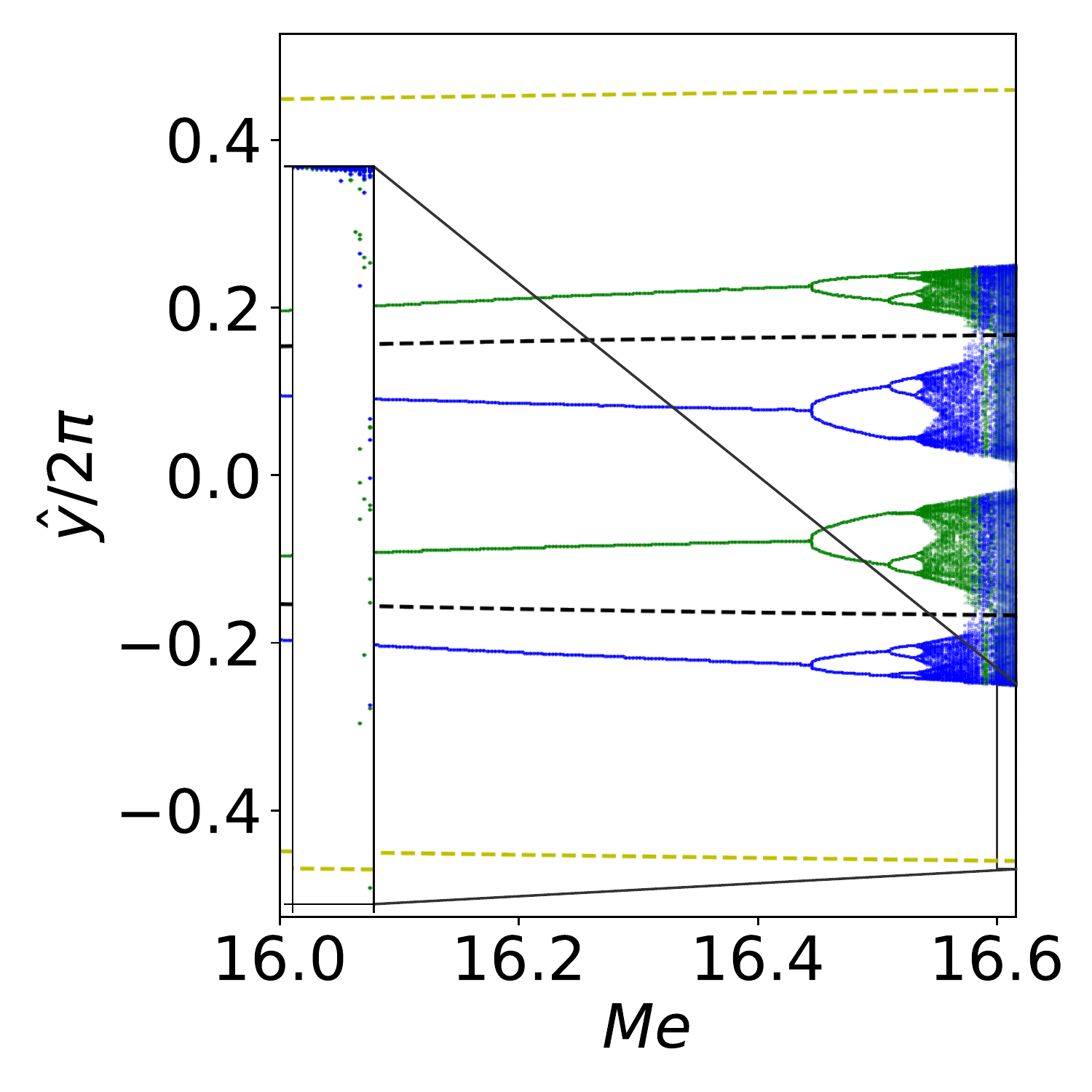}
		\put (0,0) {}
	\end{overpic}    
	\caption{
		Orbit diagram illustrating the period doubling cascade to
		chaos in the neighborhoods of $\RPO{2l}$ (green) and 
		$\sigma \RPO{2l}$ (blue). The symmetric-lemniscate 
		$\PO{2l}$ and the trefoil solutions $\RPO{1t}$ and 
		$\sigma\RPO{1t}$ are shown as black and yellow dashed-lines.
		Inset: zoom-in to the lower-right corner to illustrate the
		rare visits of chaotic dynamics to the neighborhood of 
		$\RPO{1t}$. 
		\label{f-orbitDiagRPO2}
	}
\end{figure}

The first set of period doublings take place in the neighborhoods of 
\RPO{2l} and its reflection symmetry $\sigma \RPO{2l}$. In order to 
illustrate these bifurcations, we generated the orbit diagram in
\reffig{f-orbitDiagRPO2}, where we plotted the $\hat{y}$ coordinate
of the very long trajectories' intersections with the Poincar\'e 
section \refeq{e-PoincSy}. Starting at $\memory = 16.0$ with initial
conditions on $\RPO{2l}$ and $\sigma \RPO{2l}$, we increased $\memory$
in steps of $0.001$, used the final state of the previous step as 
the initial condition of the next step, and generated a long data 
set. \refFig{f-orbitDiagRPO2} shows the intersections of these 
trajectories with the Poincar\'e section \refeq{e-PoincSy} after 
initial transients are discarded. Green markers correspond to 
orbits that started from $\RPO{2l}$ and the blue ones correspond to 
the ones that started from $\sigma \RPO{2l}$. Additionally, 
the symmetric lemniscate $\PO{2l}$ is shown as black dashed lines,
and the trefoil solutions $\RPO{1t}$ and $\sigma \RPO{1t}$ are 
drawn as yellow dashed lines. 

The first period doubling of $\RPO{2l}$ and $\sigma \RPO{2l}$ takes
place at $\memory \approx 16.437$. This bifurcation is followed by
a cascade of period doublings, which yields chaotic sets \CHAOS{l}
and $\sigma \CHAOS{l}$ in the
vicinity of $\RPO{2l}$ and $\sigma \RPO{2l}$, respectively. 
As illustrated in \reffig{f-symBreak}, these two reflection-related
regions are separated by $\PO{2l}$ (drawn dashed black 
in \reffig{f-orbitDiagRPO2}) and we observed that the 
two chaotic regions merge at $\memory \approx 16.580$, 
as they cross the border set by $\PO{2l}$. At this point, the 
chaotic attractors, which are symmetry-copies of one another merge 
through a symmetry-increasing bifurcation\rf{ChoGol1988} and result 
in a reflection-symmetric attractor 
$\CHAOS{l} = \sigma \CHAOS{l}$. 

Upon further increase of $\memory$, we observed that when 
$\memory > 16.616$, the long trajectories eventually leave this 
chaotic neighborhood and settle on \RPO{1o} or $\sigma \RPO{1o}$, 
which are stable until $\memory \approx 16.9$, see 
\reffig{f-rpoZeroDestab}. 
This suggests that the
chaotic attractor crosses the stable manifold of $\RPO{1t}$, which 
sets the boundary between the neighborhoods of $\RPO{2l}$ and 
$\RPO{1o}$. At the first glance, approach of the chaotic attractor 
towards $\RPO{1t}$ is not obvious, however, when we zoomed into the
parameter region $\memory \in [16.6, 16.616]$ (the inset of 
\reffig{f-orbitDiagRPO2}), we observed that 
the chaotic dynamics came close to $\RPO{1t}$. These visits are 
rather rare: We observed only a few instances in more than 
$3500$ intersections with the Poincar\'e section 
\refeq{e-PoincSy} at $\memory = 16.616$. The sudden transition from 
a chaotic attractor to a chaotic transient is called a ``boundary 
crisis''\rf{GOY1982,ASY1997ch10}. It is not clear from our data 
whether or not the expansion of the attractor towards $\RPO{1t}$
is continuous. Nevertheless, we believe that around 
$\memory \in (16.616, 16.617)$, a boundary crisis causes the 
destruction of this chaotic attractor. For further evidence, we 
carried out a bisection search similar to that of 
\reffig{f-LavgtInset} at $\memory = 16.6$ and found, once again, 
that $\RPO{1t}$ to be the edge state in between the neighborhoods of 
$\RPO{2l}$ and $\RPO{1o}$.

\begin{figure}
	\centering
	\begin{overpic}[width=0.23\textwidth]{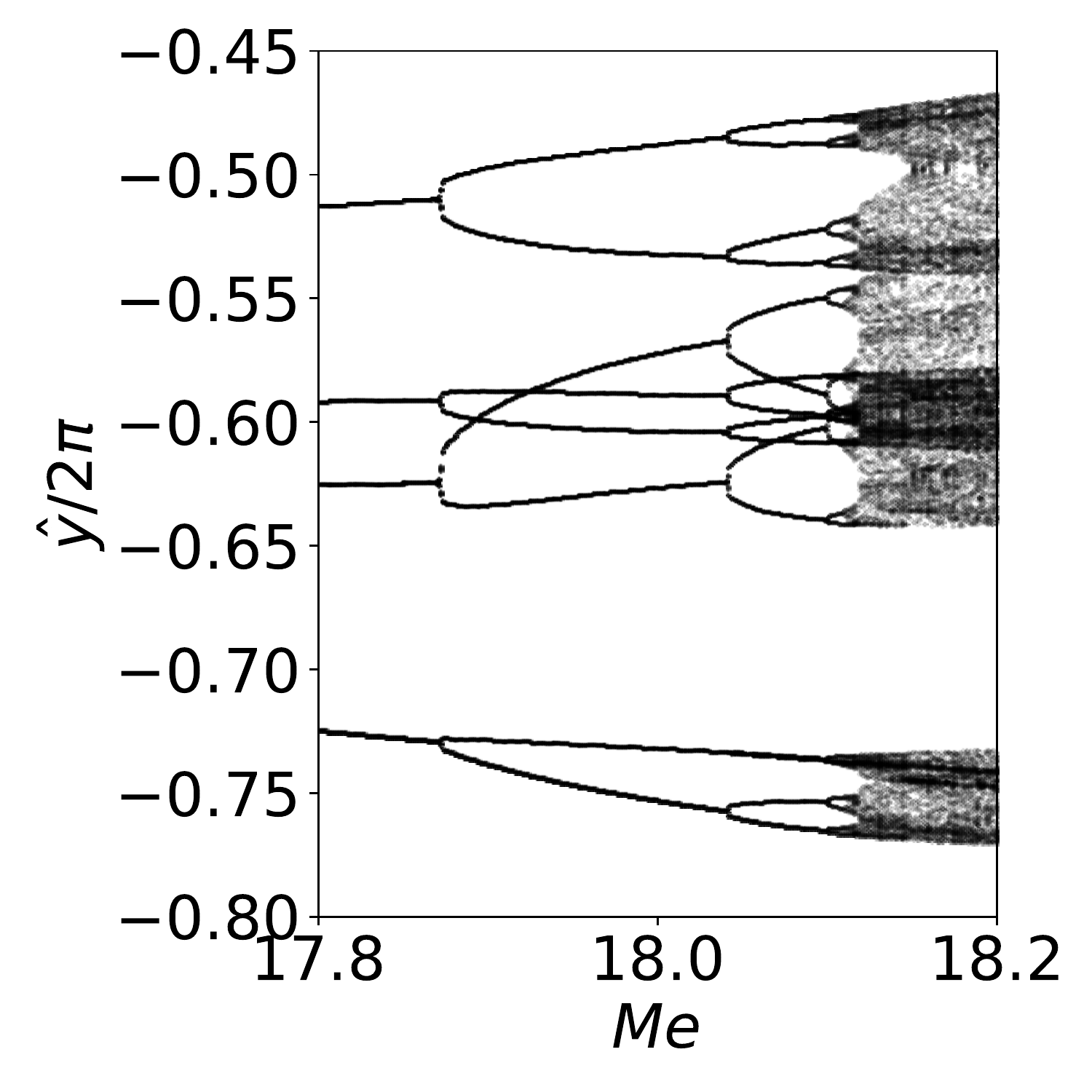}
		\put (0,2) {(a)}
	\end{overpic}                             
	\begin{overpic}[width=0.23\textwidth]{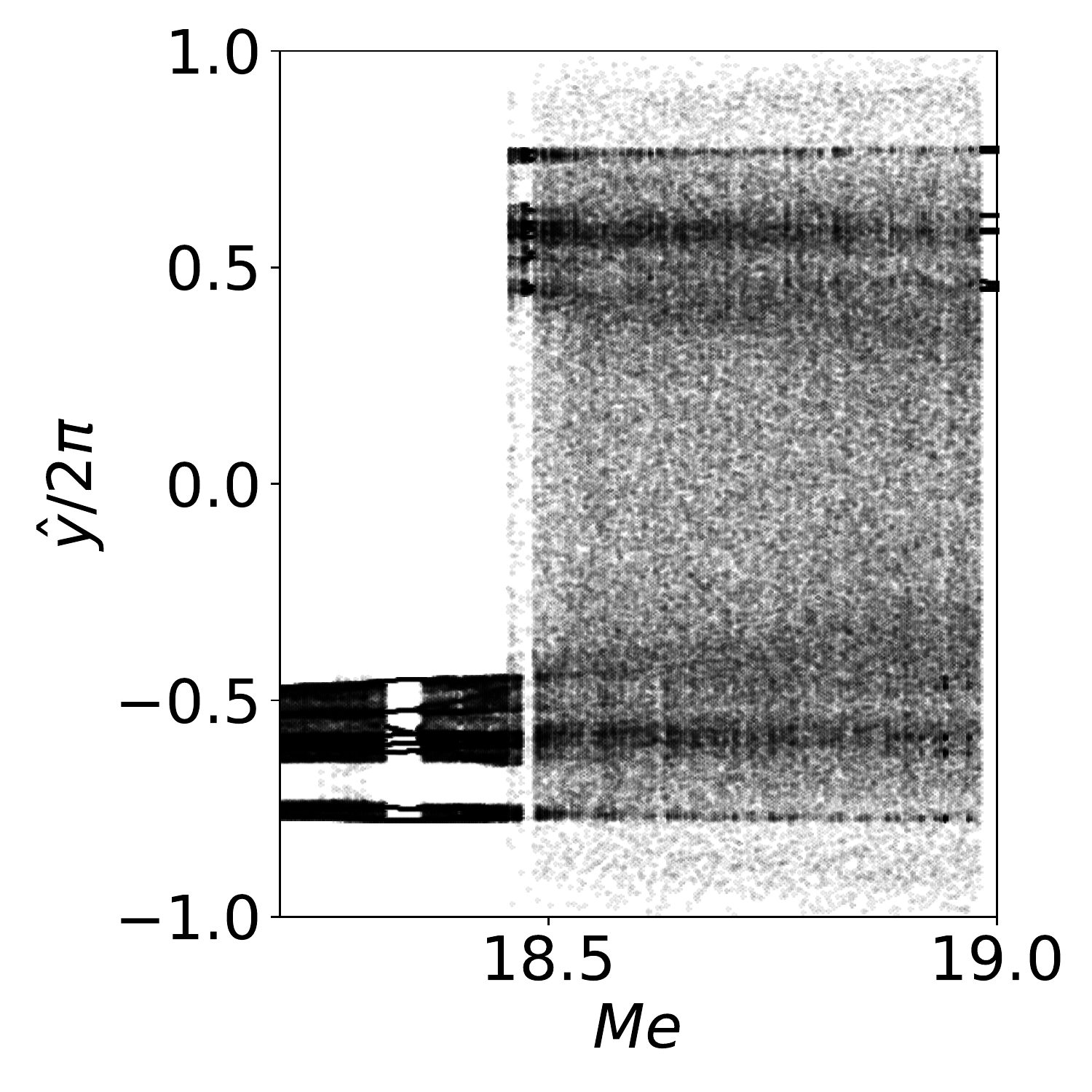}
		\put (0,2) {(b)}
	\end{overpic}
	\caption{
		Orbit diagrams visualizing
		(a) the period doubling cascade to chaos starting from 
		\RPO{4os}, 
		(b) the symmetry-increasing bifurcation at 
		$\memory \approx 18.455$ 
		leading to sudden expansion of the chaotic attractor.
		\label{f-Feigenbaum}
	}
\end{figure}

In \refsect{s-circBif}, we have shown that upon the increase of 
\memory , $\RPO{1o}$ first undergoes a Neimark-Sacker bifurcation 
and as we further increased the \memory , the invariant torus in 
the neighborhood of $\RPO{1o}$ is replaced the stable and unstable
manifolds of a pair of period-4 orbits \RPO{4os} and \RPO{4ou}. 
We ended \refsect{s-circBif} by mentioning that $\RPO{4os}$, which 
is initially stable, undergoes a period doubling bifurcation 
at $\memory \in (17.8, 17.9)$. We illustrated the following cascade
of period-doublings and the emergence of chaotic dynamics on the 
orbit diagrams of \reffig{f-Feigenbaum}

Similar to \reffig{f-orbitDiagRPO2}, we produced 
\reffig{f-Feigenbaum} by recording the intersections of 
long trajectories with the Poincar\'e section \refeq{e-PoincSy}
after discarding the initial transients for 
$\memory \in [17.8, 19.0]$.
At $\memory = 17.8$, we started with an initial condition on 
\RPO{4os} and for the following simulations, we used the final point
of the last step as the initial condition. \refFig{f-Feigenbaum}
illustrates the results of these 
simulations, where we show the $\hat{y}$ values where 
trajectories intersect \refeq{e-PoincSy} at different $\memory$. 
For clarity, we did not show the bifurcations of $\sigma \RPO{4os}$
on \reffig{f-Feigenbaum}, but one should keep in mind that the same
cascade also takes place in the vicinity of $\sigma \RPO{4os}$. 

Three consecutive period doublings are visible in 
\reffig{f-Feigenbaum}(a) and more can be found by zooming into
smaller parameter regions. This result shows that the transition 
to chaos in this system follows a typical period-doubling route 
as opposed to Ruelle-Takens-Newhouse scenario\rf{nhouse78}, 
which was previously suggested by Tambasco \etal\rf{THORB2016}.
We are going to refer to the resulting chaotic attractors as 
$\CHAOS{o}$ and $\sigma \CHAOS{o}$.

As visualized in \reffig{f-Feigenbaum}(b),
dynamics of \CHAOS{o} for $\memory > 18.2$ is chaotic with 
intermittent periodic windows, which is typical for attractors that 
follow a period doubling cascades\rf{strogb}. Notice the sudden 
expansion of the chaotic attractor at $\memory \approx 18.455$. 
This is another symmetry-increasing bifurcation where 
$\CHAOS{o}$ connects to its reflection symmetry 
$\sigma \CHAOS{o}$ yielding the final chaotic attractor of the 
pilot-wave system. As we shall now illustrate, this global 
bifurcation is indeed the result of the merging of 
$\CHAOS{o}$ and $\sigma \CHAOS{o}$ with \CHAOS{l}.

\begin{figure}
	\centering
	\begin{overpic}[width=0.23\textwidth]{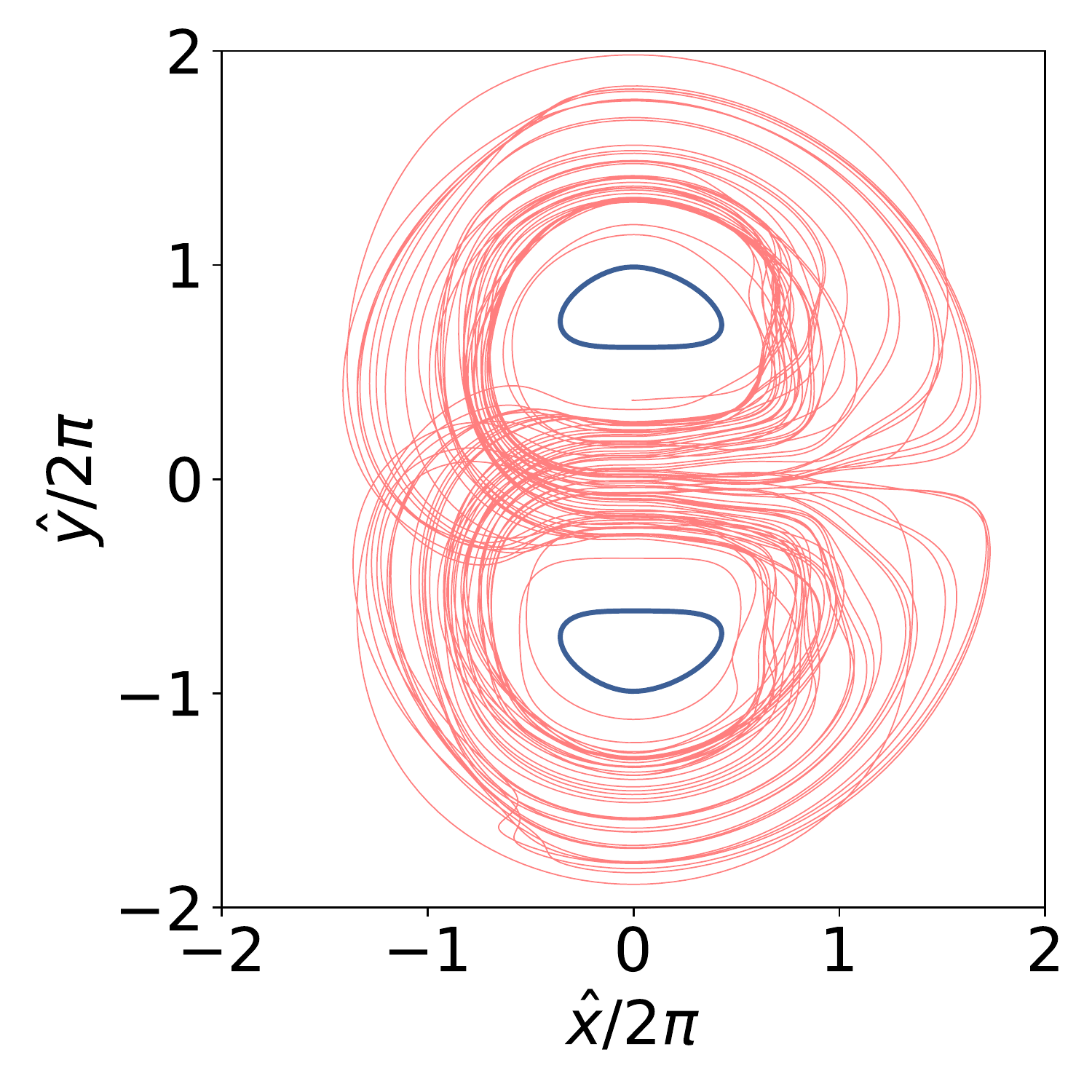}
		\put (0,2) {(a)}
	\end{overpic}                             
	\begin{overpic}[width=0.23\textwidth]{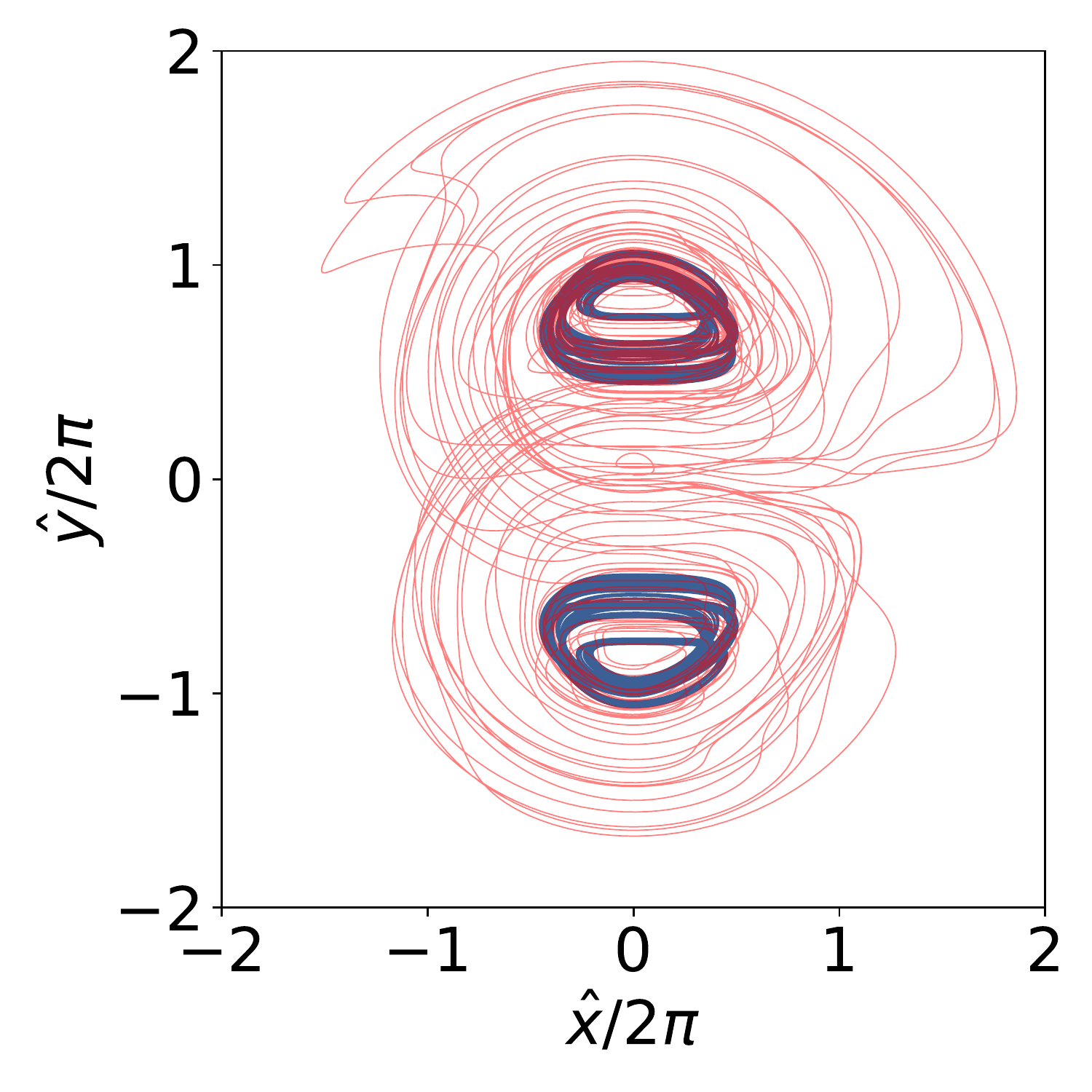}
		\put (0,2) {(b)}
	\end{overpic}
	\caption{
		(a) Chaotic edge state $\CHAOS{l}$ (red) separating $\RPO{1o}$ (blue, 
		$\hat{y}<0$) and 
		$\sigma \RPO{1o}$ (blue, $\hat{y}>0$) at
		$\memory = 16.8$.
		(b) Chaotic edge state $\CHAOS{l}$ (red) separating 
		$\CHAOS{o}$ 
		(blue, $\hat{y}<0$) and $\sigma \CHAOS{o}$ 
		(blue, $\hat{y}>0$) at $\memory = 18.45$. 
		\label{f-merging}
	}
\end{figure}

As we explained earlier, at $\memory \approx 16.617$, 
$\CHAOS{l}$ becomes unstable by merging with $\RPO{1t}$. After this 
boundary crisis, $\CHAOS{l}$ includes $\RPO{1t}$ and becomes the 
chaotic edge state between the attractors $\RPO{1o}$ and $\sigma \RPO{1o}$.
$\CHAOS{l}$ carries this role up until $\memory \approx 18.455$, when
the distinct attractors are chaotic. We illustrated this in 
\reffig{f-merging} at $\memory = 16.8$ and $\memory = 18.45$ where 
the red trajectories were obtained by edge tracking 
(\refappe{s-EdgeTrack}) between solutions 
with positive and negative angular momenta. In both cases, red 
trajectories in \reffig{f-merging} are chaotic transients that would 
fall onto one of the attractors if integrated long enough. 
\refFig{f-merging} (b) illustrates the state space picture just 
before the merging of two chaotic regions. At this point,  
blue and red regions are still dynamically disconnected and their
apparent overlap is due to the lower-dimensional projection. 

\begin{figure}
	\centering
	\begin{overpic}[width=0.23\textwidth]{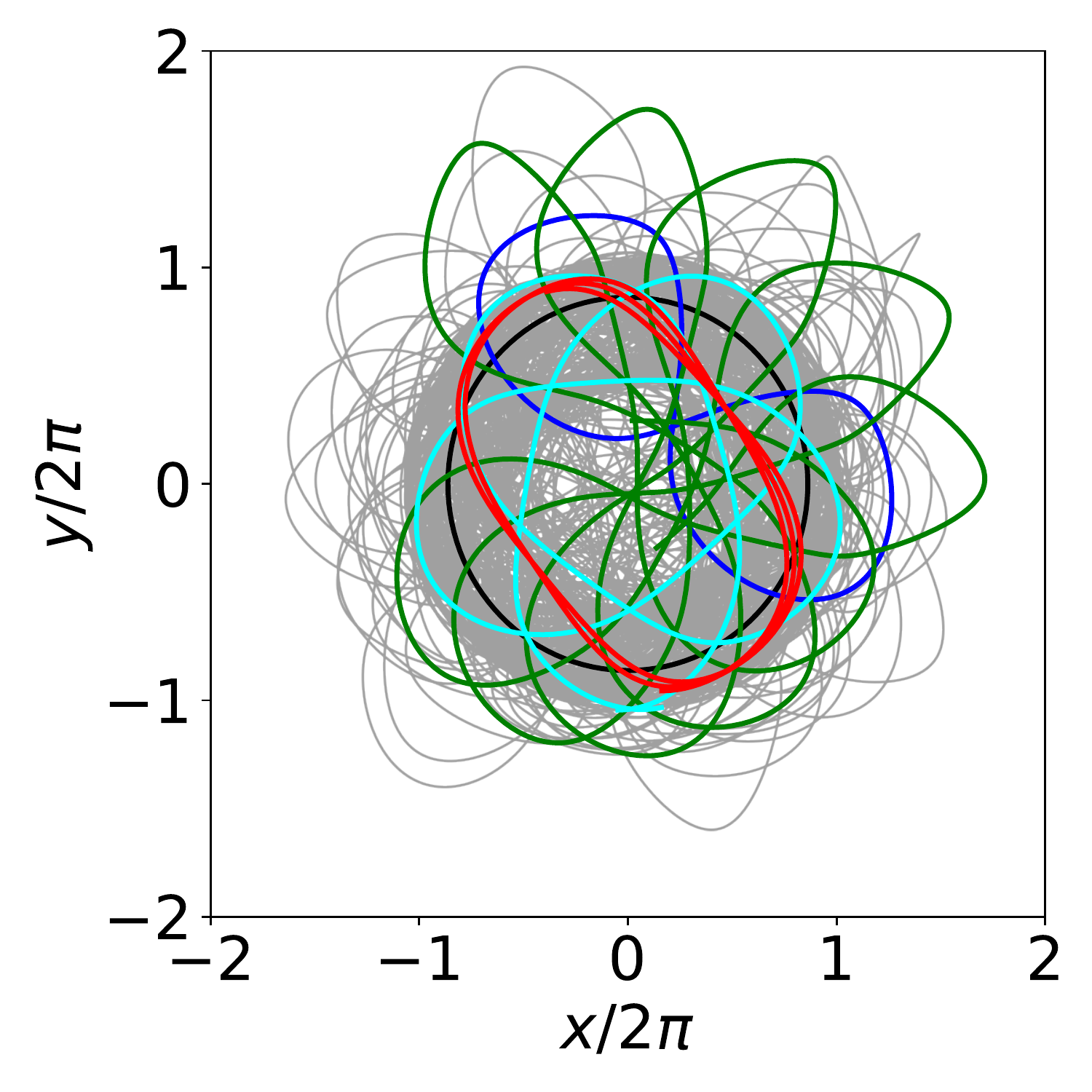}
		\put (0,2.0) {(a)}
	\end{overpic}                             
	\begin{overpic}[width=0.23\textwidth]{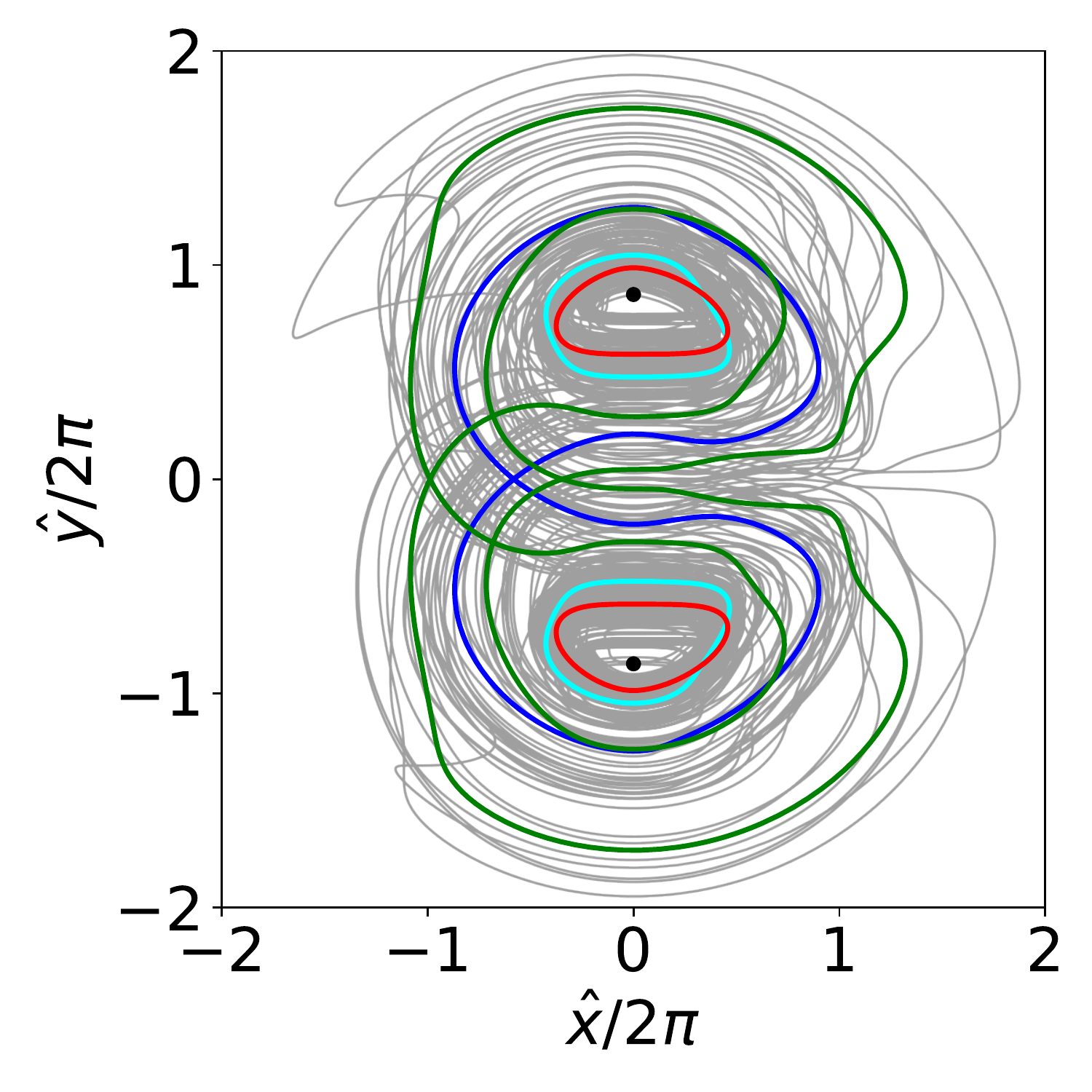}
		\put (0,2.0) {(b)}
	\end{overpic}\\
	\begin{overpic}[width=0.47\textwidth]{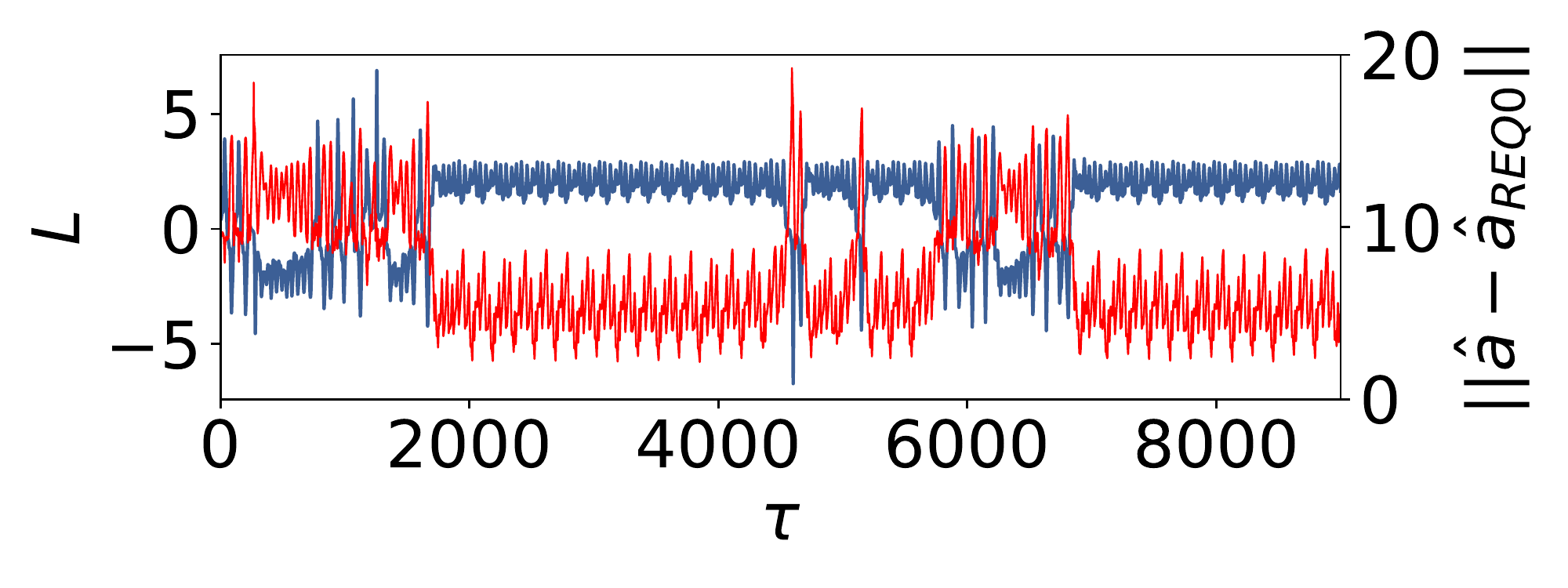}
		\put (0,2.0) {(c)}
	\end{overpic}
	\caption{
		Projections of the circular solution \REQ{0c} (black), 
		the (relative) periodic orbits 
		\RPO{1o} (red, oval), 
		\RPO{1t} (cyan, trefoil),
		\PO{2l} (blue, lemniscate), 
		\RPO{2l} (green, lemniscate), 
		and a long chaotic trajectory (gray)
		(a) in full state space onto the $(x, y)$-plane,
		(b) in the symmetry-reduced state space
		onto the $(\hat{x}, \hat{y})$-plane.
		$5$-repeats of each relative periodic orbit is shown in 
		(a). In (b), reflection-copies of the relative periodic
		orbits are also plotted with the same color.
		(c) The chaotically moving droplet's angular momentum (blue) 
		and its distance from $\REQ{0c}$(red) as functions of time.
		\label{f-chaos}
	}
\end{figure}

Finally, we are going to present the chaotic attractor following the
symmetry-increasing global bifurcation at $\memory \approx 18.455$. 
After this point, two chaotic regions \CHAOS{o} and
$\sigma \CHAOS{o}$ are connected through \CHAOS{l} and the 
trajectories intermittently switch between these neighborhoods. 
We illustrate this dynamics in \reffig{f-chaos} where we plotted a 
long chaotic trajectory along with \REQ{0c}, \RPO{1o}, 
\RPO{1t}, \PO{2l}, \RPO{2l} in the full state space 
(\reffig{f-chaos} (a)) and in the symmetry-reduced state space 
(\reffig{f-chaos} (b)) at $\memory = 18.5$.

The chaotic attractor in \reffig{f-chaos} (b), which is revealed 
after the symmetry-reduction, is qualitatively similar to the 
well-known Lorenz attractor \rf{lorenz63}. Akin to the 
counterclockwise-rotating circular solution \REQ{0c} and 
its reflection-symmetry $\sigma \REQ{0c}$, the Lorenz system at 
standard parameter values has two non-trivial equilibria that are
related by a rotation by $\pi$ and the dynamics chaotically switches 
between their neighborhoods. In the droplet system, the switchings 
between the neighborhoods of the circular solutions correspond to 
the reversal of the droplet's angular momentum. We illustrated this 
in the time-series of \reffig{f-chaos}(c), where we show the 
chaotically moving droplet's angular momentum and its distance 
from $\REQ{0c}$ as functions of time. Since $\REQ{0c}$'s angular 
momentum is positive (counterclockwise rotation), when the 
trajectory is close to it, the droplet's angular momentum is positive and
vice versa. This, in fact, is also visible from the projection of \reffig{f-chaos}(b). Since the symmetry reduction 
\refeq{e-velPolar} sets $\hat{v}_y = 0$ and 
$\hat{v}_x > 0$, the angular momentum in 
the symmetry-reduced state space is 
$L = - \hat{y} \hat{v}_x$ and its sign is the opposite of 
$\hat{y}$'s. Hence, the $\hat{y} = 0$ hyperplane separates solutions
with positive angular momenta from those with negative angular 
momenta in the symmetry-reduced state space. 

Besides the chaotic attractor, we also visualized the invariant 
solutions of the pilot-wave system in \reffig{f-chaos} (a,b). It is 
clear from \reffig{f-chaos}(b) that the oval (\RPO{1o}) and the 
trefoil (\RPO{1t}) solutions' angular momenta does not change sign 
along one period whereas the angular momenta of the lemniscates
reverse twice in one full period. The chaotic trajectories' 
angular momentum reversals correspond to the episodes during which
they are shadowed by the lemniscates.

\section{Conclusion and discussion}
\label{s-conclusion}

In this paper, we introduced a continuous symmetry-reduction method 
for hydrodynamic pilot-wave systems with rotational symmetry. In 
essence, our method is fixing the polar-angle in the  
velocity plane $(v_x, v_y)$ (as opposed to $(x, y)$-plane) in order to
obtain a transformation that is nonsingular for generic dynamics. 
We formulated this transformation in such a way that 
it can be used to simplify both experimental 
and numerical data by eliminating the symmetry degeneracy of the 
solutions. We then proceeded to reformulate this transformation in the 
framework of the method of slices. This formulation brought us a 
set of geometrical tools, which are useful mainly for theoretical and 
computational undertakings. In order to demonstrate the utility of 
the introduced symmetry-reduction technique, we applied it to the 
numerical study of Oza \etal's \rf{ORB2013} trajectory equation 
with central harmonic forcing. With the help of the 
symmetry-reduction, we analyzed the bifurcations and 
the subsequent chaotic dynamics in this system. 

The main message we would like to deliver is that the 
continuous symmetry reduction substantially simplifies the pilot-wave 
 dynamics by eliminating infinitely many copies of each 
generic solution. This is perhaps best illustrated in 
\reffig{f-chaos}, where the symmetry-reduction reveals a relatively 
simple chaotic attractor with qualitatively different regions. \\
Moreover, the symmetry reduction allowed us to numerically identify 
the unstable 
invariant solutions and their roles in organizing the state space
of the system. 

In the previous experimental\rf{PLMFC2014,PLFC2014} 
and numerical\rf{KOB2017} 
studies of hydrodynamic pilot-wave systems with central 
harmonic forcing, invariant solutions such as circles, ovals, 
lemniscates, and trefoils, were observed by tuning system 
parameters to the values at which these solutions were stable. 
These papers also demonstrated that the chaotic solutions 
transiently resembled the invariant solutions, hence they 
conjectured that in the chaotic regime, the dynamics can be 
decomposed into episodes approximated by different 
invariant solutions. Our results essentially confirm this 
conjecture: The chaotic pilot-wave dynamics is a union of two 
chaotic state space regions that are formed in the vicinity of 
the oval and lemniscate solutions and these solutions persist 
in the chaotic regime, albeit they are unstable. 
This observation can potentially be brought to a quantitative 
level: Long-term averages of the observables associated with a 
chaotic system can be predicted from its periodic 
orbits if the system and periodic orbits satisfy certain 
properties such as ergodicity and hyperbolicity. The collection of 
methods to carry out such computations is known as the periodic 
orbit theory\rf{DasBuch,Cvitanovic1992} and 
our results suggest that such a calculation can be done in this
context. Hence, the periodic orbit analysis of chaotic 
pilot-wave systems is one of our future research directions.

Symmetries of the radially-confined pilot-wave systems were recognized in previous literature at varying 
levels. For example, Perrard et al.\rf{PLMFC2014} noted that the 
symmetries of the droplet trajectories were reflected in the 
Fourier-Bessel decomposition of the bath's surface height. 
Durey \& Milewski\rf{DurMil2017} observed that the perfect 
lemniscate solution had $0$ mean angular momentum due to its 
reflection symmetry. Durey \& Milewski\rf{DurMil2017} also 
used successive maxima of radius as the averaging window in their 
data analysis. Similarly, Perrard et al.\rf{PLFC2014} used 
successive maxima of the radius in order to construct a return map.
The choice of radius as a diagnostic variable is inherently 
symmetry-related since the radius is an invariant of rotation 
and reflection. The main difference of our work against the previous
techniques is that we do not only use symmetry-invariant variables 
for extracting information but also express the 
dynamics in terms of them. Thus, our symmetry-reduced representation
retains all dynamical information.

Transition to chaos in the model studied here were also examined by 
Tambasco \etal \rf{THORB2016} who conjectured that the 
chaotic dynamics appear following a Ruelle-Takens-Newhouse 
transition scenario\rf{nhouse78}. Their conjecture was based on 
the observation of a three-frequency orbit (fig. 6d of 
\refref{THORB2016}) prior to chaos. However, as noted by the 
authors themselves, two frequencies $f_1$ and $f_2$ that are 
visible in the spectrum of this orbit are in fact commensurate, 
\ie\ $f_1 = 4 f_2$. Thus, even if the third frequency is 
incommensurate with the first two, the associated symmetry-invariant
dynamics takes place on a two-torus which does not meet the
necessary criterion for a Ruelle-Takens-Newhouse transition 
scenario\rf{nhouse78}. 
Here, we found that the transition from the stable 
period-4 orbit $\RPO{4os}$ to chaos 
is a result of a period-doubling cascade, through 
which a new commensurate frequency appears at each 
bifurcation. 
We would like to note that the numerical method\rf{OWHRB2014} 
adopted in \refref{THORB2016} is very different from ours: 
They integrate the integro-differential trajectory equation 
\refeq{e-TrajectoryOza} with \refeq{e-FieldInt} using
an explicit fourth-order Adams-Bashfort method with a fixed 
time-step. 
We, on the other hand, simulate a truncated ODE 
representation. In order to check our results' robustness against
truncation at this regime, we quadrupled number of Fourier-Bessel
modes and confirmed that we still observe period-doubled orbits
prior to the chaos.

While we focused on pilot-wave systems in this paper,
our symmetry-related methods can be extended to a larger class of 
systems in which a finite-size (or point-like) object interacts with 
a continuous environment. Similarly, it could be possible to 
generalize our method to three-dimensional space by introducing Euler
angles for polar coordinate transformations. 

In summary, we have shown that the continuous symmetry reduction 
drastically simplifies the data generated by a hydrodynamic
pilot-wave system with chaotic dynamics. To this end, we formulated
a symmetry-reduction method and applied it to the numerical 
simulations of a rotation-equivariant pilot-wave model with a 
central harmonic force. We identified
local bifurcations, the onset of chaos, and the global attractor 
merging bifurcations in the system; as well as a qualitative 
description of its state space geometry.

\section*{Acknowledgement}

We acknowledge stimulating discussions with 
Predrag Cvitanovi\'c and Matthieu Labousse.
We are grateful to George Choueiri for his critical 
	reading of an early version of this manuscript.

\appendix
      \section{Newton's method}
	  \label{s-Newton}

Let $(\sspRed_{\RPO{}}^{(i)}, T_{\RPO{}}^{(i)}$ be
the guesses for an initial condition on a 
(relative) periodic orbit and its period, which
approximately satisfy \refeq{e-rpoRed}. In order 
to bring them to a desired numerical precision, 
we utilize Newton's method as follows. Our goal
is to find the small corrections 
$(\Delta \sspRed, \Delta T)$ which yield
\beq
\sspRed_{\RPO{}}^{(i)} + \Delta \sspRed 
= \flowRed{T_{\RPO{}} + \Delta T^{(i)}}{
		  \sspRed_{\RPO{}}^i + \Delta \sspRed} \,. 
\label{e-Newton1}
\eeq
Expanding \refeq{e-Newton1} to the linear order 
in $\Delta \sspRed$ and $\Delta T$, we obtain
\bea
\sspRed_{\RPO{}}^{(i)} + \Delta \sspRed 
&\approx& \flowRed{T_{\RPO{}}^{(i)}}{\sspRed_{\RPO{}}^{(i)}} 
+ \hat{J}^{T_{\RPO{}}^{(i)}}(\sspRed_{\RPO{}}^{(i)}) \Delta \sspRed 
\continue
&+& \velRed(\flowRed{T_{\RPO{}}^{(i)}}{\sspRed_{\RPO{}}^{(i)}})
\Delta T \label{e-Newton2}
\,. 	
\eea
Defining 
$\hat{J}^{(i)} \equiv \hat{J}^{T_{\RPO{}}^{(i)}}(\sspRed_{\RPO{}}^{(i)}$), 
$\velRed^{(i)} \equiv \velRed(\flowRed{T_{\RPO{}}^{(i)}}{\sspRed_{\RPO{}}^{(i)}})$,
and 
$E^{(i)} = \flowRed{T_{\RPO{}}^{(i)}}{\sspRed_{\RPO{}}^{(i)}} - \sspRed_{\RPO{}}^{(i)} $ 
, and rearranging the terms in \refeq{e-Newton2}, we obtain 
the linear system of equations
\beq
(\mathrm{I} - \hat{J}^{(i)})
\Delta \sspRed^{(i)} 
- \velRed^{(i)} \Delta T^{(i)} 
= E^{(i)}\, \label{e-Newton3}
\eeq
for the approximate corrections 
$(\Delta \sspRed^{(i)}, \Delta T^{(i)}))$. 
\refeq{e-Newton3} provides $d$ equations for 
$d+1$ unknowns, where $d$ is the dimension of the
symmetry-reduced state space. We supplement 
\refeq{e-Newton3} with the constraint 
\beq
	\inprod{\velRed(\sspRed_{\RPO{}}^i)}{\Delta \sspRed^i}
	= 0, \label{e-Newton4}
\eeq
which disallows for updating our guesses in the flow direction.
We can now solve \refeq{e-Newton3} and \refeq{e-Newton4} for
$(\Delta \sspRed^i, \Delta T^i)$ and update our guesses as
\bea
	\sspRed_{\RPO{}}^{i+1} &=& \sspRed_{\RPO{}}^{i} 
	+ \alpha^k \Delta \sspRed^i \,, 
	  \continue
	T_{\RPO{}}^{i+1} &=& T_{\RPO{}}^{i+1} + \alpha^k \Delta T^i \,,
\eea
where we introduced the step size $\alpha^k \in (0, 1]$. 
We set $\alpha = 0.5$ and choose the smallest 
$k = 0, 1, 2 \ldots$ such that $||E_{i + 1}|| < ||E_i||$.
We iterate this procedure until the error $|| E_{i+1} ||$ 
is less than a tolerance $\epsilon$. 

At each Newton step, we sample the relative periodic orbit with
$\lfloor T_{\RPO{}}^{i} / \delta \zeit \rfloor$ steps in time, 
hence the final converged time-step $\delta \zeit_{\RPO{}}$ 
of the periodic orbit is close to the rest of the simulations.

\section{Edge-tracking algorithm}
\label{s-EdgeTrack}
Let $\langle L \rangle (\zeit)$ denote the moving 
average of the angular momentum with a time-window $\zeit_w$ and 
$\sspRed_{1}$ and $\sspRed_{2}$ be two state vectors that belong
to attractors $\mathcal{M}_1$ and $\mathcal{M}_2$, respectively. 
We assume that 
$\mathcal{M}_1$ and $\mathcal{M}_2$ have mean 
angular momenta $L_1$ and $L_2$ respectively such that 
$L_1 > L_{th, 1}$ and $L_2 < L_{th, 2}$ for some 
thresholds $L_{th, 1}$ and $L_{th, 2}$.
We iterate the following algorithm in order to
locate the edge that separates 
$\mathcal{M}_1$ and $\mathcal{M}_2$'s basins of attraction. 
\begin{algorithmic}
	\While {$||\sspRed_1 - \sspRed_2|| > \epsilon$}
	\State $\sspRed_3 \gets (\sspRed_{0} + \sspRed_{1}) / 2 $ 
	\State Set initial condition to $\sspRed_3$
	\State Compute $\sspRed (\zeit)$ and $\langle L \rangle (\zeit)$ 
	for $\zeit \in [0, \zeit_f]$ 
	\If {$\langle L \rangle (\zeit_f) > L_1$}
	\State $\sspRed_1 \gets \sspRed_3$
	\ElsIf {$\langle L \rangle (\zeit_f) < L_2$}
	\State $\sspRed_2 \gets \sspRed_3$	
	\Else
	\State Terminate
	\EndIf
	\EndWhile
\end{algorithmic}

\bibliography{walkers}{}

\begin{thebibliography}{10}

\bibitem{deBroglie1}
L.~de~Broglie.
\newblock Ondes et quanta.
\newblock {\em C. R.}, 177, 1923.

\bibitem{CFGB2005}
Y.~Couder, E.~Fort, C.-H. Gautier, and A.~Boudaoud.
\newblock From bouncing to floating: Noncoalescence of drops on a fluid bath.
\newblock {\em Phys. Rev. Lett.}, 94(17):177801, 2005.

\bibitem{CPFB2005}
Y.~Couder, S.~Proti\`ere, E.~Fort, and A.~Boudaoud.
\newblock Walking and orbiting droplets.
\newblock {\em Nature}, 437(7056):208, 2005.

\bibitem{CouderFort2006}
Y.~Couder and E.~Fort.
\newblock Single-particle diffraction and interference at a macroscopic scale.
\newblock {\em Phys. Rev. Lett.}, 97(15):154101, 2006.

\bibitem{AMRRLT2015}
A.~Andersen, J.~Madsen, C.~Reichelt, S.~Rosenlund Ahl, B.~Lautrup,
  C.~Ellegaard, M.~T. Levinsen, and T.~Bohr.
\newblock Double-slit experiment with single wave-driven particles and its
  relation to quantum mechanics.
\newblock {\em Phys. Rev. E}, 92:013006, 2015.

\bibitem{PHFB2018}
G.~Pucci, D.~M. Harris, L.~M. Faria, and J.~W.~M. Bush.
\newblock Walking droplets interacting with single and double slits.
\newblock {\em J. Fluid Mech.}, 835:1136–1156, 2018.

\bibitem{Bush2015}
J.~W.M. Bush.
\newblock Pilot-wave hydrodynamics.
\newblock {\em Annu. Rev. Fluid Mech.}, 47(1):269--292, 2015.

\bibitem{FEBMJC2010}
E.~Fort, A.~Eddi, A.~Boudaoud, J.~Moukhtar, and Y.~Couder.
\newblock Path-memory induced quantization of classical orbits.
\newblock {\em Proc. Natl. Acad. Sci.}, 107(41):17515--17520, 2010.

\bibitem{HarrisBush2014}
D.~M. Harris and J.~W.~M. Bush.
\newblock Droplets walking in a rotating frame: from quantized orbits to
  multimodal statistics.
\newblock {\em J. Fluid Mech.}, 739:444–464, 2014.

\bibitem{PLMFC2014}
S.~Perrard, M.~Labousse, M.~Miskin, E.~Fort, and Y.~Couder.
\newblock Self-organization into quantized eigenstates of a classical
  wave-driven particle.
\newblock {\em Nature communications}, 5, 2014.

\bibitem{PLFC2014}
S.~Perrard, M.~Labousse, E.~Fort, and Y.~Couder.
\newblock Chaos driven by interfering memory.
\newblock {\em Phys. Rev. Lett.}, 113(10):104101, 2014.

\bibitem{ORB2013}
A.~U. Oza, R.~R. Rosales, and J.~W.~M. Bush.
\newblock A trajectory equation for walking droplets: hydrodynamic pilot-wave
  theory.
\newblock {\em J. Fluid Mech.}, 737:552--570, 2013.

\bibitem{THORB2016}
L.~D. Tambasco, D.~M. Harris, A.~U. Oza, R.~R. Rosales, and J.~W.~M. Bush.
\newblock The onset of chaos in orbital pilot-wave dynamics.
\newblock {\em Chaos: An Interdisciplinary Journal of Nonlinear Science},
  26(10):103107, 2016.

\bibitem{Luce95}
B.~P. Luce.
\newblock Homoclinic explosions in the complex {Ginzburg-Landau} equation.
\newblock {\em Physica D}, 84:553--581, 1995.

\bibitem{SCD07}
P.~Cvitanovi{\'c}, R.~L. Davidchack, and E.~Siminos.
\newblock On the state space geometry of the {Kuramoto-Sivashinsky} flow in a
  periodic domain.
\newblock {\em SIAM J. Appl. Dyn. Syst.}, 9:1--33, 2009.

\bibitem{BudCvi15}
N.~B. Budanur and P.~Cvitanovi\'c.
\newblock Unstable manifolds of relative periodic orbits in the
  symmetry-reduced state space of the {Kuramoto-Sivashinsky} system.
\newblock {\em J. Stat. Phys.}, 167(3):636--655, 2017.

\bibitem{ACHKW11}
A.~P. Willis, P.~Cvitanovi{\'c}, and M.~Avila.
\newblock Revealing the state space of turbulent pipe flow by symmetry
  reduction.
\newblock {\em J. Fluid Mech.}, 721:514--540, 2013.

\bibitem{WFSBC15}
N.~B. Budanur, K.~Y. Short, M.~Farazmand, A.~P. Willis, and P.~Cvitanovi{\'c}.
\newblock Relative periodic orbits form the backbone of turbulent pipe flow.
\newblock {\em J. Fluid Mech}, 833:274--301, 2017.

\bibitem{BudCvi14}
N.~B. Budanur, P.~Cvitanovi\'c, R.~L. Davidchack, and E.~Siminos.
\newblock Reduction of the {SO(2)} symmetry for spatially extended dynamical
  systems.
\newblock {\em Phys. Rev. Lett.}, 114:084102, 2015.

\bibitem{ChossLaut00}
P.~Chossat and R.~Lauterbach.
\newblock {\em Methods in Equivariant Bifurcations and Dynamical Systems}.
\newblock World Scientific, Singapore, 2000.

\bibitem{Faraz15}
M.~Farazmand.
\newblock An adjoint-based approach for finding invariant solutions of
  {Navier-Stokes} equations.
\newblock {\em J. Fluid M.}, 795:278--312, 2016.

\bibitem{BudHof17}
N.~B. Budanur and B.~Hof.
\newblock Heteroclinic path to spatially localized chaos in pipe flow.
\newblock {\em J. Fluid Mech}, 827, R1, 2017.

\bibitem{BudHof18}
N.~B. Budanur and B.~Hof.
\newblock Complexity of the laminar-turbulent boundary in pipe flow.
\newblock {\em Phys. Rev. Fluids}, 3:054401, 2018.

\bibitem{DasBuch}
P.~Cvitanovi{\'c}, R.~Artuso, R.~Mainieri, G.~Tanner, and G.~Vattay.
\newblock {\em Chaos: Classical and Quantum}.
\newblock Niels Bohr Inst., Copenhagen, 2016.

\bibitem{rowley_reconstruction_2000}
C.~W. Rowley and J.~E. Marsden.
\newblock Reconstruction equations and the {Karhunen-Lo{\'e}ve} expansion for
  systems with symmetry.
\newblock {\em Physica D}, 142:1--19, 2000.

\bibitem{MB2013a}
J.~Mol{\'a}{\v{c}}ek and J.~W.~M. Bush.
\newblock Drops bouncing on a vibrating bath.
\newblock {\em J. Fluid Mech.}, 727:582--611, 2013.

\bibitem{LOPB2016}
M.~Labousse, A.~U. Oza, S.~Perrard, and J.~W.~M. Bush.
\newblock Pilot-wave dynamics in a harmonic potential: Quantization and
  stability of circular orbits.
\newblock {\em Phys. Rev. E}, 93:033122, Mar 2016.

\bibitem{KOB2017}
K.~M. Kurianski, A.~U. Oza, and J.~W.~M. Bush.
\newblock Simulations of pilot-wave dynamics in a simple harmonic potential.
\newblock {\em Phys. Rev. Fluids}, 2:113602, Nov 2017.

\bibitem{labousse2014thesis}
Matthieu Labousse.
\newblock {\em Etude d'une dynamique \`{a} m\'{e}moire de chemin: une
  exp\'erimentation th\'eorique}.
\newblock PhD thesis, Universit\'{e} Paris 6, 2014.

\bibitem{PerrardLabousse2018}
S.~Perrard and M.~Labousse.
\newblock Transition to chaos in wave memory dynamics in a harmonic well:
  Deterministic and noise-driven behavior.
\newblock {\em Chaos: An Interdisciplinary Journal of Nonlinear Science},
  28(9):096109, 2018.

\bibitem{watson1944}
G.~N. Watson.
\newblock {\em A Treatise on the Theory of Bessel Functions}.
\newblock Cambridge University Press, 1944.

\bibitem{scipy}
E.~Jones, T.~Oliphant, P.~Peterson, et~al.
\newblock {SciPy}: {Open} source scientific tools for {Python}, 2001.

\bibitem{Kuznetsov2010}
Y.~Kuznetsov.
\newblock {\em Elements of Applied Bifurcation Theory}.
\newblock Applied Mathematical Sciences. Springer New York, 2010.

\bibitem{DKK1991}
E.~Doedel, H.~B. Keller, and J.~P. Kernevez.
\newblock Numerical analysis and control of bifurcation problems (i):
  Bifurcation in finite dimensions.
\newblock {\em Int. J. Bifurcat. Chaos}, 01(03):493--520, 1991.

\bibitem{SwiWie84}
J.~W. Swift and K.~Wiesenfeld.
\newblock Suppression of period doubling in symmetric systems.
\newblock {\em Phys. Rev. Lett.}, 52:705--708, 1984.

\bibitem{GOY1982}
C.~Grebogi, E.~Ott, and J.~A. Yorke.
\newblock Chaotic attractors in crisis.
\newblock {\em Phys. Rev. Lett.}, 48:1507--1510, May 1982.

\bibitem{GOY1983}
C.~Grebogi, E.~Ott, and J.~A. Yorke.
\newblock Crises, sudden changes in chaotic attractors, and transient chaos.
\newblock {\em Physica D}, 7(1):181--200, 1983.

\bibitem{ASY1997ch10}
K.~T. Alligood, T.~D. Sauer, and J.~A. Yorke.
\newblock {\em Stable Manifolds and Crises}, pages 399--445.
\newblock Springer Berlin Heidelberg, Berlin, Heidelberg, 1997.

\bibitem{IT01}
T.~Itano and S.~Toh.
\newblock The dynamics of bursting process in wall turbulence.
\newblock {\em J. Phys. Soc. Japan}, 70:701--714, 2001.

\bibitem{TI03}
S.~Toh and T.~Itano.
\newblock A periodic-like solution in channel flow.
\newblock {\em J. Fluid Mech.}, 481:67--76, 2003.

\bibitem{SchEckYor07}
T.~M. Schneider, B.~Eckhardt, and J.~Yorke.
\newblock Turbulence, transition, and the edge of chaos in pipe flow.
\newblock {\em Phys. Rev. Lett.}, 99:034502, 2007.

\bibitem{SGLDE08}
T.~M. Schneider, J.~F. Gibson, M.~Lagha, F.~De Lillo, and B.~Eckhardt.
\newblock Laminar-turbulent boundary in plane {Couette} flow.
\newblock {\em Phys. Rev. E.}, 78:037301, 2008.

\bibitem{MMSE09}
F.~Mellibovsky, A.~Meseguer, T.~M. Schneider, and B.~Eckhardt.
\newblock Transition in localized pipe flow turbulence.
\newblock {\em Phys. Rev. Lett.}, 103:054502, Jul 2009.

\bibitem{ScMaEc10}
T.~M. Schneider, D.~Marinc, and B.~Eckhardt.
\newblock Localized edge states nucleate turbulence in extended plane {Couette}
  cells.
\newblock {\em J. Fluid Mech.}, 646:441--451, 2010.

\bibitem{ZamEck14a}
S.~Zammert and B.~Eckhardt.
\newblock A spotlike edge state in plane {Poiseuille} flow.
\newblock {\em PAMM}, 14:591--592, 2014.

\bibitem{KKSDEH2016}
T.~Khapko, T.~Kreilos, P.~Schlatter, Y.~Duguet, B.~Eckhardt, and D.~S.
  Henningson.
\newblock Edge states as mediators of bypass transition in boundary-layer
  flows.
\newblock {\em J. Fluid Mech.}, 801, 2016.

\bibitem{ChoGol1988}
P.~Chossat and M.~Golubitsky.
\newblock Symmetry-increasing bifurcation of chaotic attractors.
\newblock {\em Physica D}, 32(3):423 -- 436, 1988.

\bibitem{nhouse78}
S.~E. Newhouse, D.~Ruelle, and F.~Takens.
\newblock Occurrence of strange {A}xiom {A} attractors near quasi-periodic
  flows on ${T}^m(m=3$ or more).
\newblock {\em Commun. Math. Phys.}, 64:35, 1978.

\bibitem{strogb}
S.~H. Strogatz.
\newblock {\em Nonlinear Dynamics and Chaos}.
\newblock Perseus Books, Cambridge, MA, 2000.

\bibitem{lorenz63}
E.~N. Lorenz.
\newblock Deterministic nonperiodic flow.
\newblock {\em J. Atmos. Sci.}, 20:130--141, 1963.

\bibitem{Cvitanovic1992}
P.~Cvitanovi{\'c}.
\newblock Periodic orbit theory in classical and quantum mechanics.
\newblock {\em Chaos}, 2(1):1--4, 1992.

\bibitem{DurMil2017}
M.~Durey and P.~A. Milewski.
\newblock Faraday wave-droplet dynamics: discrete-time analysis.
\newblock {\em J. Fluid Mech.}, 821:296--329, 2017.

\bibitem{OWHRB2014}
A.~U. Oza, {\O}.~Wind-Willassen, D.~M. Harris, R.~R. Rosales, and J.~W.~M.
  Bush.
\newblock Pilot-wave hydrodynamics in a rotating frame: Exotic orbits.
\newblock {\em Phys. Fluids}, 26(8):082101, 2014.

\end{thebibliography}
\bibliographystyle{unsrt}

\end{document}